\documentclass[sensors,article,accept,pdftex,moreauthors]{Definitions/mdpi} 
\renewcommand\hl[1]{#1}
\firstpage{1} 
\pubvolume{1}
\issuenum{1}
\articlenumber{0}
\pubyear{2025}
\copyrightyear{2025}
\externaleditor{~}
\datereceived{21 November 2024} 
\daterevised{10 January 2025} 
\dateaccepted{22 January 2025} 
\datepublished{28 January 2025} 
\hreflink{https://doi.org/} 
\Title{Commissioning An All-Sky Infrared Camera Array for Detection of Airborne Objects}

\TitleCitation{Commissioning an All-Sky Infrared Camera Array for Detection of Airborne Objects}

\Author{Laura Domine $^{1,2,}$*,  
Ankit Biswas $^{2}$, Richard Cloete $^{1,2}$, Alex Delacroix $^{1,2}$, Andriy Fedorenko $^{2}$, Lucas Jacaruso $^{2}$, Ezra~Kelderman $^{2}$, Eric~Keto $^{1,2}$, Sarah Little $^{2,3,4}$, Abraham Loeb $^{1,2,\dagger}$, Eric~Masson~$^{2}$, Mike Prior $^{2}$, Forrest Schultz $^{2,5}$, Matthew Szenher $^{2}$, Wesley Andrés Watters $^{2,3}$, Abigail White $^{1,2}$}

\AuthorNames{Laura Domine, Ankit Biswas, Richard Cloete, Alex Delacroix, Andriy Fedorenko, Lucas Jacaruso, Ezra~Kelderman, Eric Keto, Sarah Little, Abraham Loeb, Eric Masson, Mike Prior, Forrest Schultz, Matthew Szenher, Wesley Andrés Watters and Abigail White}

\AuthorCitation{Domine, L.; Biswas, A.; Cloete, R.; Delacroix, A.; Fedorenko, A.; Jacaruso, L.; Kelderman, E.; Keto, E.; Little, S.; Loeb, A.; et al.}

\address{%
$^{1}$ \quad Harvard-Smithsonian Center for Astrophysics, 60 Garden Street, Cambridge, MA 02138, USA 
\\
$^{2}$ \quad Galileo Project, 
 60 Garden Street, Cambridge, MA 02138, USA \\
$^{3}$ \quad Whitin Observatory, Department 
 of Physics \& Astronomy, Wellesley College, 106 Central Street, 
 \mbox{Wellesley, MA 02481, USA} \\
$^{4}$ \quad Scientific Coalition for UAP Studies, Fort Myers, FL 33913, USA\\
$^{5}$ \quad Atlas Lens Co., Glendale, CA 91201, USA}

\corres{Correspondence: laura.domine@cfa.harvard.edu}

\firstnote{Head of the Galileo Project.}

\abstract{ To date, there is little publicly available scientific data on unidentified aerial phenomena (UAP) whose properties and kinematics purportedly reside outside the performance envelope of known phenomena. To address this deficiency, the Galileo Project is designing, building, and commissioning a multi-modal, multi-spectral ground-based observatory to continuously monitor the sky and collect data for UAP studies via a rigorous long-term aerial census of all aerial phenomena, including natural and human-made. One of the key instruments is an all-sky infrared camera array using eight uncooled long-wave-infrared FLIR Boson 640 cameras. In addition to performing intrinsic and thermal calibrations, we implement a novel extrinsic calibration method using airplane positions from Automatic Dependent Surveillance–Broadcast (ADS-B) data that we collect synchronously on site. Using a You Only Look Once (YOLO) machine learning model for object detection and the Simple Online and Realtime Tracking (SORT) algorithm for trajectory reconstruction, we establish a first baseline for the performance of the system over five months of field operation. Using an automatically generated real-world dataset derived from ADS-B data, a dataset of synthetic 3D trajectories, and a hand-labeled real-world dataset, we find an acceptance rate (fraction of in-range airplanes passing through the effective field of view of at least one camera that are recorded) of 41\% for ADS-B-equipped aircraft, and a mean frame-by-frame aircraft detection efficiency (fraction of recorded airplanes in individual frames which are successfully detected) of 36\%. The detection efficiency is heavily dependent on weather conditions, range, and aircraft size. Approximately 500,000 trajectories of various aerial objects are reconstructed from this five-month commissioning period. These trajectories are analyzed with a toy outlier search focused on the large sinuosity of apparent 2D reconstructed object trajectories. About 16\% of the trajectories are flagged as outliers and manually examined in the IR images. From these $\sim$80,000 outliers and 144 trajectories remain ambiguous, which are likely mundane objects but cannot be further elucidated at this stage of development without information about distance and kinematics or other sensor modalities. We demonstrate the application of a likelihood-based statistical test to evaluate the significance of this toy outlier analysis. Our observed count of ambiguous outliers combined with systematic uncertainties yields an upper limit of 18,271 outliers for the five-month interval at a 95\% confidence level. This test is applicable to all of our future outlier searches.}


\begin{document}


\section{Introduction}
There is currently very little publicly available scientific-grade data on unidentified aerial phenomena (UAP) whose properties and kinematics purportedly reside outside the performance envelope of known phenomena. Several U.S. federal agencies have weighed in on this issue.
The Office of the Director of National Intelligence (ODNI)'s annual UAP reports~\cite{odni2022, odni2023} have advocated since 2021 for bolstered data collection initiatives and increased resource allocation in order to scrutinize unexplained aerial phenomena that unambiguously pose aviation safety issues and potentially challenge U.S. national security.
In 2023, NASA released an independent study~\cite{nasa} which emphasized how studying the phenomena using passive sensing ``with multiple, well-calibrated sensors is paramount'' and that ``multispectral data'' need to be collected ``as part of a rigorous data acquisition campaign''. In addition, they suggested that ``machine learning algorithms could be incorporated to detect and analyze UAP in real-time.'' The Galileo Project 
is designing, building, and commissioning a passive, multi-modal, multi-spectral ground-based observatory to continuously monitor the sky and conduct an exhaustive observational long-term survey in search of measurable anomalous phenomena~\cite{watters2023scientific, mead2023multi, szenher2023hardware, cloete2023integrated, randall2023skywatch}. This long-term field observation effort fits many recommendations from the NASA study and represents ``a complex undertaking whose outcome could allow for substantial and systematic gathering of UAP data as well as a robust characterization of the background''~\cite{nasa}.

We have designed and built the first observatory at our development site in Massachusetts~\cite{watters2023scientific}, and we are now commissioning instruments for each of the sensor modalities separately (optical sensors in the infrared, visible, and ultraviolet; acoustic; radio spectrum; magnetic field strength; charged particle count; and weather) and in combination to validate and benchmark their performance characteristics. Once commissioned, we will commence collecting scientific-grade data and begin to quantitatively identify classes of objects and statistical outliers that are corroborated over time. We can generate testable hypotheses to account for any novel class discovered in this way, which, after further investigation and possible instrument refinements, may result in the discovery of scientific anomalies, i.e., statistical outliers that resist explanation in terms of prevailing scientific beliefs,
representing objects and phenomena that are unknown to science~\cite{watters2023scientific}. We can also derive upper bounds for the occurrence rate of outliers, novel classes, and anomalies at our development site. As the instruments of our observatory are commissioned, we will make copies and distribute them to additional sites. Such a network of observatories will enable us to monitor a larger volume of the sky. This will not only increase our likelihood of novel class discovery, but also open the possibility of scientific measurement of the UAP event rate and spatio-temporal distribution, which to date has never been achieved over significant time and spatial scales.

\subsection{Related Work}
For a review of relevant historical field work and related studies, we refer the reader to section~2.3 of our original design paper~\cite{watters2023scientific}, and to~\cite{ailleris2024exploring}. A variety of related contemporary work exists, but academic, instrumented, field efforts focused on UAP are rare. These include the automated, roof-top, and continuously recording system of optical and infrared cameras developed by the Interdisciplinary Research Centre for Extraterrestrial Studies at Julius-Maximilians-Universität Würzburg in Germany~\cite{vodopivec2018autonomous, kayal, kayal2019hyper}; a one week field study on and near Catalina Island in 2021, conducted by UAPx, led by a group of academic scientists from the University of Albany, using optical and thermal infrared cameras, and high-energy particle detectors~\cite{szydagis2023initial}; a long-term effort re-purposing data from existing aerial monitoring networks that is underway by ``Sigma2'' (the UAP technical commission of the learned society, Association Aeronautique et Astronautique de France), which makes use of active and passive bistatic radars, visible and thermal infrared cameras, optical and radio meteoroid surveillance networks (including FRIPON~\cite{fripon}, operated by the Institute of Celestial Mechanics and Ephemeris Calculation of the Paris Observatory), gravimeter and magnetometer networks, as well as IR and optical imaging satellites~\cite{sigma2}; and the search for fast unexplained transients in photographic plates in historic sky surveys~\cite{villarroel2021exploring}, also known as the Vanishing and Appearing Sources during a Century of Observations (VASCO) project, based at Stockholm University.

\subsection{Galileo Project Commissioning Approach}
Our observational effort is unique, with its high concentration of sensor modalities designed for long-term operation in outdoor field conditions. We require careful scientific calibration and commissioning to ensure a clear understanding of the reliability, operational environment, and detection volumes of individual instruments and combinations of instruments. Each instrument of our observatory goes through four distinct phases: development, integration and testing, commissioning, and science.
Commissioning is a final group of activities (generally performed in situ after deployment) that aims to bring a system into an operational condition by ensuring it meets key requirements and science expectations. Meeting basic system requirements is \hl{\textit{verification}}, while confirming science gathering expectations is \textit{\hl{validating}} the system. Commissioning activities include calibrations, performance characterizations, functionality checks, and data processing evaluations, and are undertaken for both instruments and infrastructure generally on a system-level basis, but may occur on subsystems or individual components as needed.  

Our approach is primarily driven by a ``top-down'' paradigm whereby our observatory goals (e.g., an all-sky survey, object triangulation, autonomous data processing with accurate object detection, remote autonomous operations, etc.) taken from our Science Traceability Matrix (STM)~\cite{watters2023scientific} will drive observation plans and then specific instrument and operational use cases. These will in turn form the basis of test cases and data collection campaigns that, when evaluated as successful, act as ``proof'' of performance. Additionally, observatory commissioning will be step-wise on an instrument-by-instrument 
basis (following the approach commonly used for NASA science missions), with the advantage of collecting scientific data as early as possible. Commissioning is also ``end-to-end'', from sensing to data product production, meaning that it encompasses the evaluation not only of sensor calibrations but also of final data processing pipelines and the accuracy and utility of the final data products.

Importantly, although we define the completion of commissioning as the demarcation between instrument development and testing vs. scientific data gathering operations, it is typically the case for scientific observatories that commissioning is just the beginning of a continuous improvement process as more is learned about how the instruments and systems perform under in situ operating conditions. Continued fine-tuning of instruments, improved data handling and processing, better algorithms for science data analysis, new and better data reduction products, better tools for performing remote observatory monitoring and control, etc., are all expected well after completion of initial planned commissioning.

 Our all-sky infrared camera array with eight uncooled long-wave-infrared FLIR Boson 640 cameras, nicknamed ``Dalek'', is the first instrument of our first observatory to enter the commissioning phase. This paper describes the commissioning process and results. In Section~\ref{sec_methods}, we describe the Dalek design, the camera calibrations, and the algorithms to detect and reconstruct object trajectories. In Section~\ref{sec_results}, we report on commissioning studies of the Dalek and the full system, including the detection pipeline. We also propose that UAP instrument-driven studies should adopt likelihood-based statistical tests to evaluate the significance of outlier searches. Section~\ref{sec:likelihood} showcases a \textit{\hl{toy}} outlier analysis, i.e., a deliberately simplistic outlier search which can be used to illustrate the application of likelihood-based statistical tests to future outlier searches. This simplified outlier analysis searches for reconstructed object trajectories with large sinuosity in our commissioning~dataset.

This paper describes the innovative commissioning steps developed and used to bring a new type of all-sky infrared camera that we designed~\cite{szenher2023hardware} into full scientific field operation in the search for aerial anomalies. These contributions can be applied to the commissioning of any all-sky infrared system, and include (a) the development of an innovative extrinsic calibration method for infrared cameras using ADS-B-equipped aircraft (stars are not imaged in infrared and the usual celestial methods cannot be applied); (b) the novel use of ADS-B-derived data to automatically establish the performance envelopes of an all-sky infrared camera array system, and as functions of atmospheric weather conditions, distance, and aircraft size; (c)  an aerial census of the sky above our test site over the commissioning period of five months, which (d) enables an evaluation of our detection and tracking pipeline on manually labeled real-world videos and connects the recorded dataset with environmental and instrumental effects; and (e) a demonstration of robust quantification of the uncertainties in our search for aerial anomalies via a generalizable likelihood-based method to measure the significance of any given outlier search.

\section{Materials and Methods}
\label{sec_methods}

The \textit{reconstruction} of an aerial object transiting in range of our observatory is defined as the process of identifying and characterizing the original object, its properties and kinematics, from the different sensors' signals and the multiple outputs of the signals analysis pipeline, which can include, for example, the reconstructed object's trajectory. This section aims to present our current methods for \textit{reconstructing} aerial objects using solely the infrared camera array, whose details are described in Section~\ref{sec:dalek_intro}. 
Section~\ref{sec:calibration} presents the different calibrations applied to the cameras, including intrinsic, extrinsic, and thermal calibration. Section~\ref{sec:yolo_sort} summarizes the reconstruction of aerial events, for which we use a mixture of machine learning and traditional object tracking algorithms, which are evaluated in this section. Also in Section~\ref{sec:yolo_sort}, we have summarized the datasets referenced throughout the paper for performance evaluations.

\subsection{The Dalek Infrared Camera Array}
\label{sec:dalek_intro}
\begin{figure}[H]
\begin{adjustwidth}{-\extralength}{0cm}
\centering
\includegraphics[width=10.8cm]{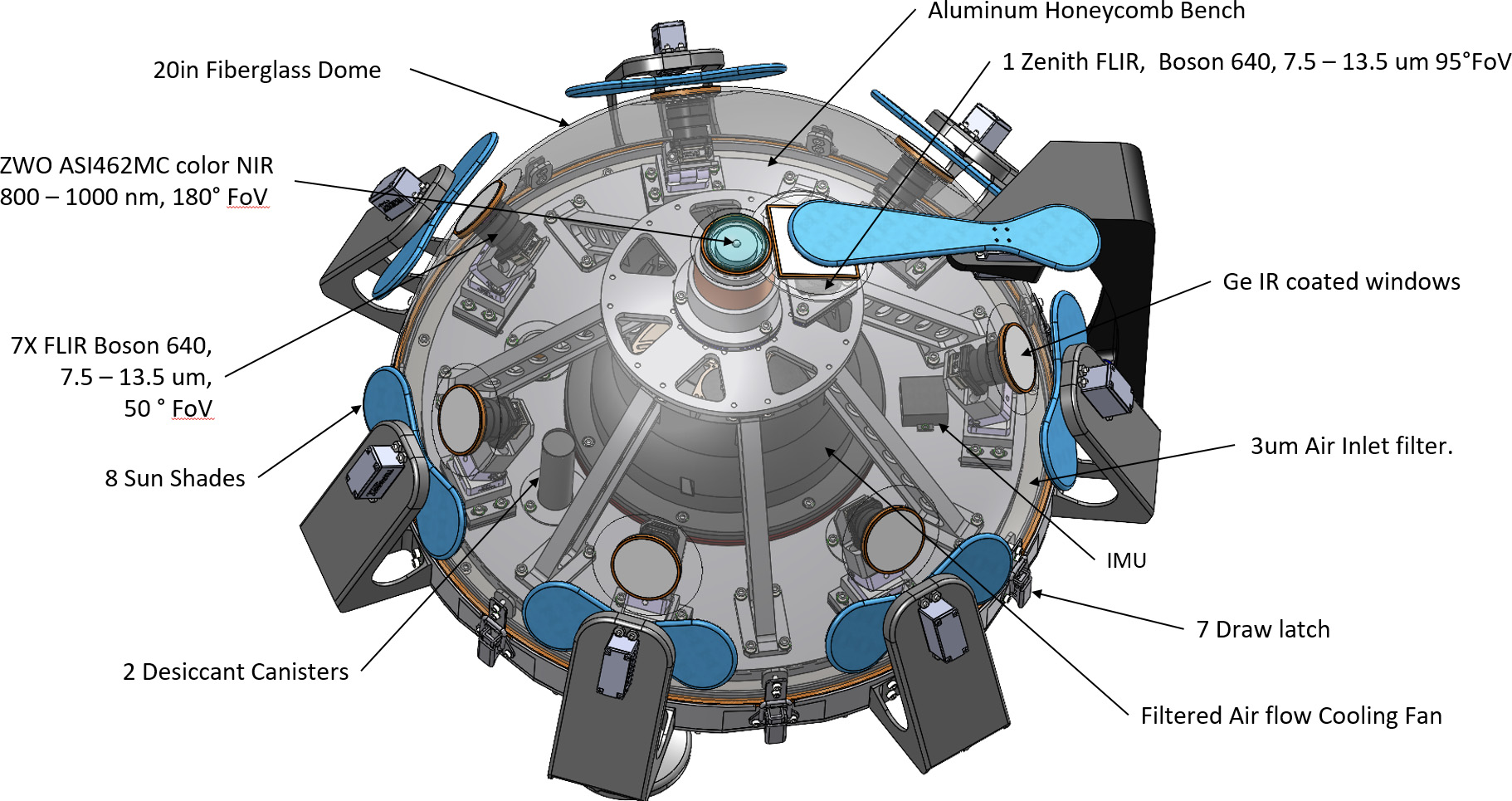}
\quad
\includegraphics[width=6.8cm]{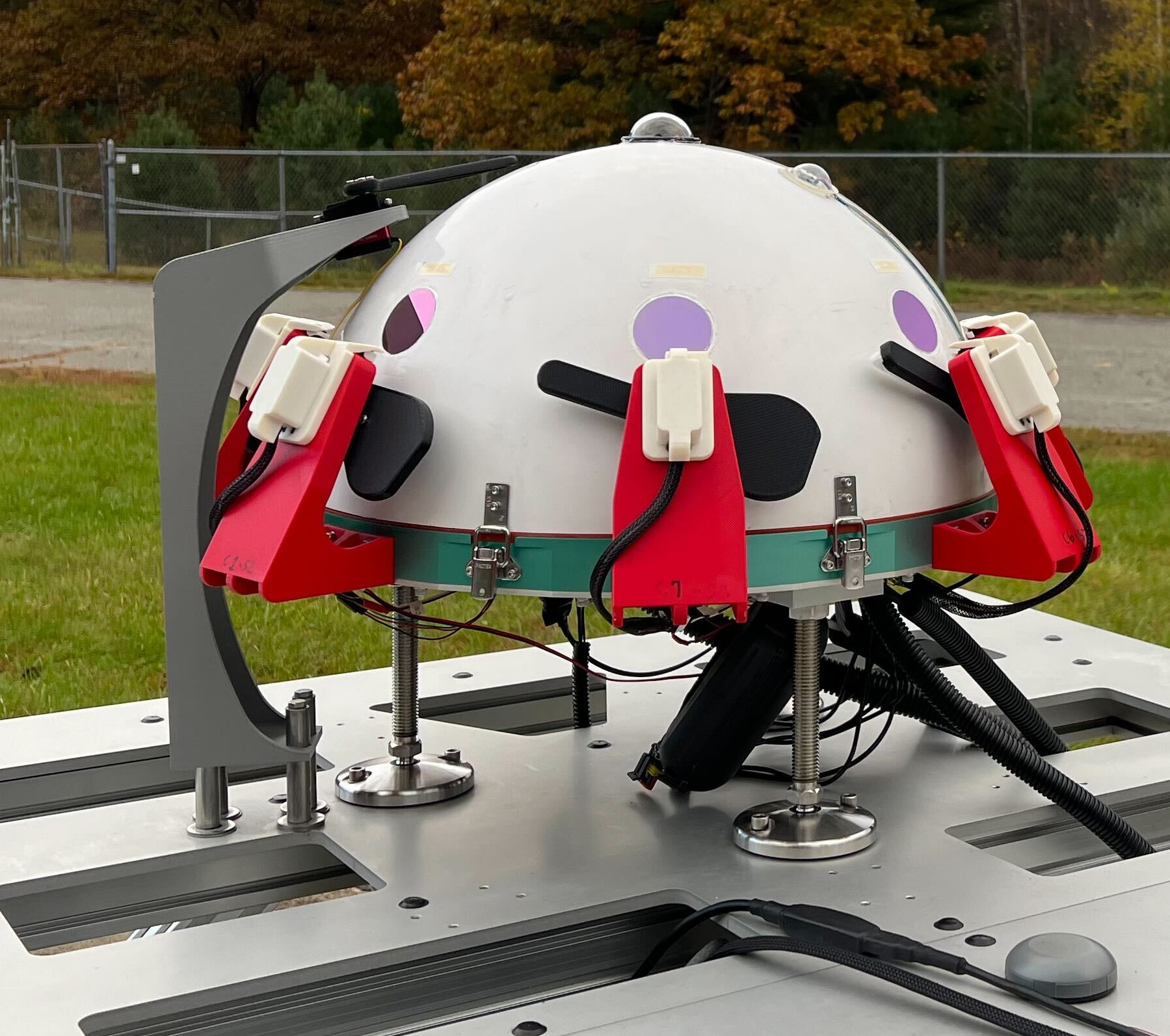}
\end{adjustwidth}
\caption{\hl{(\textbf{Left}): Mechanical} 
 design drawing of the Dalek IR camera array. (\textbf{Right}): Photograph of the Dalek as constructed at the development site.\label{fig:dalek}}
\end{figure}

One of the core modalities of the ground-based observatories is a suite of weatherproofed optical sensors which can be used to detect, track, and classify objects in the entire sky at multiple wavelengths: visible, infrared (IR), and ultraviolet (UV). One of the flagship instruments is a custom all-sky array of eight, long-wavelength-IR (LWIR) cameras. 
It is a hemispherical array of seven IR cameras with one additional IR camera pointing towards the zenith, which provide a 360$^{\circ}$ azimuth by +80$^{\circ}$ elevation view of the environment, as shown in Figure~\ref{fig:dalek}. This figure also shows a near-infrared camera ZWO ASI462MC which is built into the Dalek but is not yet being used, and thus is outside the scope of this paper. 

When designing the all-sky infrared array, we considered from four to fourteen cameras. The camera count of eight was selected to allow for high angular resolution, which translates into a larger detection volume, and for enhanced ability to resolve details, while balancing cost considerations. We are also developing a lower-cost and less complex four-camera array and we will compare its performance to the eight-camera array in the~future.

The prototype has been deployed to our development site in Massachusetts. This site is surrounded by forest and within 5 miles of a regional airport, which ensures a regular flow of airplanes above the Dalek that we can harness for both calibration and~commissioning.

\begin{figure}[H]
\begin{adjustwidth}{-\extralength}{0cm}
\centering
\includegraphics[width=10cm]{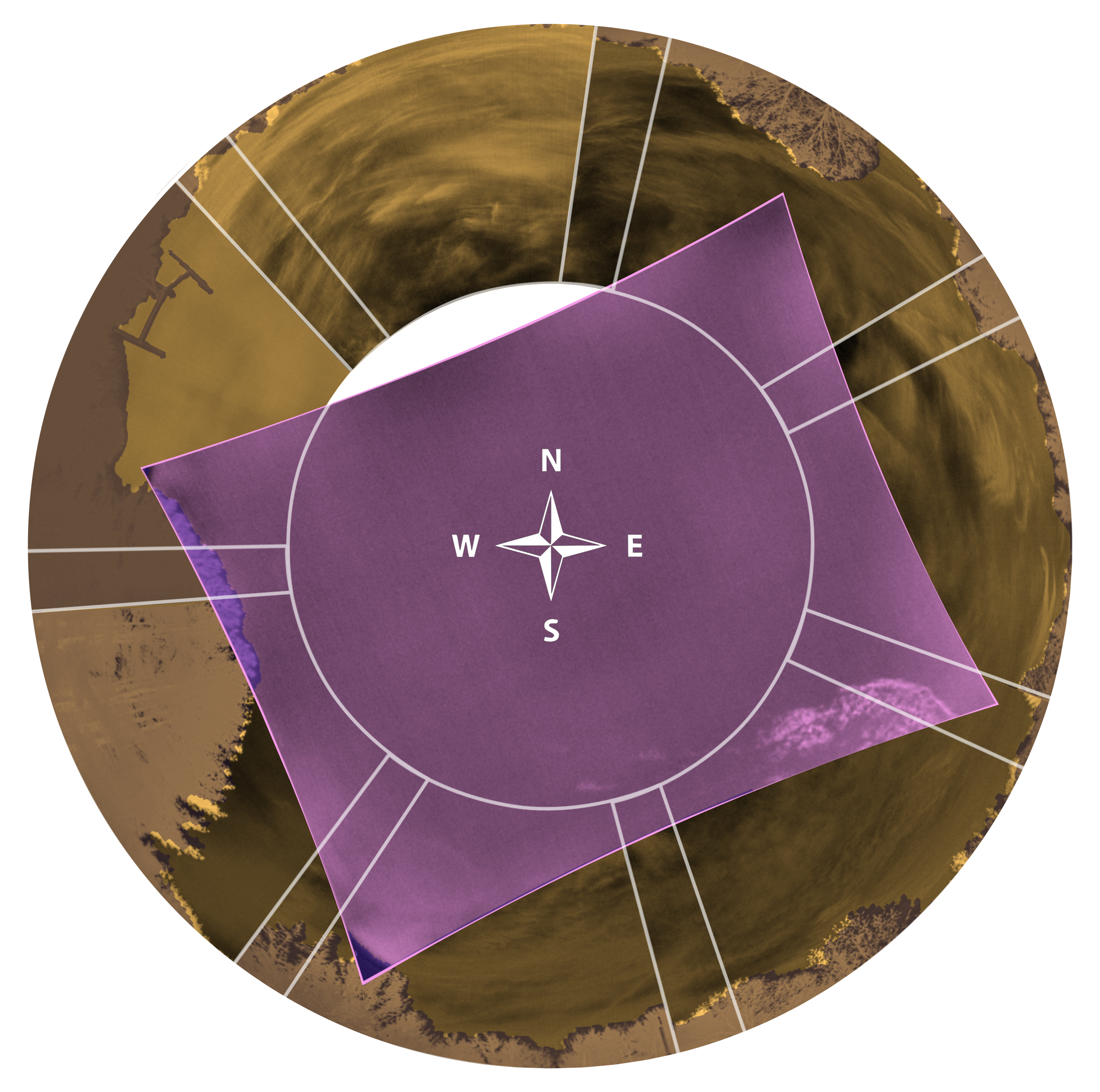}
\quad
\includegraphics[width=7cm]{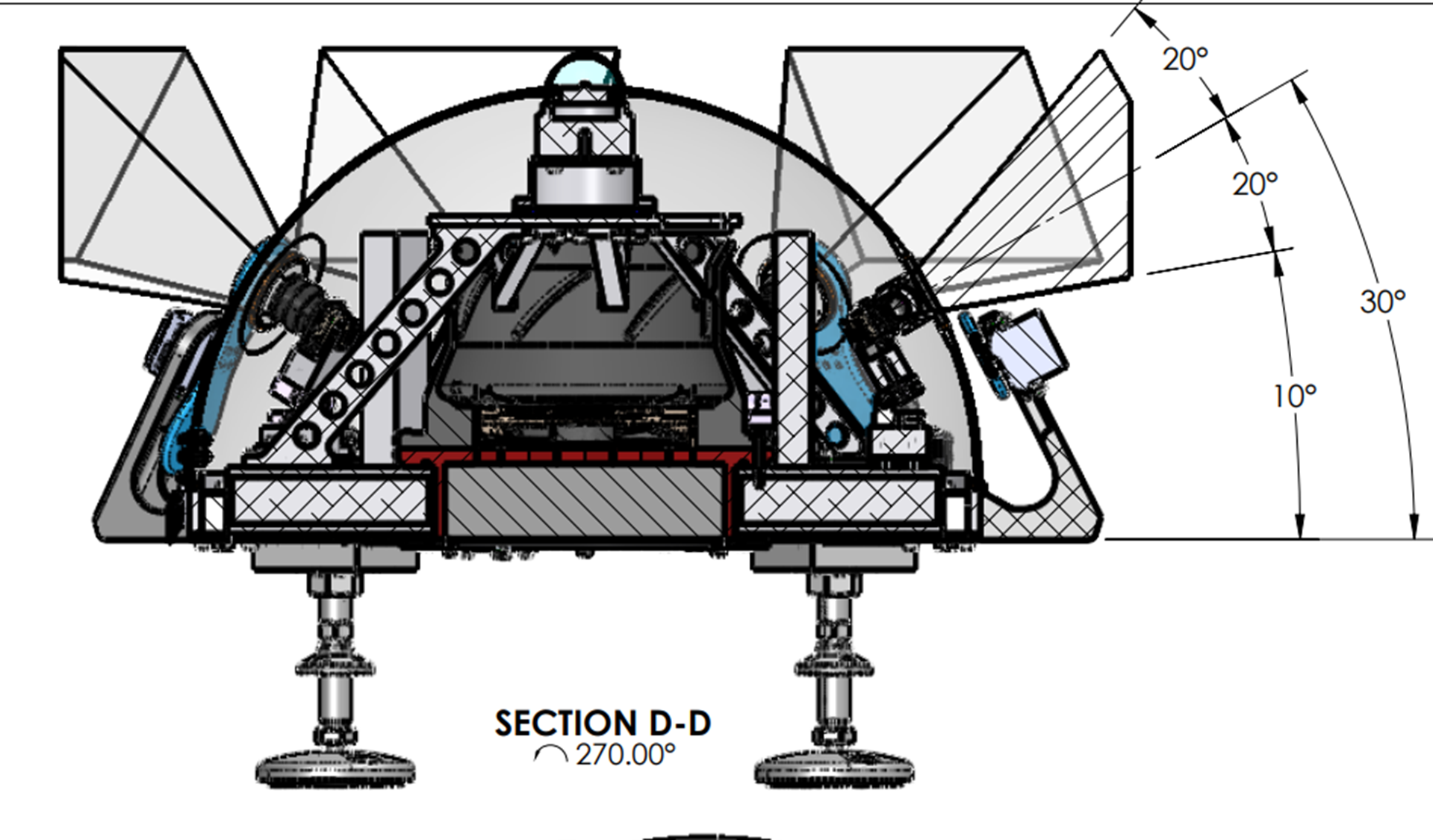}
\end{adjustwidth}
\caption{(\textbf{Left}): Illustration of fields of view (FOVs) and their overlap between the eight cameras of the Dalek. The orange areas represent the FOVs of the seven cameras arranged hemispherically. The purple area shows the FOV of the zenith camera on top of the Dalek. (\textbf{Right}): Side view of fields of view for the hemispherical Dalek cameras. As the center of their optical axes are pointing 30$^{\circ}$ above the horizon, the bottom of the images that they capture corresponds to 10$^{\circ}$ above the horizon.\label{fig:dalek_fov}}
\end{figure}

\begin{figure}[H]
\begin{adjustwidth}{-\extralength}{0cm}
\centering
\includegraphics[width=\linewidth]{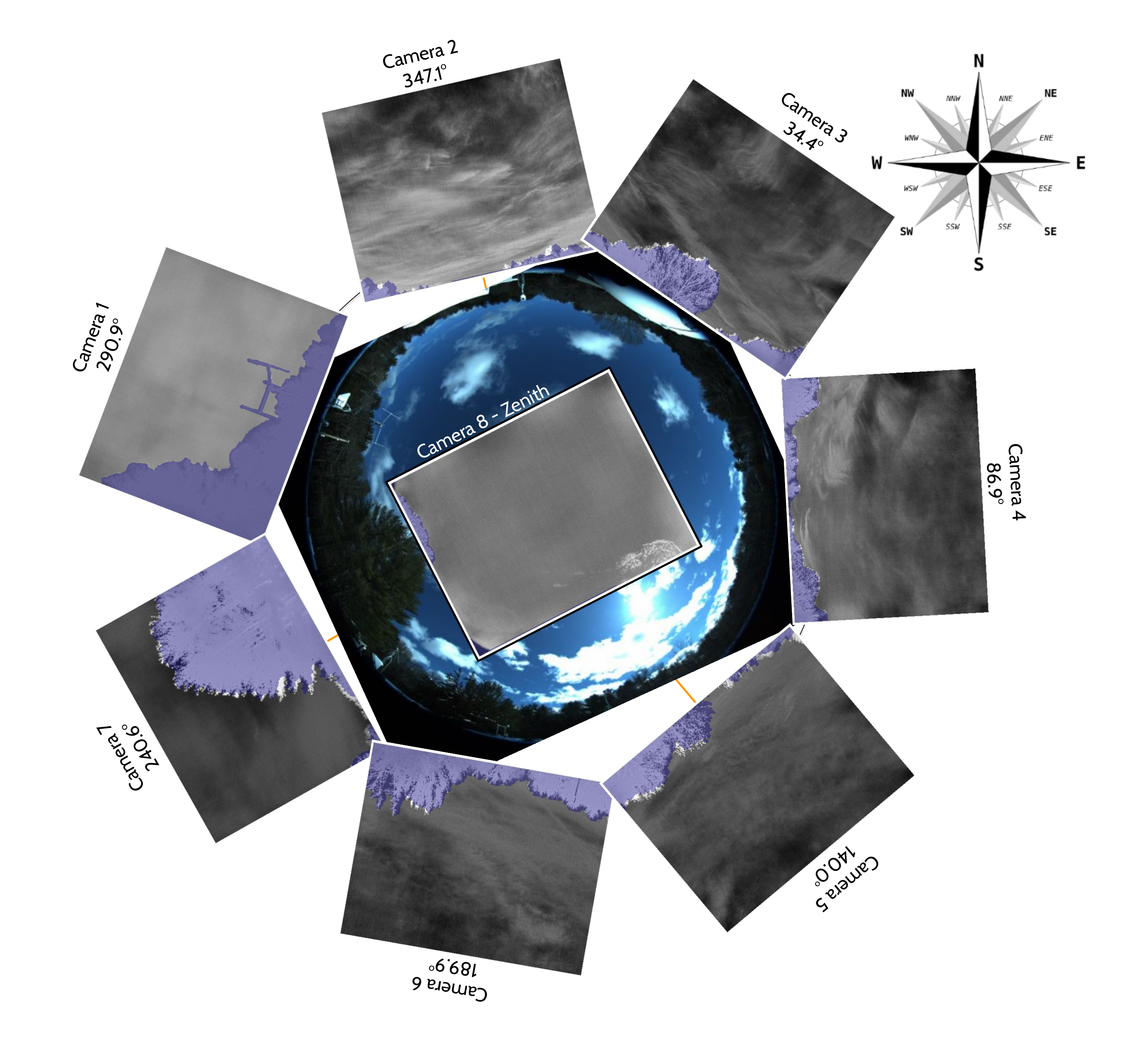}
\end{adjustwidth}
\caption{Map view of a mosaic of images from the seven hemispheric cameras and the one zenith Boson IR camera, and their orientation with respect to a visible all-sky camera photograph from the Dalek's location (background, center of the image). The shaded (purple) semi-translucent overlays show the corresponding treeline masks that are used in post-processing to ignore all but the sky area of the images. All camera frames are taken from a video recording from 7 May 2024 
except for camera 1 (3 April 2024) and camera 7 (10 May 2024).\label{fig:dalek_mosaic}}
\end{figure}

The seven IR cameras in the hemispherical array are FLIR LWIR Boson 640~$\times$~512 (spectral band 7.5--13.5~$\upmu$m, focal length 8.7~mm, inverse relative aperture of F/1). Their field of view (FOV) is $50^{\circ} \times 40^{\circ}$ (horizontal and vertical). The center of their optical axis is pointing 30$^{\circ}$ above the horizon such that the bottom of the image taken by these cameras corresponds to 10$^{\circ}$ above the horizon. As a result, adjacent cameras have overlapping FOVs, as shown in the left of Figure~\ref{fig:dalek_fov}. The zenith camera is an FLIR LWIR Boson 640~$\times$~512 (spectral band 7.5--13.5~$\upmu$m, focal length 4.9~mm, inverse relative aperture of F/1.1) with an FOV of $95^{\circ} \times 72^{\circ}$. Figure~\ref{fig:dalek_mosaic} shows a mosaic of the views of the eight cameras in the array, oriented with respect to true north in map view. Each camera has trees and vegetation in the lower part of its FOV; the same figure also overlays in semi-translucent purple the treeline masks that we use in post-processing to only analyze the sky area of the images. These masks are initially generated using the recent pre-trained Segment Anything Model (SAM)~\cite{kirillov2023segment} machine learning model, which classifies every pixel of an image into different semantic categories. The enclosure is a fiberglass dome, which provides weatherization. The IR cameras are set behind germanium windows and can be protected from direct sunlight by programmable sunshades.

Each camera is connected via a USB3 cable to an Nvidia Jetson Orin NX, which can handle edge computing tasks such as continuous recordings and, in the future, real-time object detection, tracking, and classification. One Jetson receives data from four FLIR cameras.
A \textit{Gstreamer}~\cite{gst} pipeline feeds from each camera and creates 5-minute video segments, which are stored on the Jetson. For all of the data used in this paper, we were recording at 10 frames per second, although we plan to increase this to 30 frames per second in the future to take full advantage of the Boson cameras' capabilities. The bit depth is 8-bit unless otherwise specified. The cameras have low, high, and automatic gain settings. All the cameras are set on high-gain mode. An automatic flat-field calibration (FCC) using a shutter happens regularly and interrupts the recordings, overlaying a green square in the top right of the frame while the FFC is performed. This mitigates the effects of infrared non-uniformity over long recording time periods. While we expect that certain parameters such as a higher frame rate would improve the detection and tracking performance, a quantitative analysis of the influence of these parameters would require us to collect data over a longer period of time, up to a year, to have sufficient statistics of ADS-B-equipped aircraft, which is our most abundant source of labeled real-world data to assess detection and tracking performance. We plan to write a future paper focused on comparing the influence of the array camera system parameters, such as the camera count, on the performance envelope.

The custom Dalek enclosure protects the cameras from weather conditions such as wind, rain, snow, dust, or disturbance from animals. The fiberglass dome has a maximum deformation of less than 1~mm of radial deformation when submitted to a 200 mph wind, according to a finite element analysis of its structure, which is roughly double the record wind gusts in this area. 
The enclosure contains two 72-cubic-inch desiccant containers meant to keep the enclosure dry. The desiccant changes color from blue to pink when it reaches its full moisture-absorbing capacity. Baking it releases the moisture and restores the absorbing capacity for a new cycle of use.
Finally, a fan in the dome enclosure helps to evacuate hot air from both direct sunlight and hardware heat emission. This keeps the hardware below the maximum operating temperature of 80$^{\circ}$C for the FLIR Boson 640 cameras. 
A first set of germanium windows was chosen for their high transmissivity (95\% anti-reflection/anti-reflection (AR/AR) coating) over corrosion resistance (85\% diamond-like carbon/anti-reflection (DLC/AR) coating). Prolonged outdoor exposure significantly degraded the AR/AR coating within a year. The germanium windows have since been replaced with ones having both a hydrophobic coating and a high transmissivity ($\sim$96\% hydrophobic/anti-reflection (HP/AR) coating). These have not shown as many signs of degradation.

Ideally, all eight of the Dalek's cameras would record continuously 24 h per day, seven days a week, all year. However, the IR camera sensors can be damaged if the sun is directly in their field of view for long periods, so the Dalek is equipped with a sunshade on each camera that automatically opens and closes on a per-camera schedule as the sun crosses the sky. The mechanical sunshades are mounted on servos and controlled by an Arduino micro-controller. They also have integrated brushes on the side facing the camera to periodically remove debris from the windows. When the sunshade is closed on a camera, its recording is stopped in order to save data storage space. As a result, all cameras run during the night, but during the day, each one has a different recording schedule, with the north-facing cameras recording for longer durations and the south-facing cameras for shorter. Thus, the total expected recording duration is less than 24 hrs/day for each camera. Camera 2, facing north-northwest, has the longest recording duration, and camera 6, facing south-southeast, has the shortest. The sunshade schedule and total recording time per camera per season is broken down in Table~\ref{fig:dalek_schedule}.  The sunshades are depicted in blue in the schematic on the left side of Figure~\ref{fig:dalek}, and are black under red mounts on the photograph of the Dalek as-built, shown on the right of Figure~\ref{fig:dalek}.

\begin{figure}[H]
\centering
\includegraphics[width=\linewidth]{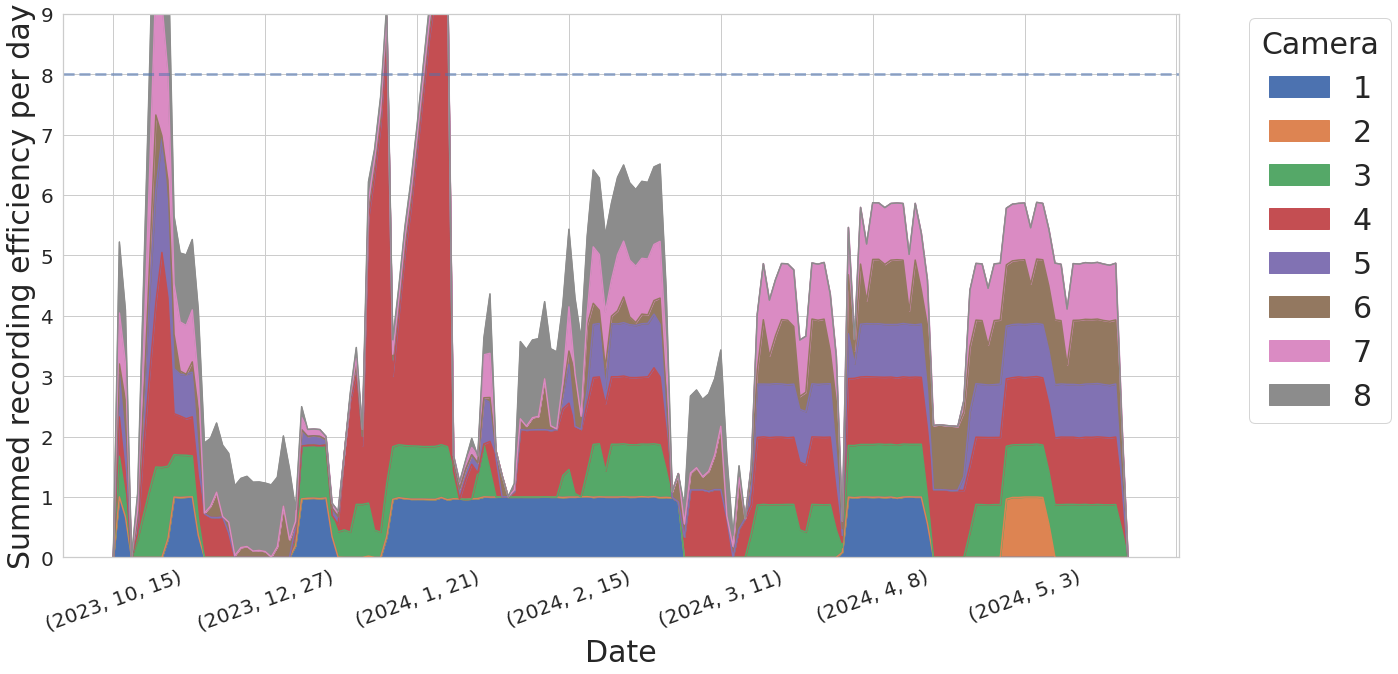}
\caption{Stacked area plot showing the evolution over time of the sum of all cameras' recording efficiencies, defined as recording duration per camera per day divided by expected duration based on recording schedule, which varies per camera. If all cameras were recording according to the schedule all the time, the summed efficiencies should add up to 8. The few data points above 8 are due to manual enabling of the recording, for testing purposes. This timeline goes from November 2023 to May 2024. Some cameras, such as cameras 6 and 7, show a drastic improvement over time. \label{fig:dalek_filecount}}
\end{figure}

Although we started recording continuously in November 2023, the system's uptime only started to stabilize in January 2024. 
Figure~\ref{fig:dalek_filecount} shows the improvement over time of the Dalek cameras' uptime, quantified by the recording efficiency per day or ratio between the total recording duration and the expected duration from the recording schedule, for each camera. Cameras 2 and 6 suffered the most downtime, and were eventually diagnosed with USB cable failure. After replacement of the USB cable, their behavior stabilized dramatically. The data points in Figure~\ref{fig:dalek_filecount} with an efficiency above 8 are due to manual enabling of recording outside of the automatic schedule, for testing purposes.

\subsection{Calibration}
\label{sec:calibration}
To relate object positions in the camera image to object positions in the sky, several camera calibration steps are required. Intrinsic calibration corrects for slight aberrations in the lens itself, both intentional (e.g., barrel distortions) and lens imperfections. Removal of image non-uniformities (INUs) involves correcting for the impact of microscopic lens aberrations on the recorded image under certain lighting conditions. Extrinsic calibration, applied last, enables a mapping from the sensor's physical location and object's location in the 2D field of view to the object's 3D world location, for example, from an image's pixel to an object's azimuth and elevation relative to the camera location.  

\subsubsection{Intrinsic Calibration}
\begin{figure}[H]

\includegraphics[width=4.9cm]{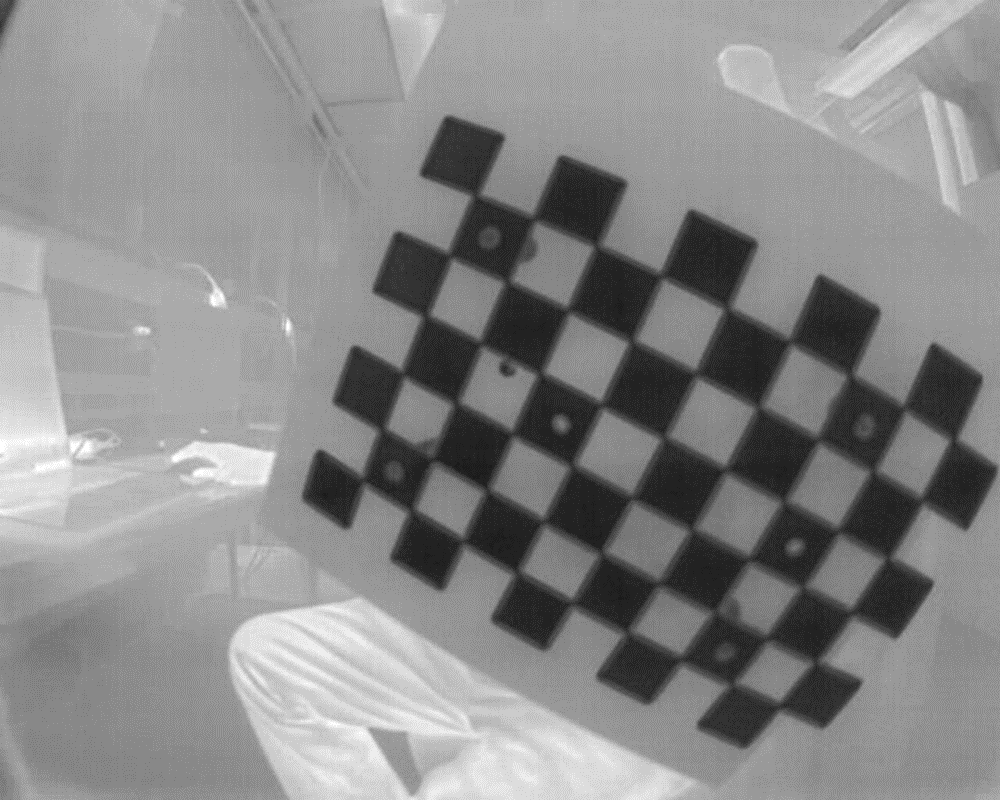}\quad
\includegraphics[width=5.3cm]{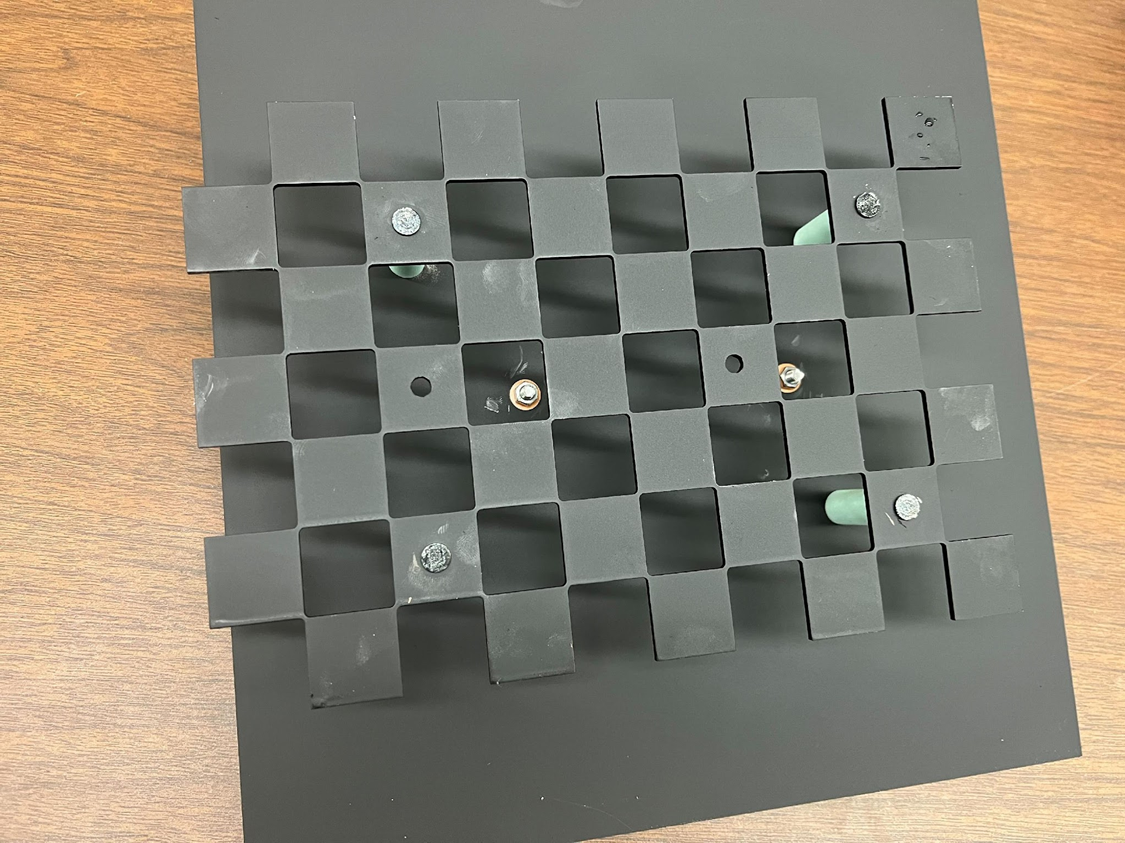}
\caption{Metal chessboard used for intrinsic calibration of the FLIR Boson 640 cameras. (\textbf{Left}): Boson camera image. \hl{(\textbf{Right})}: 
 Dark-painted metal cutout chessboard on dark-painted metal base plate. 
\label{fig:calibration}}
\end{figure}

The chessboard calibration method described in~\cite{888718} and coded in the OpenCV library~\cite{opencv_library} allows us to perform an intrinsic calibration of each camera independently.   It yields intrinsic parameters: focal length, optical image center, and distortion coefficients. These are stored in the camera configuration file. This calibration is performed in the laboratory.
Since the Boson cameras are only sensitive to long-wave-infrared wavelengths, we cut a chessboard grid out of a metal sheet. To obtain a good contrast between cold and hot, we cool it in a freezer to $-$20~$^{\circ}$C, quickly assemble it with thermal insulators as spacers to a solid background metal sheet that has been heated to 50 $^{\circ}$C, and hold it by a handle that is mounted on the back of the warm metal sheet to capture a series of calibration images covering the field of view of the camera, as shown in Figure~\ref{fig:calibration}.

\subsubsection{Removal of Image Non-Uniformities (INUs)}
\begin{figure}[H]
\begin{adjustwidth}{-\extralength}{0cm}
\centering
\includegraphics[width=4cm]{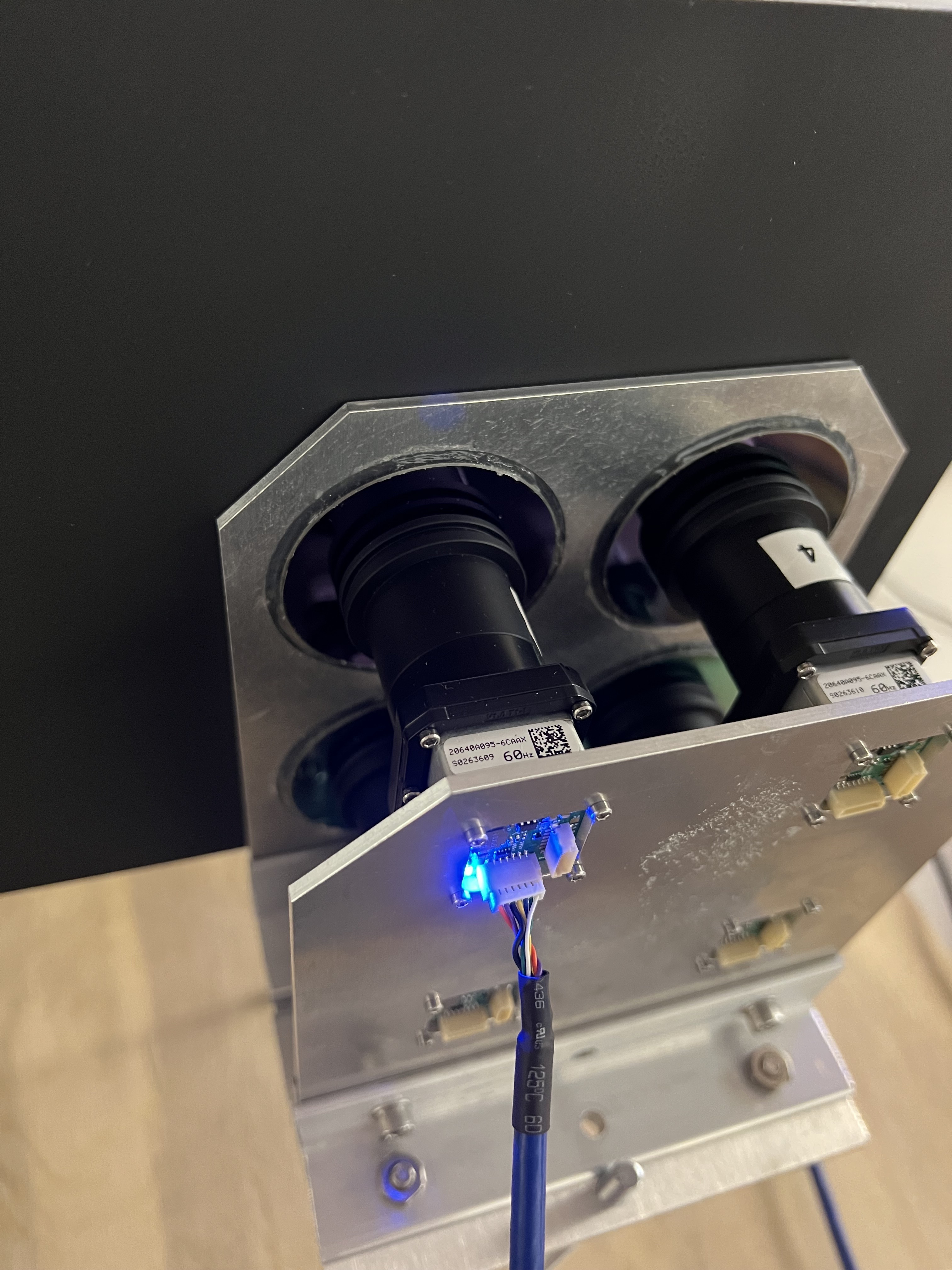}\quad
\includegraphics[width=4cm]{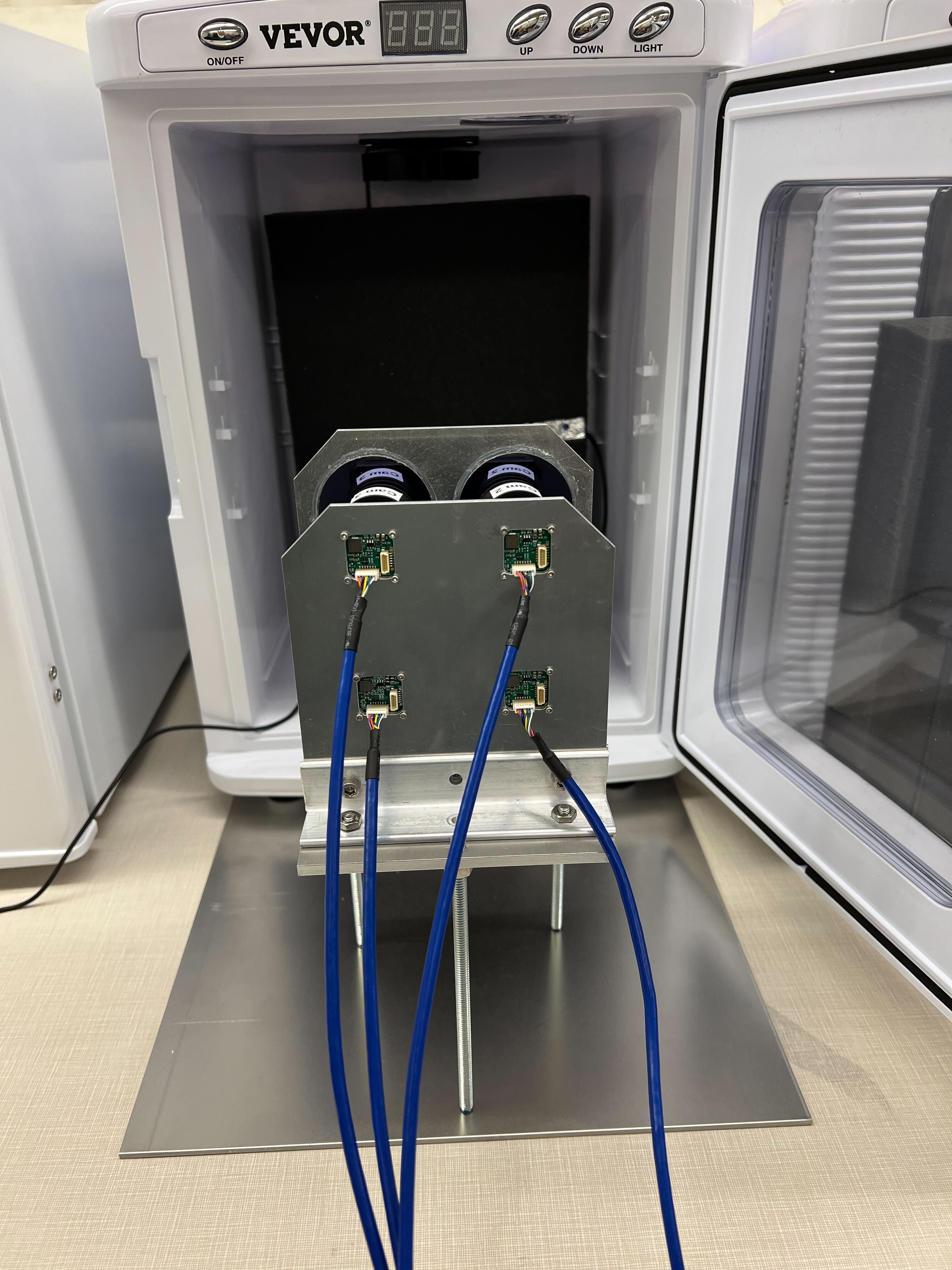}\quad
\includegraphics[width=7cm]{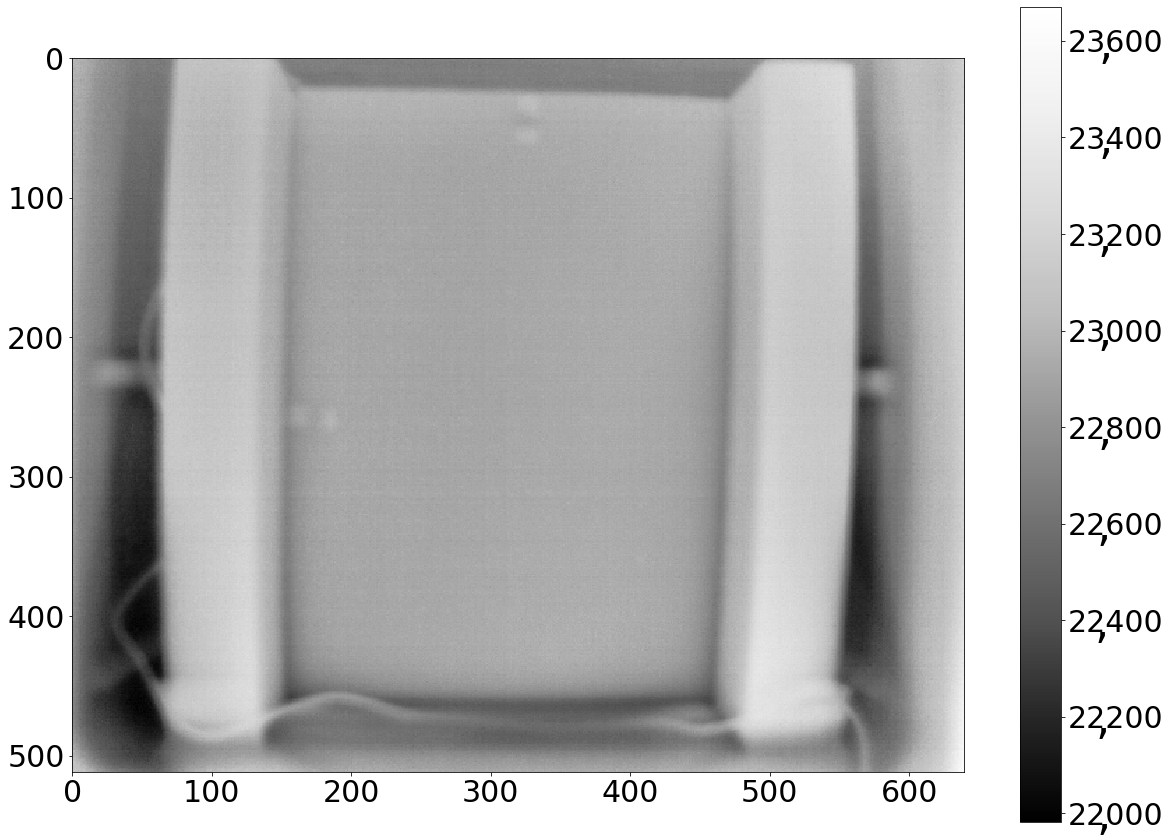}
\end{adjustwidth}
\caption{\hl{(\textbf{Left}): Four} cameras mounted on a fixture and pointed at the black metal plate used for removing INUs through lens gain correction and supplemental flat-field correction (SFFC). (\textbf{Middle}): Four cameras mounted on a fixture and pointing at the black polyurethane foam in the incubator; used for thermal calibration of four Boson cameras at the same time. (\textbf{Right}): Example of 16-bit image taken with a Boson camera using the thermal calibration setup described in Section~\ref{sec:thermal}. The two small pieces of reflective tape, pinned at the top and left side of the foam, are used to find the geometrical center of the foam in the image, where the target temperature was measured using a thermometer. The ``ears'' of foam minimize thermal reflections from the sides and door of the freezer or incubator.\label{fig:calibration2}}
\end{figure}

The Boson camera manufacturer, FLIR, defines INUs as ``a group of pixels which are prone to varying slightly from their local neighborhood under certain imaging conditions''~\cite{flir2018}. Specks of dirt, dried water droplets, or microscopic defects in the coating on the outside of the Boson camera's factory window, or small defects on the inside of the window, can cause INUs. FLIR recommends two procedures to mitigate INUs: lens gain correction followed by supplemental flat-field correction (SFFC); these should be carried out in this order~\cite{flir2017}. For lens gain correction, we image two metal plates at different but uniform temperatures. The metal plates are covered in black Krylon ColorMaxx paint to ensure thermal uniformity. The software to compute the lens gain correction is provided by FLIR. The SFFC compensates for irradiance from heat generated inside the camera body itself. We occupy the entire FOV of the camera with a black-body target (polyurethane foam) to perform the SFFC. Both the metal plates for lens gain correction and the black foam for SFFC are pictured in Figure~\ref{fig:calibration2}.

\subsubsection{Extrinsic Calibration with ADS-B-Equipped Airplanes}
\label{ADS-B}
The rotation matrix and translation vector which convert a camera's coordinate frame to world coordinates constitute the extrinsic calibration parameters of a camera. To find the translation vector, we use GPS coordinates from a cellphone positioned next to the Dalek, which has an accuracy of approximately 5~meters~\cite{Diggelen2015TheWF} in this case. Extrinsic calibration of visible-light cameras often relies on astrometry, but the LWIR Boson cameras cannot see stars. Therefore, to find the rotation matrix, we adopt the novel calibration technique described in~\cite{watters2023scientific} using airplanes, which emit at LWIR wavelengths and also reflect emissions from the warm ground. Automatic Dependent Surveillance Broadcast (ADS-B) Mode-S transmission devices are operated in most airplanes flying in US airspace, as required by the US Federal Aviation Administration~\cite{faa_14_CFR_91.225}. Airplanes send with ADS-B, in near real-time (transmission of the aircraft position must happen within 2~s from the measurement time, with up to 0.6~s of uncompensated latency, and at least once a second~\cite{faa}) and roughly once every 2 seconds, their GPS-derived latitude and longitude positions, and GPS- and barometric-derived altitudes. ADS-B receivers are widely available for retail and anyone with a software-defined radio receiver can collect local ADS-B data. Non-profits such as OpenSky Network aggregate ADS-B data, crowdsourcing individuals' ADS-B receiver data from all over the world into a single database~\cite{OpenSky}. This historical ADS-B database is made available to academic researchers. For this paper, we rely on the OpenSky historical database~\cite{OpenSky} for ADS-B records to obtain the coordinates of the many aircraft flying in the Dalek's field of view because our own ADS-B receiver was not operating continuously during the five-month period of the Dalek's commissioning.

If the camera captures $N$ images of an aircraft as it passes overhead, we can interpolate the 3D coordinates provided by the ADS-B data over this time range, and associate a 3D world location with the aircraft's camera pixel location at the timestamp of each frame. Let $W = (w_i)_{1 \leq i \leq N}$ be the vectors pointing from the camera's known origin in world coordinates, given by GPS, to the airplane's position in the same frame as given by ADS-B. We then use OpenCV's implementation of the Perspective-n-Point (PnP) algorithm~\cite{marchand:hal-01246370} to compute the camera's pose in world coordinates, giving as input the camera's intrinsic parameters, extrinsic parameters, and the airplane's image points and corresponding world pointing vectors. This returns a rotation matrix which represents the camera's pose.
This method to extrinsically calibrate a camera can be applied to both optical and IR cameras. Applying it to the Dalek data was our first demonstration of this method, and we plan to use this for other cameras in the observatory.
\subsubsection{Monitoring Camera Orientation Changes}

We automated this extrinsic calibration technique in order to repeat this at regular intervals. This opens the door to detecting unwanted changes in camera orientation; for example, this can happen during high wind weather events or hardware maintenance that requires opening the dome, or temperature-related expansion of the aluminum bench supporting the cameras. Small changes affect the treeline mask used downstream in the analysis pipeline for reconstruction of aerial events, and cause incorrect estimation of object positions in the sky, which depend on the calibration matrices. Determining this threshold and using this to detect significant changes is important for consistency and accuracy in tracking and trajectory reconstruction. 

\begin{figure}[H]
\includegraphics[width=0.5\linewidth]{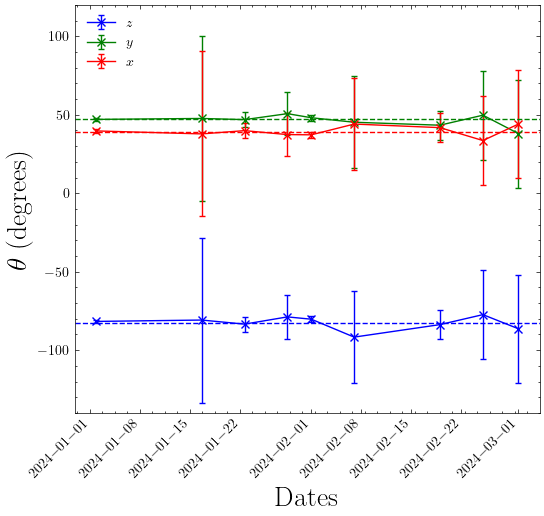}
\caption{\hl{Euler} 
 angles representing the orientation matrix of camera 1 in the Dalek over a period of three months: January to March 2024. The error bars are estimated from the inverse Hessian matrix after minimization. Large error bars are correlated with a lack of statistics. Overall the fluctuations are of the order of 1\%. The dashed lines represent the fixed Euler angle values that were used throughout this paper.\label{fig:euler_orientation}}
\end{figure}

The process we use to monitor for unexpected changes in camera orientation is as follows. We use the You Only Look Once (YOLO) machine learning (ML) object detection method~\cite{yolov5, jiang2022review}, described in Section~\ref{sec:yolo_sort}, to automatically detect aircraft in a given frame of a Boson camera video. The existing extrinsic and intrinsic calibration allows us to translate an aircraft's ADS-B GPS coordinates into camera 2D image coordinates. The ADS-B predicted positions are interpolated in a time interval of $\pm5$ seconds, to account for timing inaccuracies in both the ADS-B and the camera image frame timestamps, and matched to the closest center of the detection bounding boxes predicted by YOLO. We allow such a generous time threshold to account for the occasionally less-than-ideal timestamp accuracy of the video files recorded during this commissioning phase. For matching we use a distance threshold between the centers of the YOLO-predicted and ADS-B-derived bounding boxes of 20~pixels and a range threshold of 2~km from the observatory to the aircraft. The distance threshold accommodates latency-induced inaccuracy in the ADS-B-derived position estimation and inaccuracies in the existing extrinsic calibration. We collect data from a minimum of 20~airplanes before computing a new orientation matrix. This procedure is repeated on a weekly basis to monitor the changes over time. Figure~\ref{fig:euler_orientation} shows the actual evolution over time of the rotation matrix of one of the Dalek cameras. The rotation matrix is represented by its three Euler angles and computed over a period of at least 1 week, ensuring statistics from at least 20 airplanes. We observe slight fluctuations in the Euler angles over time, of the order of 1\%. Errors are derived from the inverse Hessian matrix, obtained using the minimization algorithm~\cite{liu1989limited}. These errors do not account for the uncertainty in bounding boxes' center positions, which could affect this measurement. This method will be improved with further refinements but it can be used in its current form to trigger an alarm if the orientation of a Boson camera changes significantly, for example, by more than one degree in any Euler angle.

\subsubsection{Thermal Calibration}
\label{sec:thermal}
The FLIR Boson 640 cameras have the ability to record images with 8- or 16-bit depth. The 16-bit data can be used for thermal radiometry. The camera model selected for the Dalek comes without factory thermal calibration, hence we performed the thermal calibration of the cameras ourselves. This requires taking images of a black-body target of known emissivity when both the camera and the target are exposed to a wide range of temperatures. This is necessary in order to correlate the 16-bit pixel values and camera temperature to the target temperature. 

\begin{figure}[H]
\begin{adjustwidth}{-\extralength}{0cm}
\centering
\includegraphics[width=9cm]{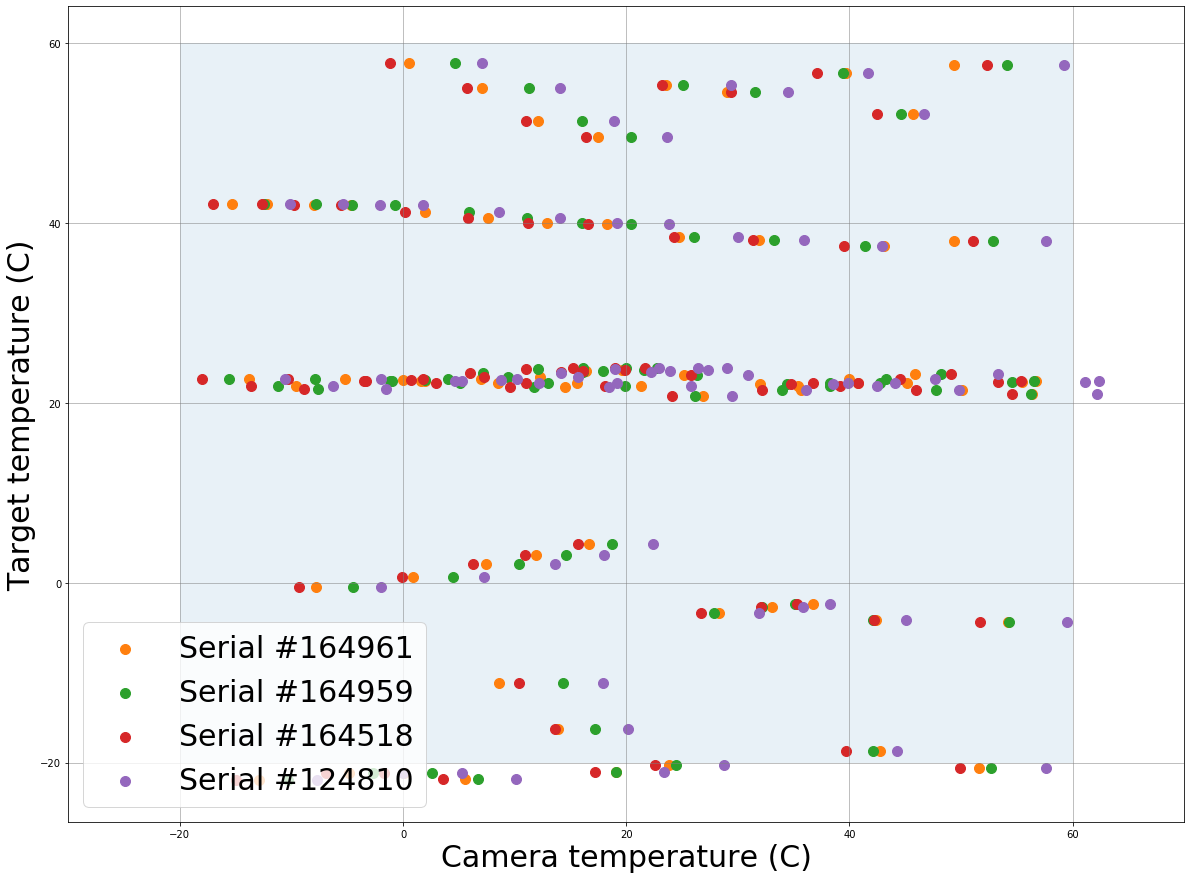}\quad
\includegraphics[width=8cm]{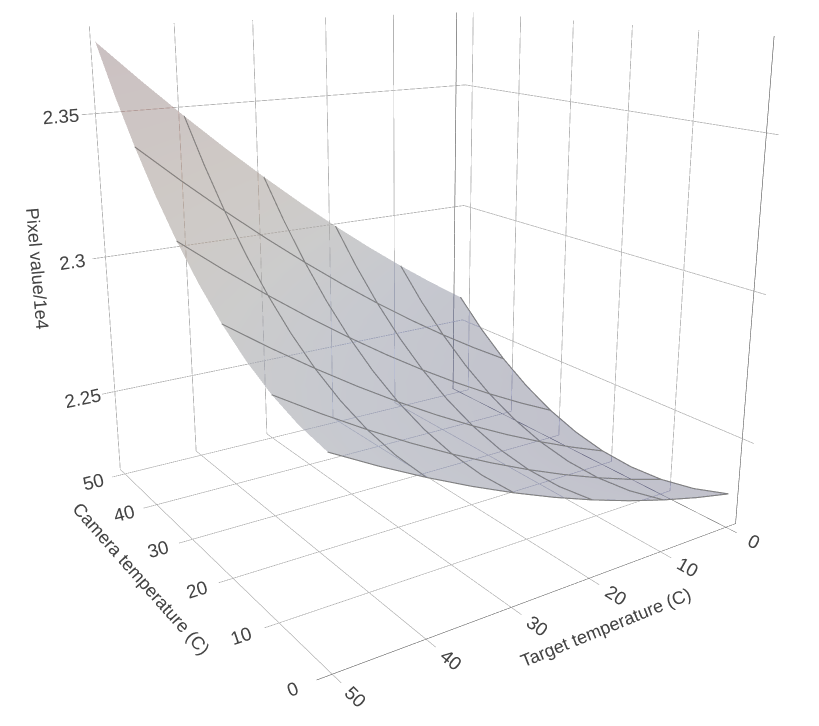}
\end{adjustwidth}
\caption{(\textbf{Left}): Map of measurements made on four cameras of the Dalek. (\textbf{Right}): Visualization of fitted polynomial. The maximum pixel values occur when the camera and target are both ``hot''. Pixel values are divided by $10^4$ on the z-axis.\label{fig:thermal_calibration}}
\end{figure}

Our target was a block of black polyurethane foam. In order to calibrate up to four cameras at the same time, we used the custom-built mechanical fixture that holds both the cameras and the germanium windows, shown previously in Figure~\ref{fig:calibration2} and used for the intrinsic calibration procedure. Since the opacity of the germanium windows is temperature-dependent, being almost opaque at 100$^{\circ}$C, it is important to include them in the calibration fixture.
We drilled holes in a plywood sheet to keep the distance and orientation of the fixture relative to the foam constant. Two pieces of foam were taped into position, one inside an incubator and one inside a freezer. Figure~\ref{fig:calibration2} shows an example of this setup, where the fixture faces the foam target in the incubator. A thermal probe was inserted in the foam to measure the temperature at the center of the block. The position of the probe tip inside the foam was marked using horizontal and vertical reflective tape indicators at the edge of the foam. We used the temperature sensor onboard the camera chips to record the camera temperatures. 
We explored a 2D grid of target and camera temperatures in the range of $[-20\text{, }60]~^{\circ}$C. For a given target temperature, the process was as follows:
\begin{enumerate}
\item Warm or cool down the foam to the target temperature.
\item Warm up the camera to 60~$^{\circ}$C using an incubator.
\item Place it in front of the foam target and capture images at regular intervals while it is cooling down back to room temperature. 
\item After cooling down the camera to $-$20~$^{\circ}$C with the help of a freezer, place it in front of the target and capture images while it is warming back up to room temperature. 
\end{enumerate}
Ambient humidity proved to be a challenge in this process, as it created condensation droplets on the camera lenses and germanium windows. We had to maintain the air humidity in the lab below 40\% in order to take meaningful measurements.
Figure~\ref{fig:thermal_calibration} shows the temperatures sampled for different cameras following this process.
Once all the images are taken, i.e., once sampling the 2D grid of camera and target temperature is completed, we find the pixel coordinates of the centers of both foam targets (the foam in the incubator and the foam in the freezer) for each individual camera. We record the mean pixel value in a $32\times32$-pixel square at these coordinates.
We grid the pixel values using inverse distance weighting interpolation and $3\times3$ smoothing. The result is fitted to a bi-quadratic polynomial, where the two variables are $x = $ camera sensor temperature and $y = $ target temperature, and the output is $z = $ pixel value. Figure~\ref{fig:thermal_calibration} has an example of a fitted polynomial that can be used to relate a 16-bit pixel value and camera temperature to the target radiance, assuming it is a black body. The black-body emission of the target is related to its temperature through Planck's equation, which we numerically integrate for the wavelength band of the Boson camera, i.e., from $\lambda_1 = 7.5~\upmu$m to $\lambda_2 = 13.5~\upmu$m:
\[
\mathcal{L}(T) = \frac{\epsilon}{\pi} \int_{\lambda_1}^{\lambda_2} d\lambda \frac{C_1}{\exp(C_2/(\lambda T)-1} \lambda^{-5}\quad \text{[W/m}^2\text{/sr]}
\]
\hl{where} 
 $C_1 = 3.7415 \times 10^{8}~\text{W}\cdot\text{m}^{2}\cdot\upmu\text{m}^{4}$ and $C_2 = 1.43879 \times 10^{4}$ $\upmu$m$\cdot$K.

\subsubsection{Object Temperature Measurement}
This section demonstrates a first-order example of using the thermal calibration of the previous section to estimate the temperature of an aerial object.
A surface emits thermal radiance depending on both the surface temperature and the surface emissivity. To use the calibrations from the previous section on real images captured with a Boson camera, we need first to measure the emissivity $0 \leq \epsilon \leq 1$ of the foam used in the calibration process. We let the foam equilibrate to room temperature. We measure the temperature of the foam through a surface temperature probe which yields 23$^{\circ}$C$~\pm~0.5~^{\circ}$C. Simultaneously, an external laser infrared thermometer set to an emissivity of 1.0 measures 22$^{\circ}$C$~\pm~1.5~^{\circ}$C at the site of the temperature probe.  We deduce an estimate of the foam emissivity from the ratio of the two temperatures in kelvin units, $\epsilon = 0.997 \pm 0.007$.

For the thermal calibration, we assumed that, owing to the short distance between the camera and the foam target, any atmospheric effects would be negligible. In general, however, another parameter is involved in the measurement of an object's temperature through our cameras in the field: the atmospheric transmission, which quantifies the attenuation of thermal radiation emitted by the object due to water vapor and carbon dioxide along the transmission path. This depends on the relative humidity, the temperature of the atmosphere, and the distance from the camera to the object. We use the Passman--Larmore tables~\cite{passman1956, minkina2016atmospheric}, which were derived experimentally, to compute the atmospheric transmission coefficient $0 \leq \tau \leq 1$. For a wavelength of $\lambda = 10.5 ~\mu$m (the center of the FLIR Boson camera spectral band), a distance $d = 5000$ m to the object, relative humidity $\omega = 44$\%, and atmospheric temperature $T_{atm} = 22~^{\circ}$C, we use equation (6), table 1, and table 2 
from~\cite{minkina2016atmospheric} to compute the water vapor absorbance $P_{H_2O} = 0.544$ and carbon dioxide absorbance $P_{CO_2} = 0.978$. Then, we can approximate the atmospheric transmission coefficient with~\cite{gaussorgues} $\tau \sim P_{H_2O} P_{CO_2} \sim 0.532$.

\begin{figure}[H]
\includegraphics[width=6.5cm]{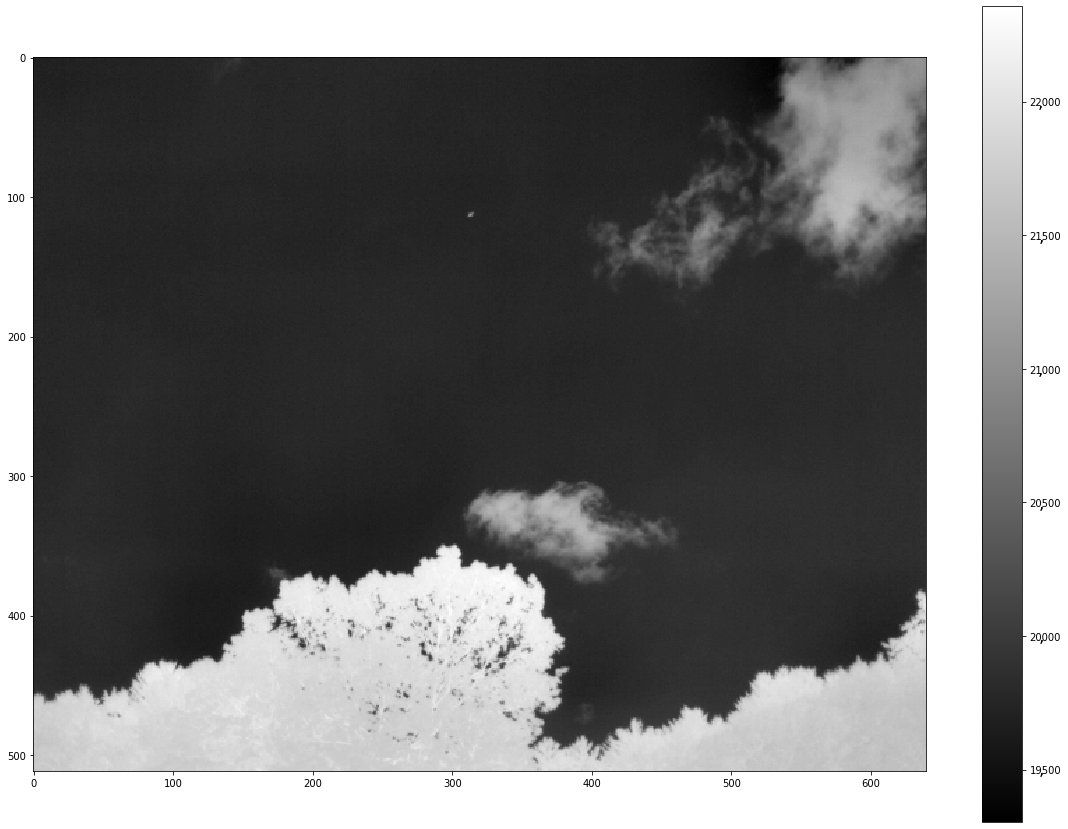}\quad
\includegraphics[width=5.5cm]{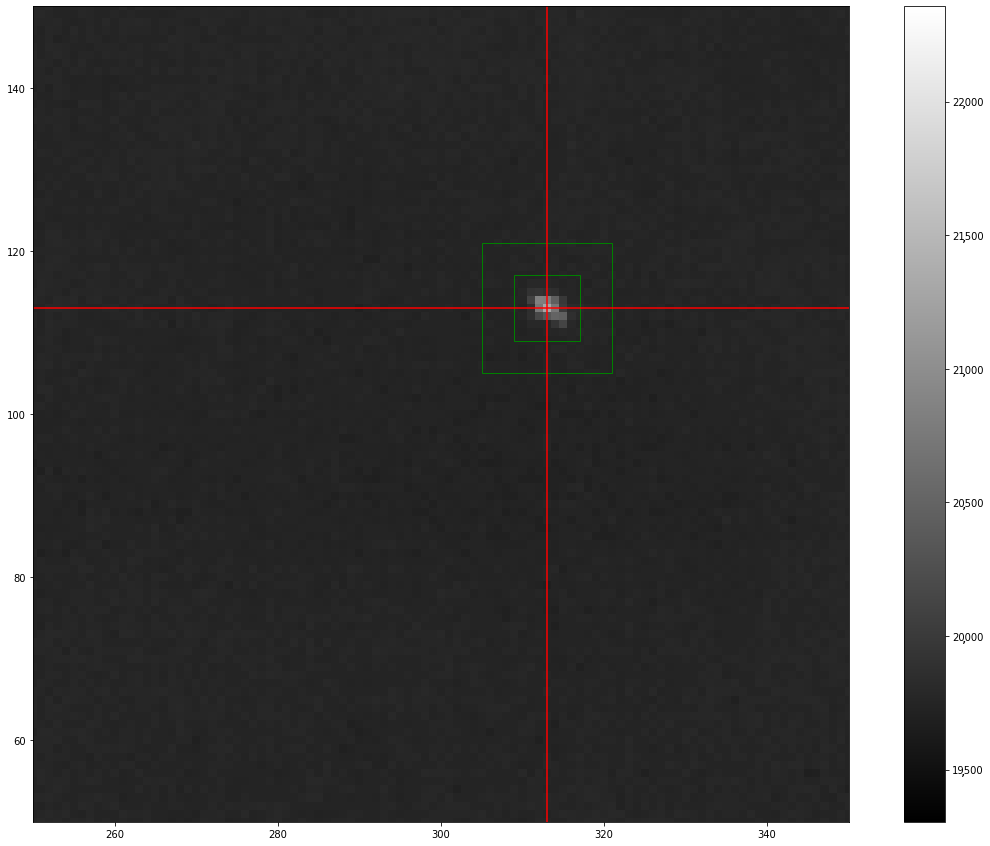}
\caption{\hl{(\textbf{Left}): 
 Example} of 16-bit infrared image of an aircraft taken with a Boson camera. (\textbf{Right}): Zoom in on the airplane in the scene. The red lines cross at the pixel of maximum brightness. The area in between the green rectangles is used to estimate the ambient brightness.\label{fig:temperature_measurement}}
\end{figure}

Given the emissivity of the object $\epsilon$ and the atmospheric transmission $\tau$, we see a brightness of
\[
\mathcal{L}_{\text{observed}} = \epsilon\tau\mathcal{L}_{\text{object}} + \tau(1-\epsilon)\mathcal{L}_{\text{ambient}} + (1-\tau)\mathcal{L}_{\text{atmosphere}}
\]
where $\mathcal{L}_{\text{ambient}}$ is the brightness emitted by the object's surroundings and reflected by the object, $\mathcal{L}_{\text{atmosphere}}$ is emitted by the atmosphere, $\mathcal{L}_{\text{observed}}$ is what we measure from the camera sensor, and $\mathcal{L}_{\text{object}}$ is emitted by the object, i.e., the quantity we wish to determine~\cite{minkina2009infrared, minkina2016atmospheric}. Figure~\ref{fig:temperature_measurement} shows an example of a 16-bit frame from camera~3 of the Dalek during an airplane fly-by. The aircraft is an Airbus A321 taking off from BOS airport at a slant range of 5.3~km from the observatory. We measure the ambient reflected brightness in units of pixel values, and then use our calibration to convert to temperature. We average the pixel values immediately surrounding the object. For measuring the atmospheric brightness, we average the pixel values in the top left quarter of the frame, which is empty sky. We obtain 19,748 and 19,765. 
By plugging in the values for the object emissivity and atmospheric transmittance, we can infer the actual object brightness of 22,672 from the observed brightness of 21,307. 
This would be the pixel value for a piece of foam in the same conditions as it was during the thermal calibration. The fitted relationship between brightness and radiance that we obtained from the calibration in the previous section yields an estimate of the object's radiance: assuming a sensor temperature of 30~$^{\circ}$C, given the ambient temperature at the time the image was taken, the target radiance is estimated at 53~W/m$^2$/sr. Next, we set equal the radiant flux emitted from a surface of $\sim$8~$\times~8$ m$^2$ (assuming an average airliner has a wingspan of 50 m, at a distance of 5.3 km, one pixel of the camera corresponds to $\sim$8 m of the object), which sees the camera sensor from a solid angle of $\sim$7.10$^{-6}$sr at $\sim$5 km distance, and the radiant flux emitted by an object at $\sim$0.1 m distance, with $\sim$0.25 m$^2$ surface area, and which sees the camera sensor from a solid angle of $\sim$0.001 sr; from this, 
we find the equivalent radiance of the actual target, the aircraft at 5 km, to be $\sim$37~W/m$^2$/sr. Planck's law, assuming an emission coefficient of $\epsilon = 0.7$ for an aluminum and stainless steel object, yields a corresponding temperature of 24~$^{\circ}$C.

We compare this result to a temperature estimate based on aircraft altitude. The standard adiabatic lapse rate is a temperature decrease of 2~$^{\circ}$C for every 1000 feet increase in elevation, so we expect a temperature of $\sim$2~$^{\circ}$C at the aircraft barometric altitude of 11,450 feet. Since the aircraft has just taken off, to estimate its body temperature we use Newton's law of cooling for an aluminum (specific heat capacity 903 J/kg/K, heat transfer coefficient $\sim$5 W/m$^2$/K) sphere of $\sim$83 metric tons going from a temperature on the ground of $\sim$20~$^{\circ}$C to the temperature of 2~$^{\circ}$C at the aircraft's current altitude within 5 min. We find that we should expect a temperature of $\sim$18~$^{\circ}$C for this specific aircraft, which would mean an error of $\sim$6~$^{\circ}$C in this example.

This temperature estimate is included in this study as a proof-of-concept demonstration. During the commissioning period we collected only 8-bit recordings and did not yet have triangulation in place for range estimation. The rest of this paper is not concerned with objects' temperatures. In the near future, we plan to enable the continuous recording of 16-bit infrared videos with the Dalek, which will help characterize the temperatures of objects in combination with range estimation derived from triangulation, for example.

\subsection{Reconstruction of Aerial Objects Using YOLOv5 and SORT}
\label{sec:yolo_sort}

The primary design purpose of the Dalek system is object detection and trajectory reconstruction. A combination of machine learning and traditional algorithms suited for this purpose are applied to the IR images obtained from the Dalek video recordings.
You Only Look Once (YOLO)~\cite{yolov5, jiang2022review} is a common machine learning architecture for real-time object detection. We train version 5 of YOLO using both synthetic images and the Dalek's real-world images~\cite{cloete2023integrated}. The datasets are described in Section~\ref{sec:datasets} and the training outcomes are evaluated in Section~\ref{sec:manual_benchmark}.
YOLOv5 marks detections on each frame individually. To reconstruct the trajectories of detected objects within a single camera's field of view, we apply a multi-object tracking (MOT) algorithm to track individual objects in a video sequence. We use the popular Simple Online Real-time Tracking (SORT)~\cite{SORT}, which combines a Kalman filter with the Hungarian algorithm. Section~\ref{sec:sort_benchmark} contains more details about our implementation and a performance benchmark for this stage.

\subsubsection{Datasets for Training and Evaluation}
\label{sec:datasets}
\paragraph{Synthetic Image Dataset}
An object detection algorithm such as YOLOv5 requires supervised training and tens or hundreds of thousands of labeled training samples. We developed a synthetic image generation tool called \textit{AeroSynth}, which leverages the Python bindings of Blender~\cite{blender}, a free and open-source 3D modeling software.
\hl{Ref.} 
 \cite{cloete2023integrated} contains more details about AeroSynth and the training process of a YOLOv5~\cite{yolov5, jiang2022review} architecture using synthetic datasets. Our synthetic dataset contains 25 object bounding boxes (labels) per image, sampled from 40 different 3D models including airplanes, balloons, drones, birds, and helicopters, but excluding some natural objects such as leaves and clouds, totaling $\sim$800k objects spread over $\sim$32k + 8k (train + test) images. 

\paragraph{Mixed Synthetic and Real-World Image Dataset}
In addition to bootstrapping the training using a synthetic dataset, we also selected real-world images from Boson cameras which contained highly confident (score of $ 0.9$) detections from the YOLOv5 model trained using exclusively synthetic data. We included these images, using these detections as truth bounding boxes or labels, to fine-tune the model on real-world images as well. This mixed dataset contains $\sim$424k objects spread over $\sim$45k real-world images. Roughly 10\% of these images are background, empty images. On average, there are 9 object bounding boxes (truth labels) per image. In Section~\ref{sec:manual_benchmark}, we compare these two training approaches (using the synthetic dataset and using the mixed dataset) in a benchmark.

\begin{figure}[H]
    \fbox{\includegraphics[width=0.3\textwidth]{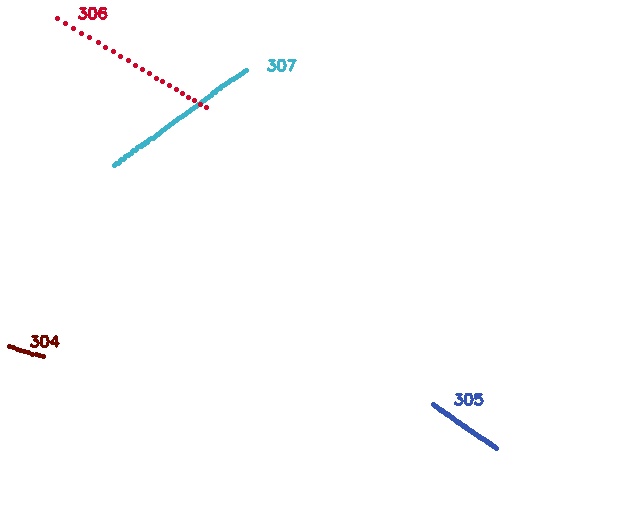}}\quad
    \fbox{\includegraphics[width=0.3\textwidth]{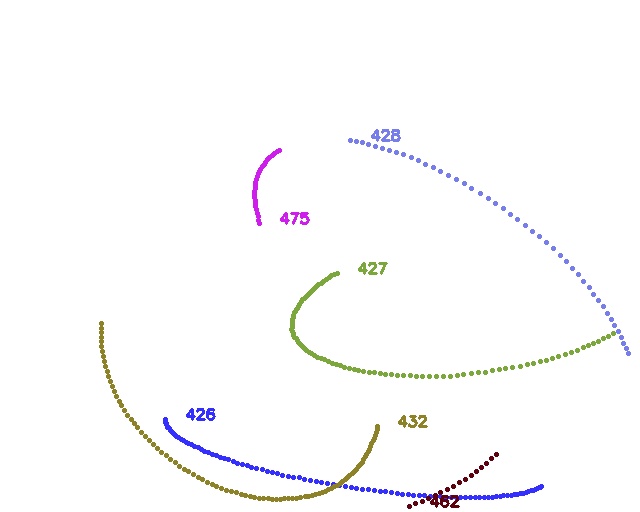}}\quad
    \fbox{\includegraphics[width=0.3\textwidth]{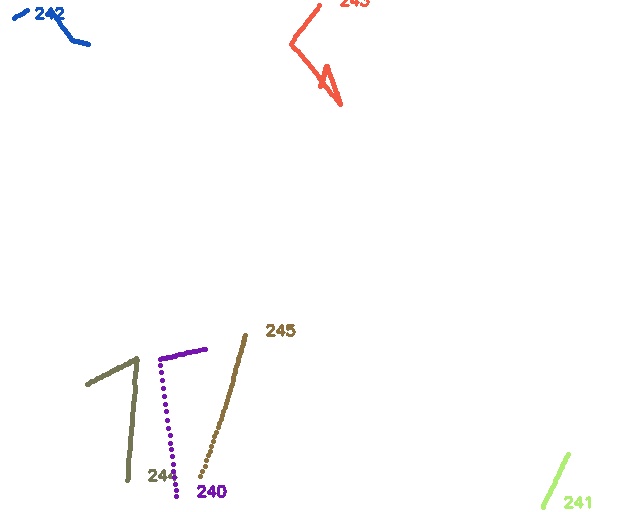}}
    \caption{Comparison of straight (\textbf{left}), curved (\textbf{middle}), and piecewise (\textbf{right}) 3D trajectories from the synthetic video dataset. Trajectories are identified by color and unique number.}
    \label{fig:synthetic_trajectory_comparison}
\end{figure}

\paragraph{Synthetic Video Dataset}
\label{synthetic_videos_dataset}
A video is a set of frames or images that are correlated in time. In order to evaluate tracking algorithms on synthetic, controlled trajectories,  we need to generate consecutive images of synthetic objects that, when stitched together, form a video of object trajectories.
We again rely on AeroSynth~\cite{cloete2023integrated} to render videos of 3D models that mimic the video recordings by individual Boson infrared cameras. Each synthetic video is a collection of 100~images or frames, which are snapshots of the 3D models in the scene at successive points in time, separated by 0.1~seconds. The rate of 10 frames per second is chosen to match the Dalek commissioning dataset frame rate. We include the same 3D models that were used to create the synthetic image dataset used to train YOLOv5.
The synthetic trajectories fall into three categories: straight, simple curve, and piecewise. Straight trajectories are given a starting point chosen randomly in the 3D space within the field of view of the camera and follow a straight line in a random, uniformly sampled direction. Curved trajectories are given two points chosen randomly in the 3D space within the field of view of the camera and follow a circular arc between the two points using a random radius uniformly sampled in the range of $[0, 50]$~meters. Piecewise trajectories are given a starting point in 3D space within the camera field of view and follow a sequence of straight lines in randomly selected directions. They can have 1 to 5 inflection points. For all trajectories, the speed is uniform across the trajectory and sampled in the range of $[0, 100]$~meters per second, which means that the trajectory's points are uniformly sampled. 
We generate $\sim$1600 unique trajectories. Each video contains up to 5 moving objects at the same time. Figure~\ref{fig:synthetic_trajectory_comparison} shows visualizations of each type of trajectory.

\paragraph{Manually Labeled Real-World Image Dataset}
We manually labeled each frame of a subset of Dalek video recordings by drawing bounding boxes on visually identified aerial objects in each frame (airplanes, birds, leaves, etc.) to create truth labels for object bounding boxes using the free and open-source Computer Vision Annotation Tool (CVAT). For each of the eight cameras, we sampled up to three five-minute videos per day during January 2024: one at midnight, one at noon, and one at 6 PM Eastern Standard Time. Data availability determined the final dataset, comprising 314 videos, which amount to 904,257 individual frames. A total of 36,036 frames in this dataset (approximately 4\%) contain objects. Of these, approximately 94\% contain one object, 5\% contain two objects, and 1\% contain three or more objects. The highest number of objects observed in a single frame is 10. Ultimately, this dataset contains 40,268 individual object annotations (labels) which include the bounding box center coordinates, width, and height, but not object class. Throughout the process, we strove for tight bounding box definitions with minimal surplus margin, consistent labeling of all visible objects, and complete enclosure of objects within bounding boxes (even in cases of partial object occlusion). 

\paragraph{ADS-B-Derived Real-World Dataset}
\label{adsb_dataset}
We expect the fraction of aircraft not transmitting ADS-B to be small at our site (military aircraft only infrequently fly overhead). 
Using ADS-B data (see Section~\ref{ADS-B}) from the OpenSky database, we select ADS-B records whose latitude and longitude are within a N-S, E-W square of side 10~km centered on the observatory. Additionally, for each selected ADS-B record, we use the camera's extrinsic and intrinsic calibration to translate the 3D coordinates of the aircraft to 2D coordinates in the frame of the camera images. In that way, we can determine whether the aircraft's transmitted ADS-B location is within the field of view of at least one camera and, if so, whether it is above the treeline of the camera in which it is visible. The timestamp of the ADS-B record can be compared to the Dalek recording periods to determine whether the camera was recording when the aircraft was in view. We note that in addition to inaccuracies in the camera recording timestamps, there are also latencies in the ADS-B records timing and update frequencies that require relaxing any time-matching thresholds when working with this dataset: for example, \hl{ref.}~\cite{verbraak2017large} found a mean delay between the aircraft on-board time and the ground receiving time of 0.2~s, which corresponds to 2 frames in the Dalek recordings, while~\cite{ali2017evaluation} reported that 13\% of the ADS-B updates happened at intervals greater than 2~s. This dataset thus contains the trajectories of all aircraft that passed over the site during the five-month commissioning period, and for each of them, which cameras both were recording and had the aircraft in their unobstructed field of view (if not necessarily visible or detectable) at that time. In addition, the hourly weather information corresponding to every aircraft ADS-B transmission is downloaded from the Open-Meteo API~\cite{openmeteo} and joined to the dataset.

\subsubsection{YOLOv5 Benchmark on Manually Labeled Dataset}
\label{sec:manual_benchmark}
The first stage of the Dalek detection pipeline is a frame-by-frame object detection machine learning algorithm, YOLOv5~\cite{yolov5, jiang2022review}. Here, we present some basic metrics on the performance of the YOLOv5 model that we use for the rest of this paper. In particular, we explain why we decided to use the YOLOv5 model trained on the synthetic dataset alone rather than the model trained on the mixed synthetic and real-world dataset.
We apply both of the YOLOv5 models in evaluation mode, i.e., run their inference, on each frame of the 314 videos on the manually labeled dataset of Dalek real-world images. 
Next, we calculate the \textit{intersection over union} (IoU) between each detection bounding box and each ground truth bounding box. The IoU is a ratio of areas which measures the overlap between the predicted bounding box and the ground truth bounding box, providing a metric for the agreement between detections and ground truth bounding boxes. We set a low IoU threshold of 0.01 to define a match between \textit{predicted and ground truth bounding boxes}, i.e., a true positive. We consider detections with an IoU below this value as false positives.
Following the IoU calculation, we evaluate the detection performance using the following key metrics: 
\begin{itemize}
    \item \textit{Detection errors}: True positive (\hl{$TP$}), 
    false positive ($FP$), true negative ($TN$), and false negative ($FN$), in counts.
    \item \textit{Precision (Prcn)}: 
     The ratio of true positives to the sum of true positives and false positives: $Prcn = \frac{TP}{TP+FP}$.
    \item \textit{Recall (Rcll)}: The ratio of true positives to the sum of true positives and false negatives: $Rcll = \frac{TP}{TP+FN}$.
    \item \textit{Accuracy}: The ratio of correct detections (both true positives and true negatives) to the total number $TP + FP + TN + FN$.
    \item \textit{F1-score}: The harmonic mean (appropriate for finding the average of two rates) of precision and recall.
\end{itemize}

As illustrated in Table~\ref{tab:model_comparison}, the model trained on both real-world and synthetic data achieves a high precision of 85.6\%, indicating that when it makes a positive prediction, it is likely correct. However, its recall is extremely low at 6.60\%, meaning it fails to detect a large portion of the actual objects, as evidenced by the high number of false negatives (36,412). This poor recall significantly impacts the model’s overall accuracy (6.60\%) and F1-score (12.3\%), suggesting that this model struggles with generalization or may be overly conservative in making detections.
In contrast, the model trained solely on synthetic data shows a much better balance between precision and recall. While its precision (73.7\%) is slightly lower than the mixed-data model, it achieves a much higher recall (62.7\%), leading to a significantly improved F1-score (67.8\%) and accuracy (51.3\%). This model detects a greater number of true positives (24,465) and substantially fewer false negatives (14,535), indicating it is more effective at identifying objects, though it also produces more false positives (8728).

\begin{table}[H]
\caption{Comparison of metrics for YOLOv5 models trained on real-world + synthetic data versus synthetic-only data.}
\begin{tabularx}{\textwidth}{lcC}
\toprule
\textbf{Metric}       & \textbf{Real-World + Synthetic Data} & \textbf{Synthetic-Only Data} \\ \midrule
True positive (TP) count   & 2588                                & 24,465                       \\ \midrule
False positive (FP) count  & 435                                  & 8728                        \\ \midrule
False negative (FN) count  & 36,412                               & 14,535                       \\ \midrule
Precision             & 85.6\%                               & 73.7\%                       \\ \midrule
Recall                & 6.60\%                               & 62.7\%                       \\ \midrule
Accuracy              & 6.60\%                               & 51.3\%                       \\ \midrule
F1-score              & 12.3\%                               & 67.8\%                       \\ \bottomrule
\end{tabularx}
\label{tab:model_comparison}
\end{table}

\begin{figure}[H]
\begin{adjustwidth}{-\extralength}{0cm}
\centering
\includegraphics[width=\linewidth]{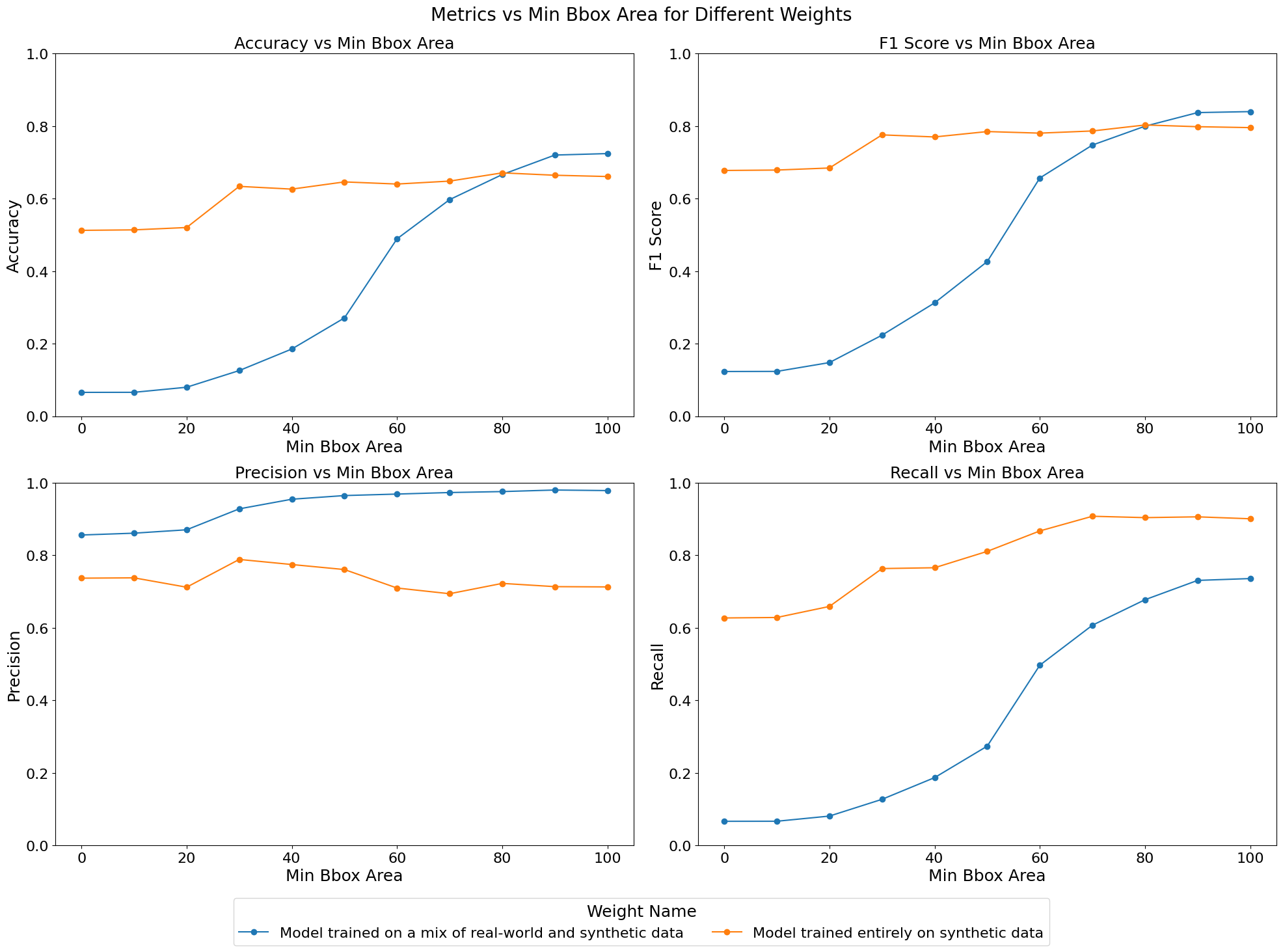}
\end{adjustwidth}
\caption{Comparison of YOLOv5 model trained on synthetic-only dataset versus model trained on a mix of synthetic and real-world datasets. The horizontal axis units are pixels square. The model trained on synthetic-only data has better overall performance and is the one used for commissioning in this paper.\label{fig:synth_vs_mix}}
\end{figure}

Given these results, we hypothesize that for larger objects, the model trained on a mix of real-world and synthetic data might perform better than its synthetic-only counterpart. Indeed real-world data could provide more accurate representations of larger objects, which might be underrepresented or inaccurately simulated in synthetic datasets.
To further explore this, we repeat the inference multiple times for each model, adjusting the bounding box area minimum threshold incrementally from 0 to 100~pixels square. The results, as shown in Figure~\ref{fig:synth_vs_mix}, reveal how the performance metrics—accuracy, F1-score, precision, and recall—vary as the minimum bounding box area increases.

For the model trained on both real-world and synthetic data, accuracy and recall improve significantly as the minimum bounding box area increases, with the most notable gains occurring between thresholds of 40 and 80 pixels square. This suggests that the model becomes more reliable when detecting larger objects, which aligns with our hypothesis. The F1-score also increases steadily, reflecting the improved balance between precision and recall as the model becomes more selective in its detections.
On the other hand, the model trained solely on synthetic data shows relatively stable performance across varying bounding box thresholds. While precision remains consistently high, recall improves slightly with higher thresholds, resulting in a stable F1-score. The accuracy of this model, however, begins at a higher value and increases more gradually, indicating that it is generally more consistent in detecting objects of varying sizes, but it might not show as much performance increase from increased object size as the mixed-data model.

The YOLOv5 stage is the first one on our object detection and tracking pipeline. False positives can still be weeded out downstream, but recovering from false negatives is challenging. Thus, the metric we care about the most in the context of the Dalek instrument is recall. Overall, the synthetic-only model demonstrates superior performance across most metrics, particularly in terms of recall and overall accuracy, making it a more reliable choice for object detection in this context. All further studies involving YOLOv5 in this paper use the synthetic-only model. In the future, further tuning and possibly the inclusion of more diverse real-world data, including more smaller-sized objects and overlays of real-world background images with synthetic objects, could help improve the performance of the mixed-data model. We expect that training with real-world background images, which have more textured skies due to weather conditions and camera optics compared to synthetic plain backgrounds, will help to decrease the number of false positives triggered by clouds, for example, and to increase the overall detection performance.

\subsubsection{Benchmark for SORT on the Synthetic Video Dataset}
\label{sec:sort_benchmark}
SORT~\cite{SORT} is a multi-object tracking algorithm which connects individual frame-by-frame object detections made by YOLOv5 into object trajectories. It relies on a Kalman filter and the Hungarian algorithm.
The Kalman filter is used in SORT to predict the future location of objects based on their past states (e.g., position, kinematics), under the assumption of constant apparent velocity in the image. The covariances in the Kalman filter represent the uncertainty associated with the predicted state. The Hungarian algorithm, on the other hand, is used to solve the assignment problem, where the goal is to associate new detections with existing tracks. The algorithm takes in a cost matrix, where each element represents the ``cost'' or mismatch distance (in our case, using the IoU) of associating a detection with a track. The Hungarian algorithm then finds the optimal matching that minimizes the total assignment cost, ensuring that each detection is paired with the most appropriate existing track.

We analyze the performance of our SORT implementation on the synthetic video dataset, which has three different types of trajectories: straight, curved, and piecewise, as shown in Figure~\ref{fig:synthetic_trajectory_comparison}. We set an IoU threshold of 0.5 to define a match between bounding boxes predicted by the Kalman filter and the ground truth bounding boxes. Each reconstructed trajectory is assigned an identifier (ID). We report the following standard metrics, in addition to the ones defined in Section ~\ref{sec:manual_benchmark}, for tracking performance~\cite{Bernardin2008,Milan2016,Yuan2009,ErgysRistani2016}: 
\begin{itemize}
    \item \textit{Identity switches (IDSs)}: Also known as association error, it is the count of  how many times the reconstructed ID associated with a ground truth trajectory changes, i.e., where objects are incorrectly re-identified as new objects.
    \item \textit{Multiple object tracking accuracy (MOTA)}: A metric compiling tracking errors over time: $MOTA = 1 - \frac{\sum_t FN_t + FP_t + IDS_t}{\sum_t TP_t + FN_t}$.
    \item \textit{Multiple object tracking precision (MOTP)}: A measure of localization accuracy: $MOTP = \frac{\sum_{t, i} d_{t, i}}{\sum_t c_t}$, where $c_t$ is the number of matches in frame $t$, and $d_{t, i}$ is the bounding box overlap of target $i$ with its corresponding ground truth box.
    \item Track quality metrics relative to ground truth trajectories that have been successfully tracked, i.e., have matched Kalman filter predictions, regardless of their predicted ID: 
    \begin{itemize}
        \item \textit{Mostly tracked (MT)}: For at least 80\% of their life; 
        \item \textit{Mostly lost (ML)}: For less than 20\% of their life;
        \item \textit{Partially tracked (PT)}: For between 20\% and 80\% of their life; 
        \item \textit{Track fragmentations (FMs)}: A count of how many times a ground truth trajectory goes from tracked to untracked status.
    \end{itemize}
    \item Identification metrics establishing a one-to-one match between ground truth trajectories and reconstructed trajectories: 
    \begin{itemize} 
        \item \textit{ID precision (IDP)}: The fraction of reconstructed trajectories which have a match;
        \item \textit{ID recall (IDR)}: The fraction of true trajectories which have a match; 
        \item \textit{ID F1-score (IDF1)}: The harmonic mean of IDP and IDR.
        \end{itemize}
\end{itemize}

\paragraph{SORT parameter optimization}
Our implementation of SORT uses three standard parameters:
\begin{itemize} 
    \item $max\_age = 100$ frames, which determines how long a reconstructed track can exist without a match before being deleted from the list of current reconstructed track;
    \item $min\_hits = 3$, which requires a track to have a minimum number of consecutive matches before being confirmed;
    \item $iou\_threshold = 0.3$, which sets the minimum IoU needed for a detection to be associated with a track;
\end{itemize}
and a custom parameter to mitigate the intermittent lack of YOLOv5 detections:
\begin{itemize}
\item $scale\_factor = 4$, which is a multiplicative factor to scale all bounding box widths and heights proportionally before running SORT.
\end{itemize} 

We confirm that the values we used for $iou\_threshold$ and $scale\_factor$ are optimal or close to optimal, balancing tracking accuracy and efficiency in our specific application.

\begin{figure}[H]
\includegraphics[width=\linewidth]{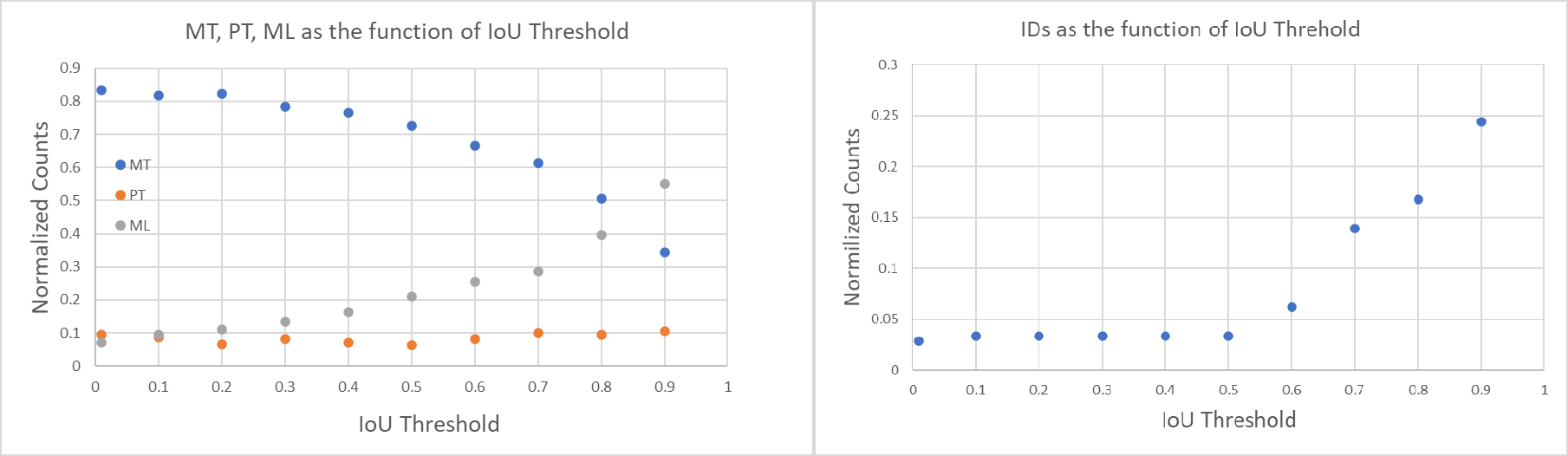}
\caption{(\textbf{Left}): Track quality metrics MT, PT, and ML as a function of IoU threshold; values are normalized by the total number of ground truth trajectories (209) and together add up to 1. Higher values of MT (mostly tracked) are better. (\textbf{Right}): IDSs normalized by the total number of ground truth trajectories (209) as a function of IoU threshold. A lower number of IDSs (track identity switches) is~better. \label{fig:sort_iou}}
\end{figure}

First, we tuned the $iou\_threshold$ by varying it from 0.01 to 0.90 and computing both tracking errors and track quality metrics. The right of Figure~\ref{fig:sort_iou} shows the number of identity switches (IDSs) divided by the total number of ground truth trajectories, across different IoU thresholds. As the threshold increases from 0.01 to 0.90, a rise in identity switches was observed, particularly at IoU thresholds of 0.60 and 0.90, where the normalized IDS peaks at 0.06 and 0.24, respectively. Lower thresholds ($\leq$0.30) exhibited far fewer identity switches, indicating fewer instances where objects were incorrectly re-identified as new objects.


The relationships between IoU thresholds and the tracking quality metrics MT, PT, and ML are shown on the left side of Figure~\ref{fig:sort_iou}. The number of ground truth trajectories remains constant at 209. The MT fraction is highest at low IoU thresholds (0.01--0.30) and worsens as the threshold increases. At thresholds $\geq$0.4, MT begins to drop monotonically, indicating that at higher IoU thresholds, fewer ground truth objects are being correctly tracked throughout their entire trajectory. The PT fraction remains relatively stable across all thresholds, with values fluctuating between 0.06 and 0.10, suggesting that partial tracking performance is not influenced by the IoU threshold. In contrast, the ML fraction increases monotonically with higher IoU thresholds, from a minimum of 0.07 at an IoU of 0.01 to a maximum of 0.55 at an IoU of 0.90. 

The results of these two track quality analyses suggests that an IoU threshold around 0.3 provides the optimal balance between minimizing identity switches (low IDSs) and maintaining accurate tracking (high MT), as it results in normalized IDSs of 0.03, MT of 0.79, ML of 0.12. In contrast, higher IoU thresholds ($\geq$0.8) lead to excessive re-identification errors, while lower IoU thresholds ($\leq$0.2) risk over-tracking due to lenient matching criteria between the Kalman filter prediction and the ground truth bounding box. In all the following studies, we use $iou\_threshold$ = 0.3.

\begin{figure}[H]
\includegraphics[width=\linewidth]{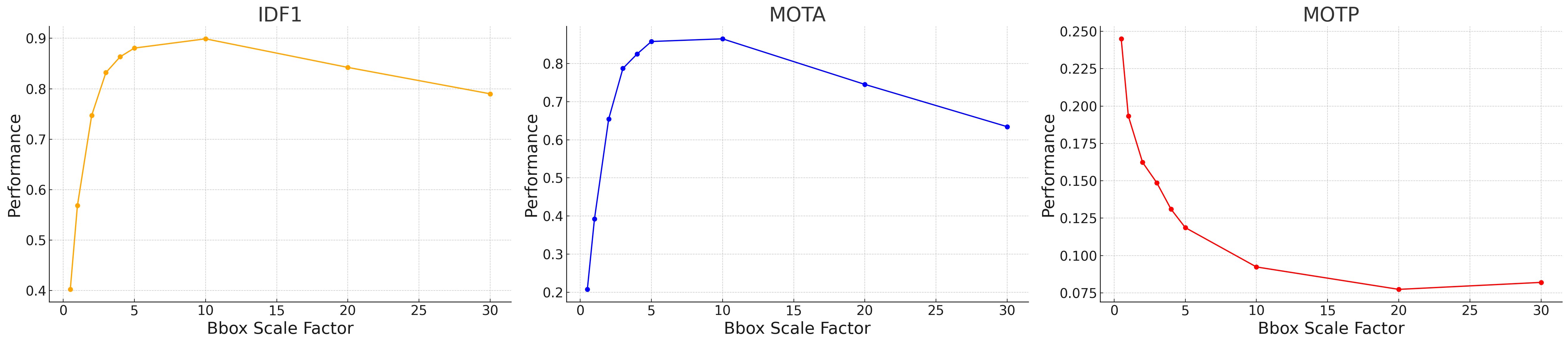}
\caption{\hl{MOTA, IDF1}, and MOTP tracking performance metrics for our SORT implementation on synthetic trajectories, as a function of the BBox scale factor.\label{fig:sort_scale}}
\end{figure}

In our implementation of SORT, instead of tuning the default parameters of the Kalman filter, we artificially increase the bounding box (BBox) size by multiplying both the height and width by a scale factor. This adjustment addresses the challenge of tracking objects with varying bounding box size, such as objects moving towards or away from the camera. Initially we picked a scale factor of 4 based on visual inspection, but the synthetic video dataset enables us to check that this choice is close to optimal.
Figure~\ref{fig:sort_scale} shows the MOTA, IDF1, and MOTP metrics as a function of the BBox scale factor, which ranges from 0.5 to 30.0. MOTA in the middle of Figure~\ref{fig:sort_scale} improves rapidly at first, peaking around a BBox scale factor of 5--10. The IDF1 graph on the left measures the tradeoff between identity precision and recall for tracking. We see a sharp rise in performance initially, with a peak around a scale factor of 5--10. Beyond a scale factor of 10, the IDF1 metric declines. The MOTP graph on the right 
shows that as the BBox scale factor increases, the MOTP decreases. 
The smallest MOTP values are observed at a scale factor of 10--20. 
This confirms that our original, heuristic choice for the BBox scale factor of 4 is not too far from optimal. Overall it results in a 40\% increase in the number of correctly matched tracks, compared to a BBox scale factor of 1.
For all the following studies, we use a BBox scale factor equal to 4.

\paragraph{Benchmark on the Synthetic Video Dataset}
 With these parameters in our SORT implementation, we also analyzed performance on the synthetic trajectories as a function of trajectory type: straight, curvy, and piecewise. The advantage of synthetic trajectories is that we control and can isolate the trajectory characteristics influencing the tracking performance.
 
The results in Table~\ref{tab:synthetic_data_metrics_transposed} show that across all three types of trajectories, we find an MOTA of 0.87. It also demonstrates that SORT exhibits strong identification precision (IDP = 95\%) and high object recall (Rcll = 90\%) across synthetic data, while maintaining a high tracking accuracy, with MOTA = 0.87. However, a moderate level of fragmentation (FM = 935) suggests that identity switches and interruptions occurred during the tracking of objects, likely due to the challenges posed by rapid or unpredictable object movements in the synthetic environment, specifically for curved trajectories. The near-perfect precision (Prcn = 98\%) indicates that the model had very few false positives. 
We investigate further the tracking performance for three types of trajectories, and as seen in Table~\ref{tab:synthetic_data_metrics_transposed}, the model's performance varies depending on the nature of the object trajectories. For the 604 straight trajectories, the model achieved its highest overall performance, reflected by an IDF1 score of 94\% and an MOTA of 0.90, due to the simplicity of tracking linear motion. The precision and recall metrics for straight trajectories are also high, with Prcn = 98\% and Rcll = 92\%. These results suggest that the model efficiently tracks objects following predictable, straightforward paths, minimizing identity switches (IDSs = 15) and fragmentation (FM = 202).
For piecewise trajectories, which involve sudden changes in object direction, the algorithm performs better than on curvy trajectories, with an IDF1 of 91\% and MOTA of 0.90. The slight improvement compared to curvy trajectories is likely due to the model's ability to adjust more easily to segmented motion changes rather than continuous curving movements. However, the number of identity switches (IDSs = 149) and fragmentation events (FM = 353) is noticeably higher for piecewise trajectories, reflecting the difficulty the model faces in consistently associating object identities across sudden directional changes.

\begin{table}[H]
\caption{\hl{SORT tracking} performance metrics for different trajectory types from the synthetic video~dataset.} 
\begin{tabularx}{\textwidth}{lCCCC}
\midrule
\textbf{Metric}       & \textbf{Curved}     & \textbf{Piecewise} & \textbf{Straight} & \textbf{All} \\ \midrule
\textbf{\hl{FP}} count     & 1757                & 1053     & 800       & 3610             \\ \midrule
\textbf{\hl{FN}} count     & 9871                & 5208     & 3258     & 18,337            \\ \midrule
\textbf{\hl{Recall}}       & 87\%                & 92\%     & 92\%     & 90\%             \\ \midrule
\textbf{\hl{Precision}}    & 97\%                & 98\%     & 98\%     & 98\%             \\ \midrule
\textbf{\hl{IDS}} count    & 122                 & 149      & 15       & 286              \\ \midrule
\textbf{\hl{MOTA}}         & 0.84                & 0.90     & 0.90     & 0.87             \\ \midrule
\textbf{\hl{MOTP}}         & 0.14                & 0.13     & 0.14     & 0.14             \\ \midrule   
\textbf{\hl{FM}} count & 380      & 353      & 202      & 935 \\ \midrule 
\textbf{\hl{IDP}}      & 95\%     & 94\%     & 97\%     & 95\%             \\ \midrule
\textbf{\hl{IDR}}      & 85\%     & 88\%     & 91\%     & 87\%             \\ \midrule
\textbf{\hl{IDF1}}     & 89\%     & 91\%     & 94\%     & 91\%             \\ \midrule
\end{tabularx}
\label{tab:synthetic_data_metrics_transposed}
\end{table}

\section{Commissioning Results}
\label{sec_results}

In this section, we focus on the Dalek as a whole system: its eight IR cameras and the pipeline from data to reconstruction of aerial events. We present basic commissioning checks on the recorded dataset and corresponding reconstructed quantities such as individual frame-by-frame detections or object trajectories. After these, we further characterize the performance of the Dalek and the detection pipeline as a function of atmospheric conditions and object characteristics using two distinct datasets: an ADS-B-derived dataset, which focuses on airplanes in real-world data, and a synthetic dataset.

\subsection{Basic Checks on Recorded Dataset}



\subsubsection{Environment Effects on Spatial Distribution of Detection Counts per Camera}
\begin{figure}[H]
\begin{adjustwidth}{-\extralength}{0cm}
\centering
\includegraphics[width=5.3cm]{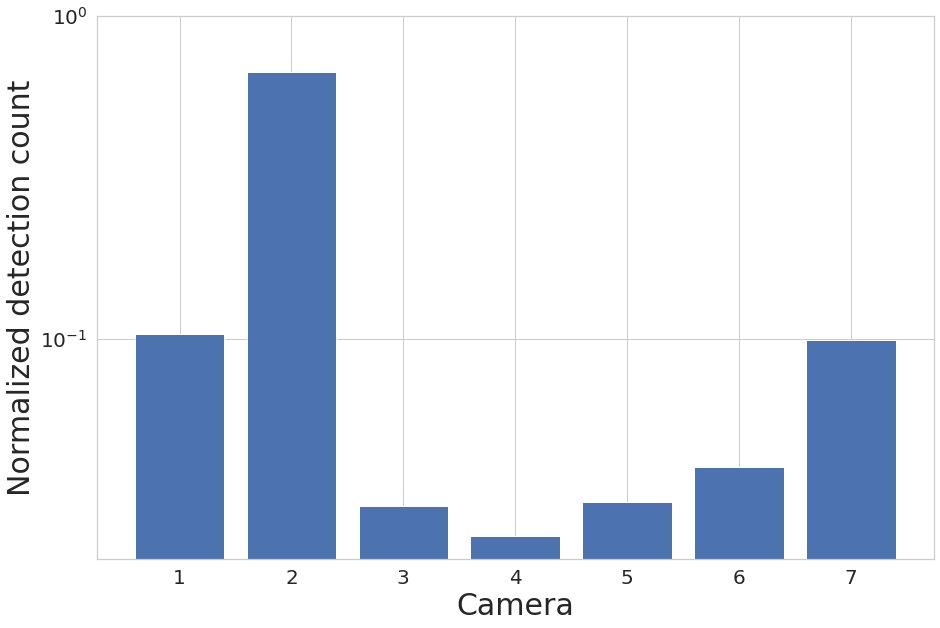}\quad
\includegraphics[width=5.3cm]{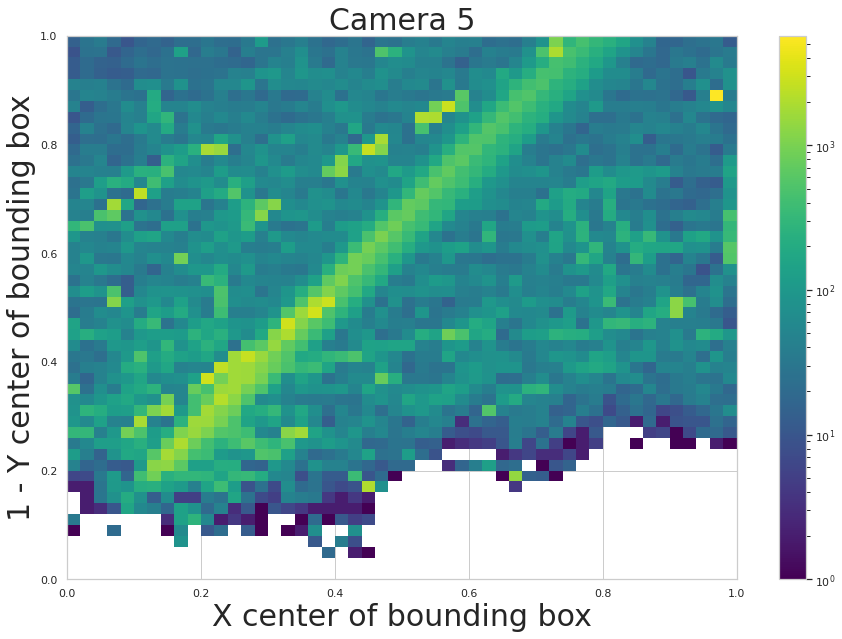}\quad
\includegraphics[width=5.3cm]{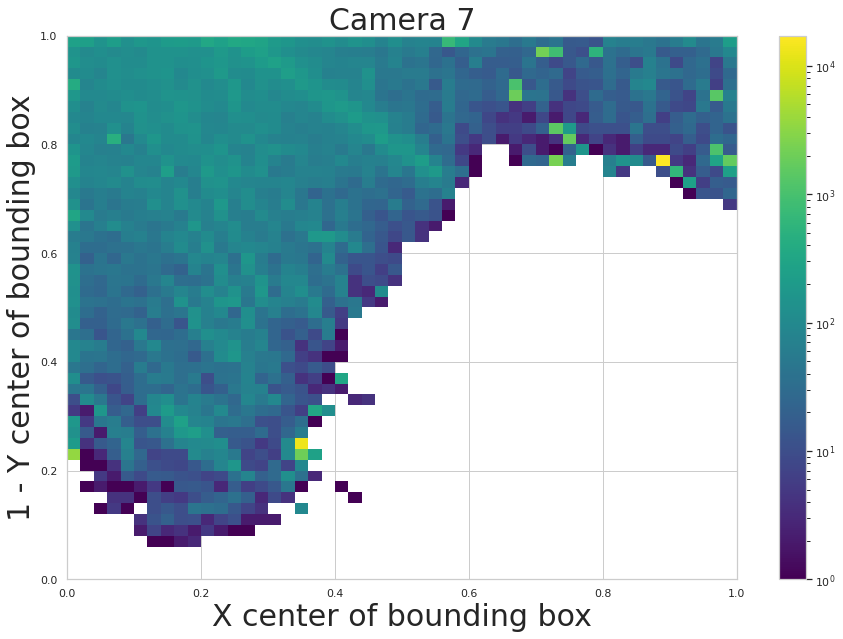}
\end{adjustwidth}
\caption{(\textbf{Left}): YOLO detection count per camera, first divided by the camera's visible sky area and total recordings count, and then normalized across cameras so all values sum up to 1. This histogram includes all detections of any object from a subset of the five months of commissioning data. Camera 8 was offline during this interval. (\textbf{Middle} and \textbf{Right}): 2D histograms showing examples of spatial distribution of YOLO detections with confidence score >0.5, for the month of May 2024, for camera 5 (SE) and 7 (SW), respectively. A regular commercial aircraft route is clearly visible for camera 5 (center).\label{fig:detection_space}}
\end{figure}

 We start with the detection counts per camera, and attempt to correlate any long-term spatial inhomogeneities with known conditions at the observatory site.  
We check each camera for its spatial distribution of YOLO detections to rule out any potential bias. 
ADS-B historical records confirm that cameras 5 and 7 capture airplane high-traffic routes. This is shown in the center of Figure~\ref{fig:detection_space}, where a commercial airline approach route to a major airport is visible when plotting purely detection counts. We also find that cameras 1 and 4 often capture the Moon at night. 
 which triggers a detection by YOLOv5. Additionally, detections caused by birds' preferential perches or nests have small spatial signatures at the edges of the treeline.
New leaves and growth of the trees beyond the treeline image mask that we utilize may trigger an unusually high count of detections in certain months. Some spurious points may be due to dust on the lens of the Boson camera or imperfections in the coating of the germanium windows. Blowing leaves and the edges of clouds are a source of many detections, but we expect them to be randomly distributed over time scales of months. And finally, the cameras regularly display a green square in the top right corner of the image during the automatic flat-field calibration, which can be detected by YOLO and shows up as an unusually high bin in this histogram, clearly visible for camera 5, Figure~\ref{fig:detection_space} (center image).

\subsubsection{Cross-Camera Check} We check cross-camera variations of the detection counts by normalizing the detection count by area of the sky above the treeline, and by total recorded time; some cameras pointing north are constantly in recording mode with their sunshades open, while others alternate between recording and pause mode, sunshade open and closed, depending on the sun position in the sky. Figure~\ref{fig:detection_space} shows on the left that the average normalized detection count is not uniform across cameras. We expect that air traffic routes or the Moon, mentioned previously, can bias this metric for certain cameras. 

\subsubsection{Bounding Box Properties}
\begin{figure}[H]
\begin{adjustwidth}{-\extralength}{0cm}
\centering
\includegraphics[width=7cm]{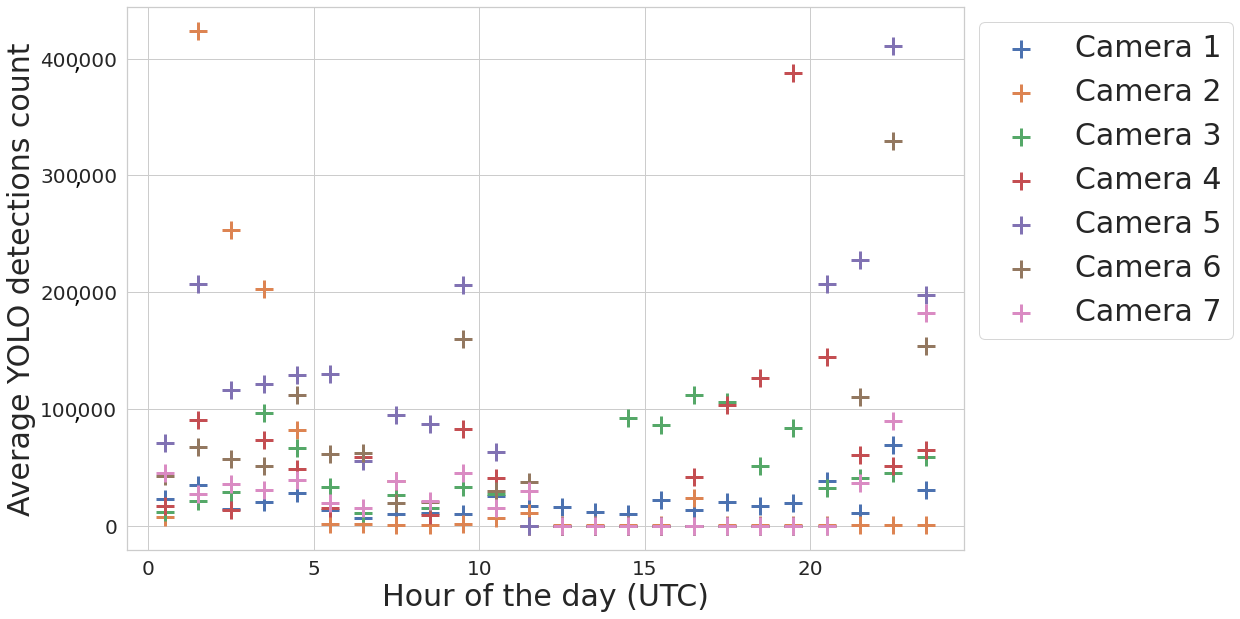}\quad
\includegraphics[width=5cm]{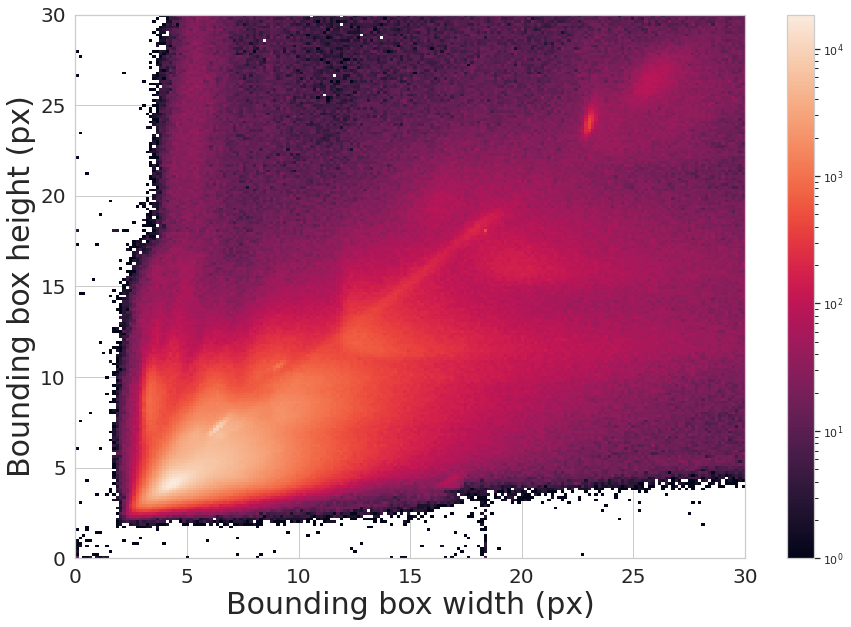}\quad
\includegraphics[width=5cm]{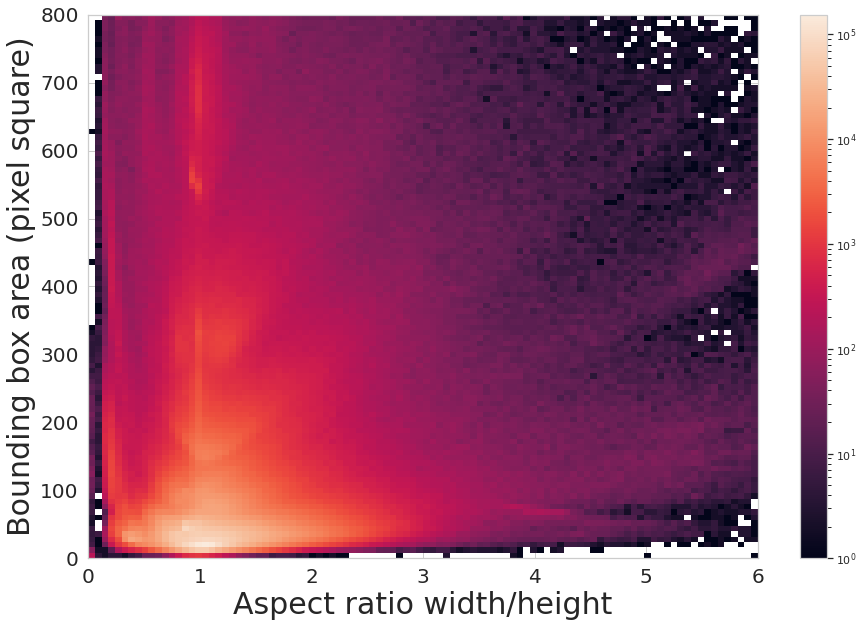}
\end{adjustwidth}
\caption{\hl{(\textbf{Left}): Average} hourly detection count for different cameras throughout the day. Middle: A 2D histogram of the detections' bounding box width and height, showing pronounced clustering that may be useful for object classification. (\textbf{Right}): A 2D histogram of the detections' bounding box area and aspect ratio, again with strong clustering. The histograms include all detections of any object from a subset of the five months of commissioning data.\label{fig:detection_time}}
\end{figure}

The middle of Figure~\ref{fig:detection_time} is a histogram of YOLOv5's detection bounding box width versus height and highlights a potentially useful feature, where distinctly different populations of detections appear as elongated clusters. The histogram 
represents all detections, not just airplanes, and currently we can only speculate about which object falls into which cluster. For example, the population cluster close to a bounding box aspect ratio of 1 would correlate with the regular Moon detections. Clustering also appears in the right-hand side of Figure~\ref{fig:detection_time}, which compares bounding box aspect ratio with its area. The variety in bounding box size and aspect ratio suggests that they may be useful features in the future to help classify objects and search for outlier populations.

After these basic checks on the Dalek detections, we further studied the performance of the Dalek and the detection pipeline, including as a function of atmospheric conditions and object characteristics, on two distinct datasets: an ADS-B-derived dataset, which focuses on airplanes in real-world data, and a synthetic dataset.

\subsection{Performance Evaluations Using ADS-B-Equipped Aircraft}
In this section, we evaluate the physical performance envelope of the full Dalek system using data related to the many ADS-B-equipped airplanes (Section~\ref{ADS-B}) that fly over our development site. 
We use the latitude, longitude, and altitude from an ADS-B-derived real-world dataset as the aircraft's true position, project these onto each 2D image frame, and compare these true points to the objects predicted by YOLOv5's detection algorithm.
\subsubsection{Evaluation Methodology}
For this analysis we define sets of ADS-B entries using the following sequential selection criteria:
\begin{enumerate}
    \item Within a square of side 10~km centered on the observatory;
    \item Within the field-of-view of at least one camera;
    \item Above the treeline of the camera from which they should be visible;
    \item At a time when there is a recording by the relevant camera; 
    \item Are detected by YOLOv5.
\end{enumerate} 

For each selected ADS-B record, we use the camera's extrinsic and intrinsic calibration to translate the 3D coordinates of the aircraft to 2D coordinates in the frame of the camera images. We match ADS-B records with existing bounding box detections using a temporal threshold of 5~seconds and a spatial threshold of 30~pixels. These thresholds are generous to account for the imprecision of the recording timestamps in the time period we selected and the potential error in the reported ADS-B positions. 

To define performance metrics, we first identify the number of aircraft within range of our site (criterion 1) and define these as \textit{in range}. From this set, we identify those that were in the right location to have been viewed by the Dalek, that is, within the effective field of view of at least one camera (criteria 2 and 3). The set of aircraft that meet the first three criteria outlined above is defined as \textit{viewable}, and it is independent of airplane size, aspect, distance (under 10 km), ambient IR conditions, cloud cover, and camera uptime. We then identify the subset of these viewable aircraft who were also in the camera's recording window (criterion 4), that is, they were in the right place and at the right time. The set of aircraft that meet the first four criteria is defined as \textit{recorded}.

The \textit{acceptance} is related to instrument uptime and is defined here as the number of \textit{recorded} aircraft divided by the number of \textit{viewable} aircraft. In other words, it is the fraction of viewable aircraft that the Dalek managed to record; these aircraft were ``accepted'' into the analysis pipeline. 
The set of image frames with recorded ADS-B aircraft is then compared with YOLOv5 detections for the same frames (criterion 5). Our final set consists of aircraft matching the time and place of objects detected by YOLOv5, meeting all five criteria. We define this set as \textit{detected}.

The \textit{efficiency} is related to YOLOv5's performance, and is defined here as the number of \textit{detected} aircraft divided by the number of \textit{recorded} aircraft. In other words, it is the fraction of aircraft meeting the first four selection criteria that are also detected by the YOLOv5 object detection stage. If our pipeline is efficient, it converts most of the objects that enter the pipeline into detections, with a minimum of loss. This metric can depend on aircraft size, aspect, distance, and atmospheric conditions. 
 If our Dalek system pipeline and site location were perfect, all aircraft within our site range limit would be in our \textit{detected} set. Our pipeline is not perfect, and these sequential metrics help us find areas on which to focus improvements. We now examine these metrics. 



\subsubsection{Performance Results}
Overall, we counted 27,467 airplanes that flew within a radius of 10~km centered on the observatory, of which 8550 met the four selection criteria outlined above, and, of these, 3678 were  matched in at least one frame to a YOLOv5 detection bounding box,  meeting criteria 1--5. 
We use this number to estimate the fraction of all objects reconstructed during the commissioning period that is made up of aircraft. Taking the total count of reconstructed object tracks from Section~\ref{sec:observations}, which includes aircraft, birds, leaves, clouds, etc., we estimate that $\sim$0.7\% of all reconstructed object tracks were associated with ADS-B-equipped aircraft. 

\begin{figure}[H]
\begin{adjustwidth}{-\extralength}{0cm}
\centering
\includegraphics[width=11cm]{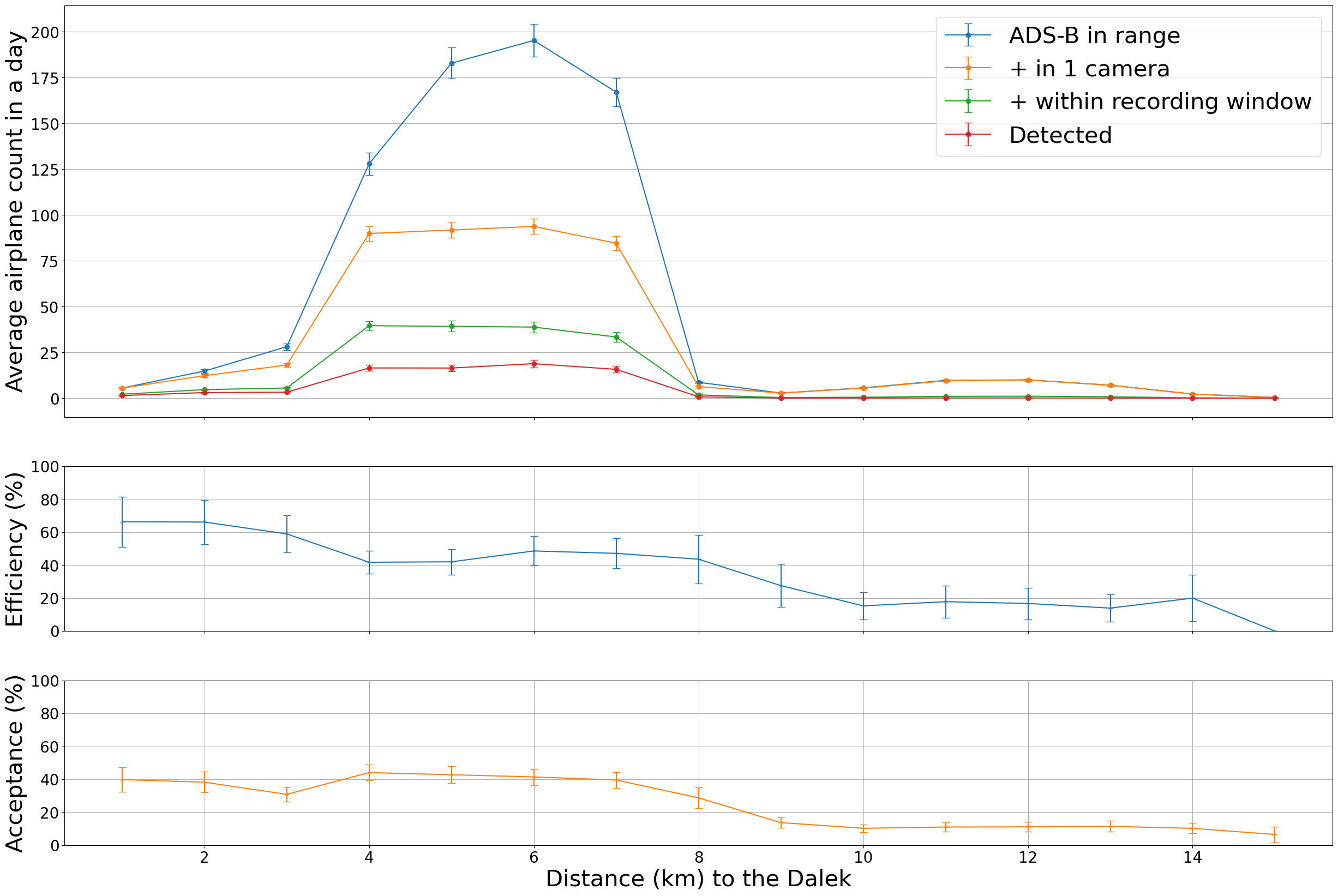}\quad
\includegraphics[width=5.5cm]{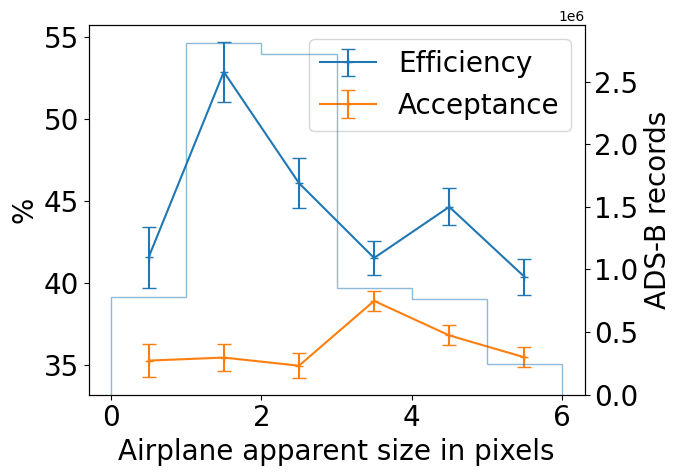}
\end{adjustwidth}
\caption{Upper left: For the commissioning period, average daily count vs. distance (km) from observatory of ADS-B-equipped airplanes in range of site (criterion 1: \textit{in range}); of those, also within the effective field of view of at least one camera (criteria 1--3: \textit{viewable}); of those, also within that camera's recording time window (criteria 1--4: \textit{recorded}); and, of those, also detected by YOLOv5 (criteria 1--5: \textit{detected}). Middle left: The number \textit{detected} divided by the number \textit{recorded} (\textit{efficiency}) vs. distance (km). Lower left: The number \textit{recorded} divided by the number \textit{viewable} (\textit{acceptance}) vs. distance (km). Right:  \textit{Acceptance} and \textit{efficiency} as functions of apparent airplane size. The histogram shows the number of ADS-B records that contributed to each point on the graph. Error bars are computed by propagating statistical errors from all ADS-B counts, assuming Poisson distributions.\label{fig:adsb_eff}}
\end{figure}

We use the five months of data from the commissioning period to examine our ability to reconstruct aerial objects such as aircraft from their physical parameters using only our Dalek system. We look at the sequence of processed information, from all the aircraft that passed overhead, to the detections that come out of YOLOv5. We can quantify the falloff from the average daily count of ADS-B airplanes within range of our site, to those also within the effective field of view of at least one camera, to those also within that camera's recording time window, and finally to those also detected by YOLOv5. 
For example, the upper left panel of Figure~\ref{fig:adsb_eff} shows this  progression in average daily counts. For a given aircraft, we bin its ADS-B records according to their distance to the site, before counting unique aircraft in each distance bin. For example, we expect, on average, within $\sim$10 km of the site, $\sim$210 airplanes to be viewable in a day, of which $\sim$76 pass by within a recording window and $\sim$31 are detected. At a distance of 10~km or less, the mean efficiency is 43\%, the rest of the recorded airplanes at this distance are not detected by YOLOv5. The detection efficiency decreases with increasing distance of the airplane from the observatory. We also see from examining the ADS-B data that there are flight lanes to regional airports within 2--5~km horizontal distance to the observatory, which, accounting for typical aircraft altitudes, correlates with the peak of aircraft in range being around $\sim$5--7~km.

On the right of Figure~\ref{fig:adsb_eff}, we examine the effect of pixel size on both acceptance and efficiency.
The aircraft's apparent size in pixels can be inferred from the track angle (heading), reported in most ADS-B records, the 3D location of the aircraft with respect to the observatory, the true aircraft size, and the camera sensor's field of view. We take the maximum between the projected width and projected length. If the track angle is not available, we assume that the aircraft track is orthogonal to the line of sight.

\begin{figure}[H]
\begin{adjustwidth}{-\extralength}{0cm}
\centering
\includegraphics[width=6cm]{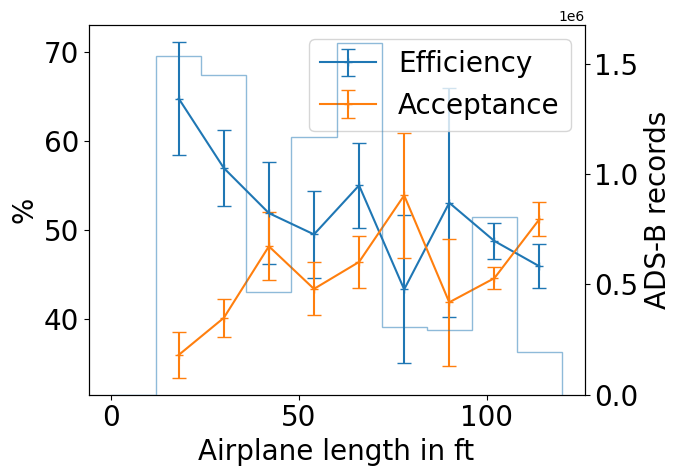}\quad
\includegraphics[width=5cm]{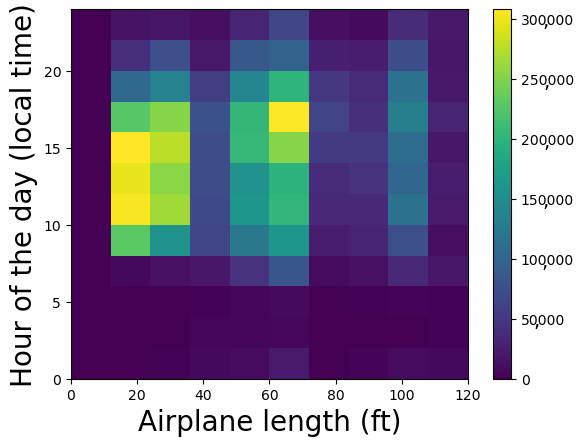}\quad
\includegraphics[width=5cm]{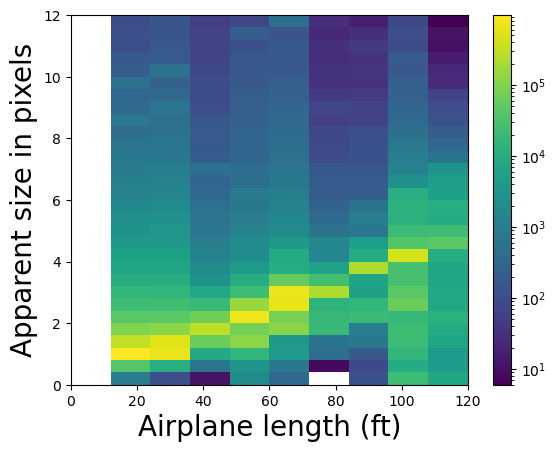}
\end{adjustwidth}
\caption{\hl{(\textbf{Left}): Efficiency} as a function of actual airplane size. Error bars are computed by propagating statistical errors from all ADS-B counts, assuming Poisson distributions. (\textbf{Middle}): Actual airplane size distribution throughout the day. (\textbf{Right}): Actual airplane size distribution compared to projected size in the image. \label{fig:adsb_size}}
\end{figure}

We see that acceptance varies slightly with apparent size in pixels. It is counter-intuitive for there to be a relationship  between the \textit{recorded} set and pixel size, unless airplane apparent pixel size is related to field of view or recording time interval. Recording intervals and aircraft flight schedules both depend on the time of day, and aircraft sizes are related to their schedules. Large aircraft, on regular commercial flights on approach in the evening, small to mid-sized private jets arriving at the end of the day, and small propeller planes operated by a nearby flight school at mid-day, all have their own distinctive flight paths and schedules. For example, a flight path used by large commercial jets on approach to a major airport can be seen in the center of Figure~\ref{fig:detection_space}, with many of these flights arriving in the early evening, when they are captured and detected in camera 5 (faces south-east with sunshade open). The correlation between aircraft size and time of day is shown in the center of Figure~\ref{fig:adsb_size}. We see from the efficiency curve in the right of Figure~\ref{fig:adsb_eff} that YOLOv5 detections, on the other hand, are least efficient at one pixel, increase with pixel size until a maximum at two pixels, and then drop again until six pixels. This matches the known behavior of YOLOv5 related to pixel size, as seen in trials using synthetic data.  

\vspace{-6pt}

In Figure~\ref{fig:adsb_size}, we examine the actual size of aircraft and its effect on acceptance and efficiency in the Dalek dataset, using length as a proxy for size. Again, the acceptance is correlated with the airplane's real size. We speculate this is because, again, both aircraft size and sunshade coverage depend on the time of day. The 2D histogram in the middle of Figure~\ref{fig:adsb_size} shows that small airplanes, many of which are associated with a nearby flight school, are more likely than large airplanes to fly in the middle of the day, when we have more camera shades closed to avoid exposure of the sensors to the sun. 
The efficiency, on the other hand, decreases when the airplane real size increases, which is also counter-intuitive at first. The right plot in the same figure shows that, on average, small and large airplane lengths correlate with a larger apparent size distribution width, which is likely a consequence of the specific air traffic routes that are visible from our site. Small planes tend to fly at lower altitudes over our site and the large planes fly at higher altitudes, such that these planes preferentially fall into the optimum apparent size for YOLOv5 detections of between two and four pixels.  

\begin{figure}[H]
\begin{adjustwidth}{-\extralength}{0cm}
\centering
\includegraphics[width=5.5cm]{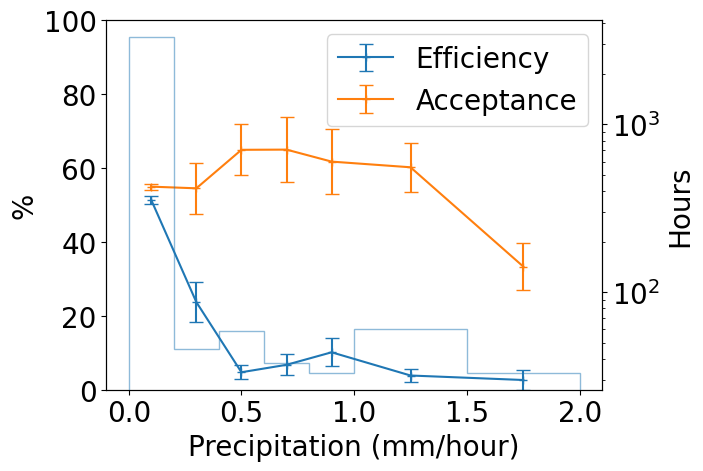}\quad
\includegraphics[width=5.5cm]{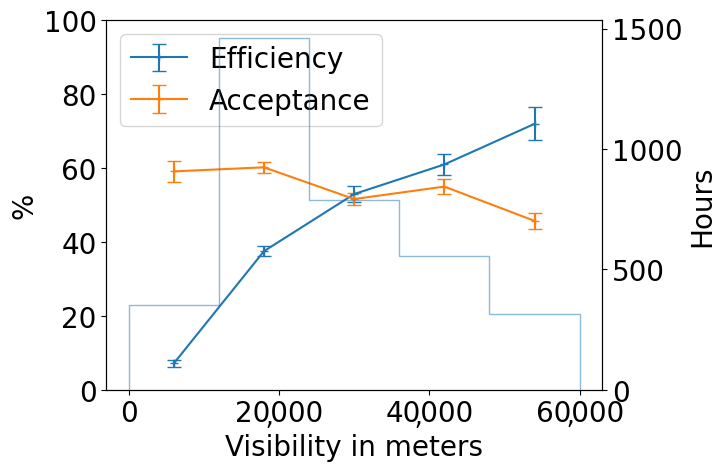}\quad
\includegraphics[width=5.5cm]{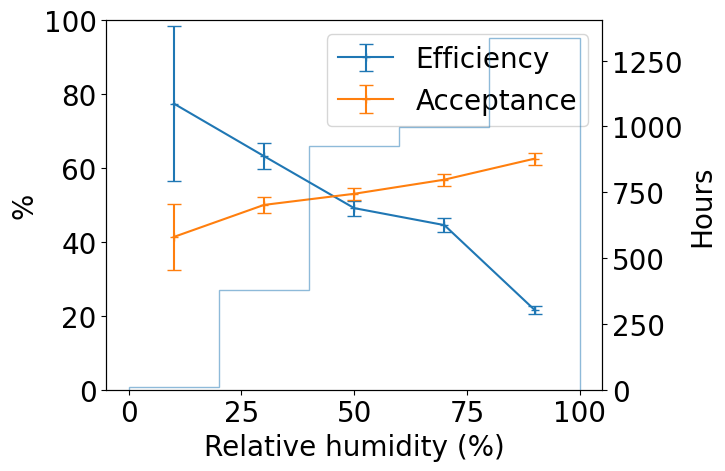}
\end{adjustwidth}
\caption{\hl{(\textbf{Left}): Efficiency as} a function of precipitation. (\textbf{Middle}): Efficiency as a function of visibility. (\textbf{Right}): Efficiency as a function of relative humidity. Error bars are computed by propagating statistical errors from all ADS-B counts, assuming Poisson distributions.\label{fig:adsb_cloud}}
\end{figure}

\begin{figure}[H]
\includegraphics[width=5.5cm]{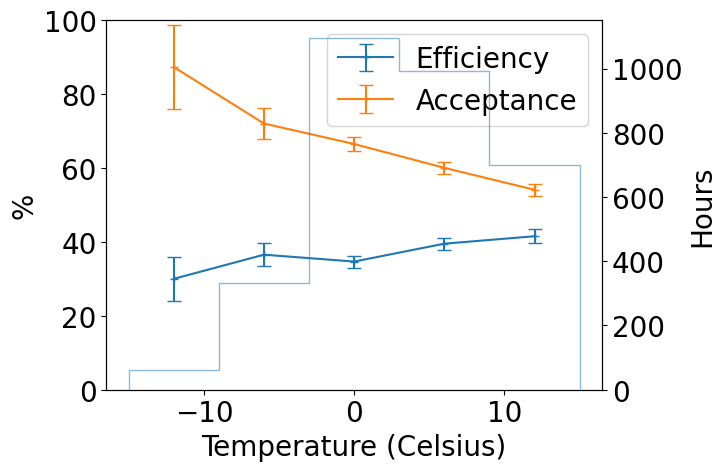}\quad
\includegraphics[width=5.5cm]{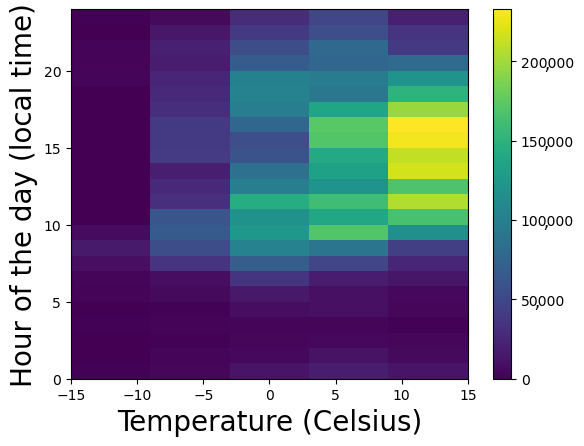}
\caption{\hl{(\textbf{Left}): Efficiency} as a function of temperature. (\textbf{Right}): Uneven temperature distribution throughout the day in our dataset. Error bars are computed by propagating statistical errors from all ADS-B counts, assuming Poisson distributions.\label{fig:adsb_temp}}
\end{figure}

Using weather data that are correlated with ADS-B records (see Section \hl{\ref{sec:datasets}}), 
we examine the effects of precipitation, visibility, relative humidity, and temperature on the acceptance (recording) and efficiency (detection) of airplanes. As seen in Figure~\ref{fig:adsb_cloud}, precipitation, visibility, and relative humidity have only a slight impact on acceptance, as might be expected. However, a clear relationship between temperature and acceptance is shown in Figure~\ref{fig:adsb_temp}, contrasting with the other weather parameters, with acceptance decreasing monotonically as temperature increases. This unexpected relationship is due to the fact that both the daily temperature cycle and the sunshade schedule depend on the sun's daily cycle, and acceptance is dependent on the sunshade schedule. We postulate that the warmer it is, the more likely that the south-facing sunshades are closed, and it is a time when the airport to the south is active.

The same figures show that increased precipitation, decreased visibility, and increased humidity have a marked adverse effect overall on efficiency. This is consistent with how moisture affects IR atmospheric transmission and therefore object image quality: we know that 
the atmosphere's relative humidity impacts  transmission in the infrared wavelengths. 
Temperature mildly affects efficiency, as shown in Figure~\ref{fig:adsb_temp}.

\begin{figure}[H]
\begin{adjustwidth}{-\extralength}{0cm}
\centering
\includegraphics[width=5.5cm]{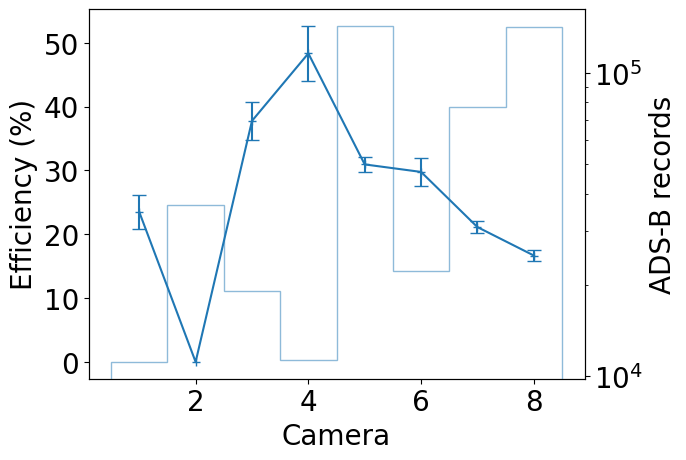}\quad
\includegraphics[width=5.5cm]{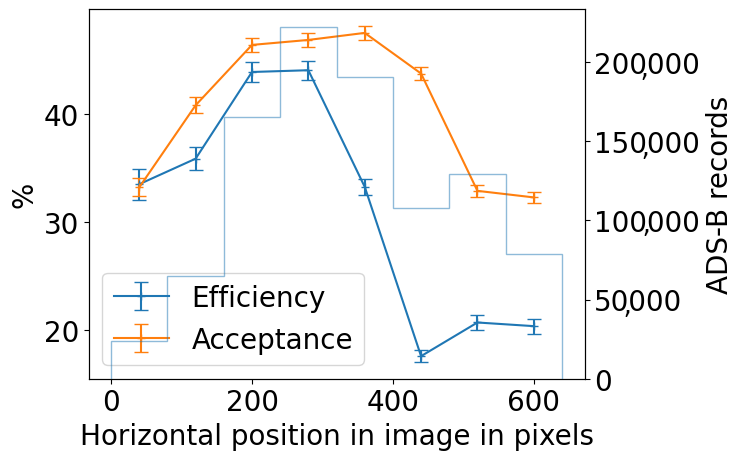}\quad
\includegraphics[width=5.5cm]{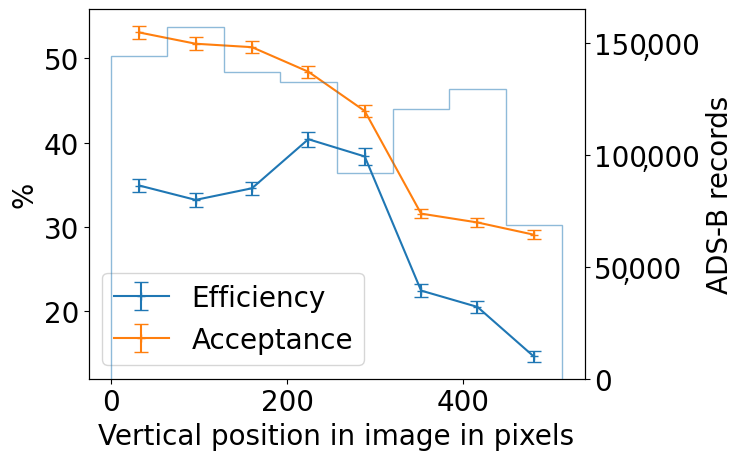}
\end{adjustwidth}
\caption{\hl{(\textbf{Left}): Efficiency} as a function of camera. (\textbf{Middle}): Efficiency as a function of horizontal position. Zero corresponds to the left of the image. (\textbf{Right}): Efficiency as a function of vertical position. Zero corresponds to the top of the image. Error bars are computed by propagating statistical errors from all ADS-B counts, assuming Poisson distributions.\label{fig:adsb_pos}}
\end{figure}

We now look at efficiency as a function of camera and pixel location within the image frame, in Figure~\ref{fig:adsb_pos}. Camera~8 has a wider FOV and more fish-eye lens distortion compared to the other seven cameras, and we see that the detection efficiency in camera 8 is lower than most cameras, with the exception of camera 2. Camera 2's extremely low efficiency in detecting aircraft that were in view and recorded is difficult to explain. This is the most northerly facing camera and perhaps this creates a lighting condition, possibly related to illuminated clouds (see Figure~\ref{fig:dalek_mosaic}, that is not optimal for detection of aircraft and will need to be explored further.
Looking at all the cameras at once, we see a dip in efficiency in the bin closest to the image frame horizontal position of 400~px. This may be related to the fact that five of the eight cameras have significant treeline masks at this position, including a radio antenna tower in camera 1, whose silhouette is included in the treeline mask. Since detection bounding boxes which overlap with masks are removed from this analysis, an airplane passing nearby may not have matching detections in these areas even if it is successfully detected elsewhere in the image. In the vertical axis (the top of the lens is at vertical position 0), the efficiency tends to drop as the view drops to lower elevation angles, where planes are typically more distant. The bump in efficiency at the center of the field of view may be due to the fact that the center of the germanium window is generally the cleanest and least corroded part of the protective window. Again, these issues will need to be explored further. 

\begin{figure}[H]
\includegraphics[width=\linewidth]{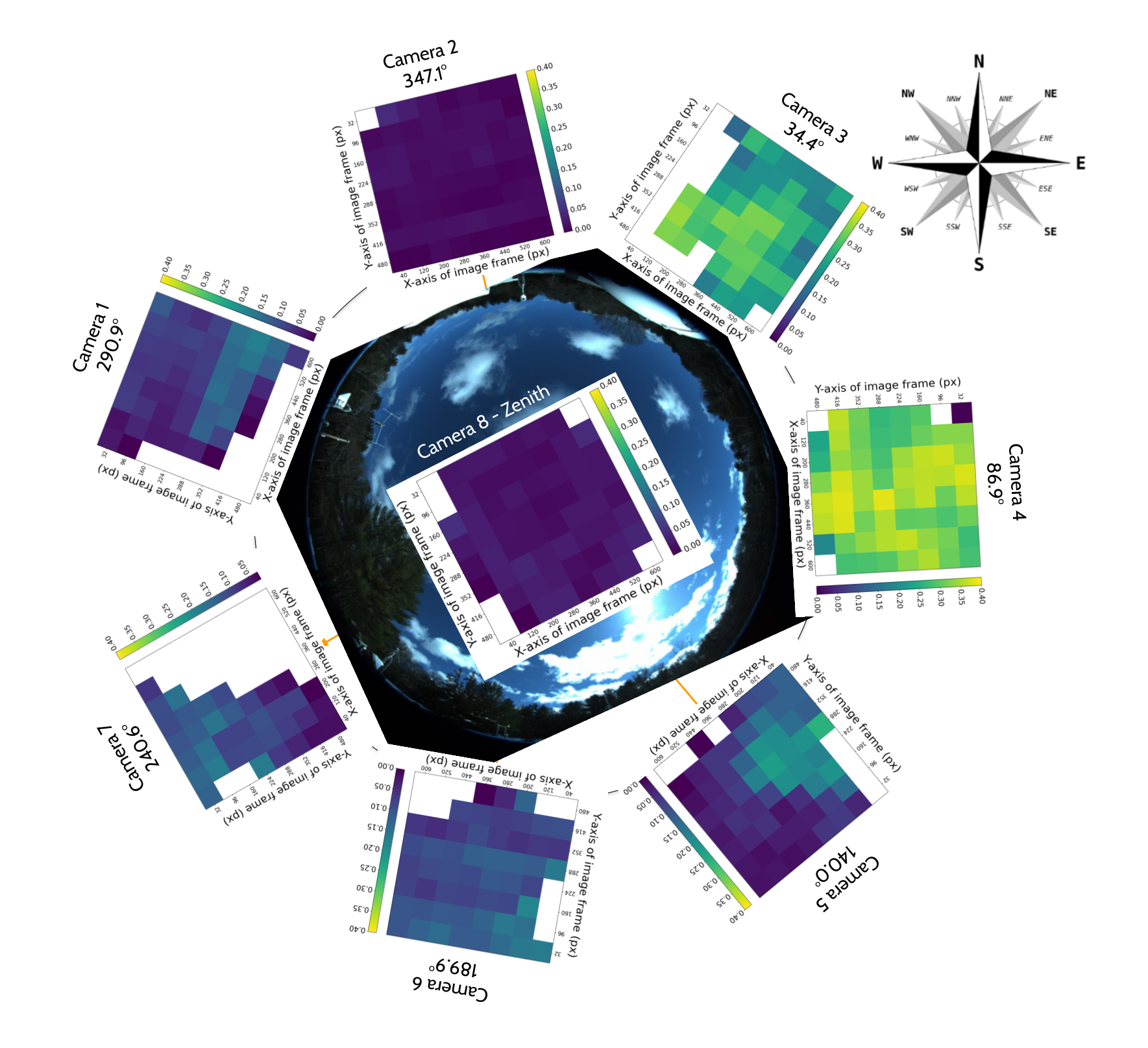}
\caption{\hl{Fraction} 
 of aircraft \textit{in range} which are \textit{detected} for each camera and spatial location in the camera image frame. Each bin represents an area of size $80\times64$~px in the original camera image. The origin of both axes is set to the upper left corner, following the computer vision convention. White bins are either entirely included in the treeline mask, where we do not look for detections, or happen to lack any aircraft in range during the commissioning period.\label{fig:adsb_mosaic}}
\end{figure}

Overall, the Dalek frame-by-frame object detection stage has an acceptance of 41\% for ADS-B-equipped aircraft within a square of side 10~km centered on the observatory, and a mean detection efficiency of 36\%. Acceptance is heavily dependent on range, field of view, and camera uptime. Efficiency depends very sensitively on weather conditions such as precipitation, atmospheric visibility, and relative humidity, as well as the condition of the lens, dust, and raindrops. It is also not perfectly uniform over camera images due to boundary effects of the detection algorithm and interference from treeline growth. Figure~\ref{fig:adsb_mosaic} summarizes the fraction of aircraft \textit{in range} which are \textit{detected} per camera and spatial location within the camera image frame. This visualization convolves the effects of recording schedule, downtime, weather conditions, detection efficiency, and underlying aircraft statistics, such as flight lanes from the BOS airport.

\subsection{Evaluation Using a Synthetic Dataset}
While ADS-B records allow us to evaluate the performance of the full pipeline on real-world data, airplanes, which have predictable trajectories and a limited range of kinematics and characteristics, are not the only objects that we detect using the Dalek. To estimate how reconstruction performance might extend to known objects with different kinematics and to quantify this based on the object trajectory properties, we generate a synthetic video dataset using a variety of objects exhibiting straight, curved, and piecewise trajectories, as described in Section~\hl{\ref{sec:datasets}}.  

 We first run YOLOv5 for object detection on the individual frames of the videos, and keep only bounding boxes larger than $3\times3$ pixels from both true (labels) and predicted (detections) bounding boxes. 
 We consider it a match if a true and predicted bounding box have an intersection over union (IoU) greater than 10\%. A total of 14\% of the true trajectories do not contain any points with a bounding box matched to a YOLOv5 detection. 
 
 In the next step, we run SORT on YOLOv5's output to form reconstructed trajectories from the frame-by-frame detections. We find that 27\% of object detections, although matched to a true bounding box of an object that has a true trajectory, are not then associated with any reconstructed trajectory. Conversely, 
we are also interested in the fraction of true trajectories which contain at least one matched bounding box, but where these individual detections are not reconstructed into a predicted trajectory. We find that 8\% of the true trajectories in this dataset are not reconstructed even though SORT was given matched detections from YOLOv5. Table~\ref{tab:yolo_sort_synthetic} breaks down these metrics for each of the three types of trajectory.

\begin{table}[H]
   \caption{Breakdown of object detection and trajectory reconstruction metrics for different types of trajectories in the synthetic dataset.}
    \label{tab:yolo_sort_synthetic}
   
\begin{adjustwidth}{-\extralength}{0cm}
 \begin{tabularx}{\fulllength}{Lccc}
        \toprule
        \textbf{Trajectory Type} & \textbf{Curved} & \textbf{Straight} & \textbf{Piecewise} \\
        \midrule
        Unique true trajectories & $\sim$430 & $\sim$740 & $\sim$440 \\
        \midrule
        Detection precision \textit{(fraction of detections which are matched to a true bounding box)} & 50\% & 41\% & 51\% \\
        \midrule
        Detection recall \textit{(fraction of true bounding boxes which are matched to a detection)} & 92\% & 54\% & 91\% \\
        \midrule
        Fraction of matched object detections without an associated reconstructed trajectory & 29\% & 29\% & 24\% \\
        \midrule
        Fraction of true trajectories which have at least one matched bounding box & 99.8\% & 81\% & 99\% \\
        \midrule
        Fraction of true trajectories which have at least one matched bounding box but no associated reconstructed trajectory & 7\% & 15\% & 5\% \\
        \midrule
        Efficiency \textit{(fraction of true trajectories matched to at least one reconstructed trajectory)} & 94\% & 74\% & 96\% \\
        \midrule
        Purity \textit{(fraction of reconstructed trajectories matched to a true trajectory)} & 90\% & 83\% & 90\% \\
        \bottomrule
    \end{tabularx}
\end{adjustwidth}
    
\end{table}
We compute the efficiency or fraction of true trajectories which are matched to at least one reconstructed trajectory, as well as the purity or fraction of reconstructed trajectories which match with a true trajectory. A match between a true and a reconstructed trajectory is declared when at least three individual bounding boxes from the reconstructed trajectory are matched to true object bounding boxes, i.e., with an IoU greater than 10\%. This means that a true trajectory can be matched to multiple reconstructed trajectories if it was fragmented during the reconstruction, which can make the purity higher than it would be if we were performing strict one-to-one matching. The efficiency and purity metrics per trajectory type are shown in Table~\ref{tab:yolo_sort_synthetic}. Overall, we find an efficiency of 81\% and a purity of 87\%. 

\begin{figure}[H]
\begin{adjustwidth}{-\extralength}{0cm}
\centering
\includegraphics[width=5.5cm]{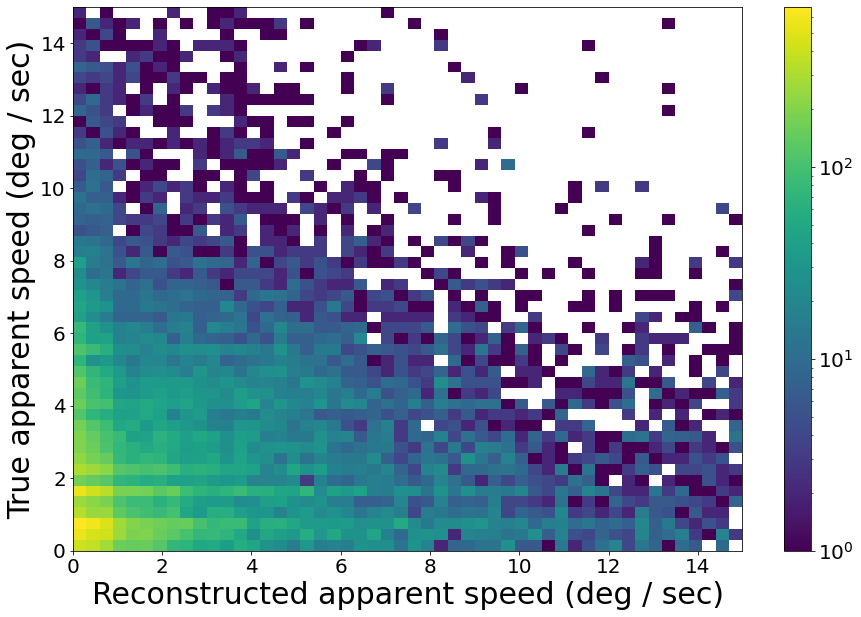}\quad
\includegraphics[width=5.5cm]{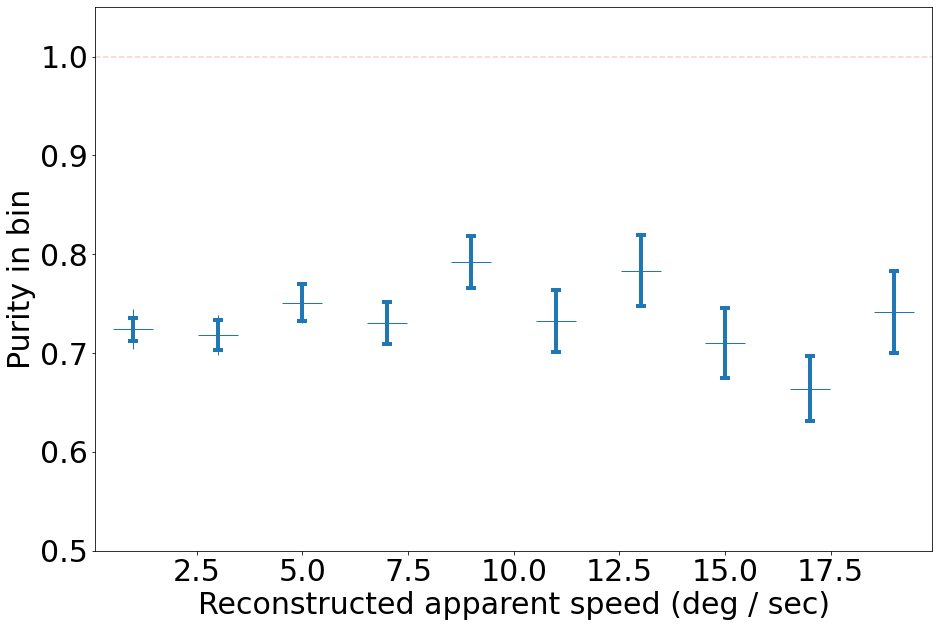}\quad
\includegraphics[width=5.5cm]{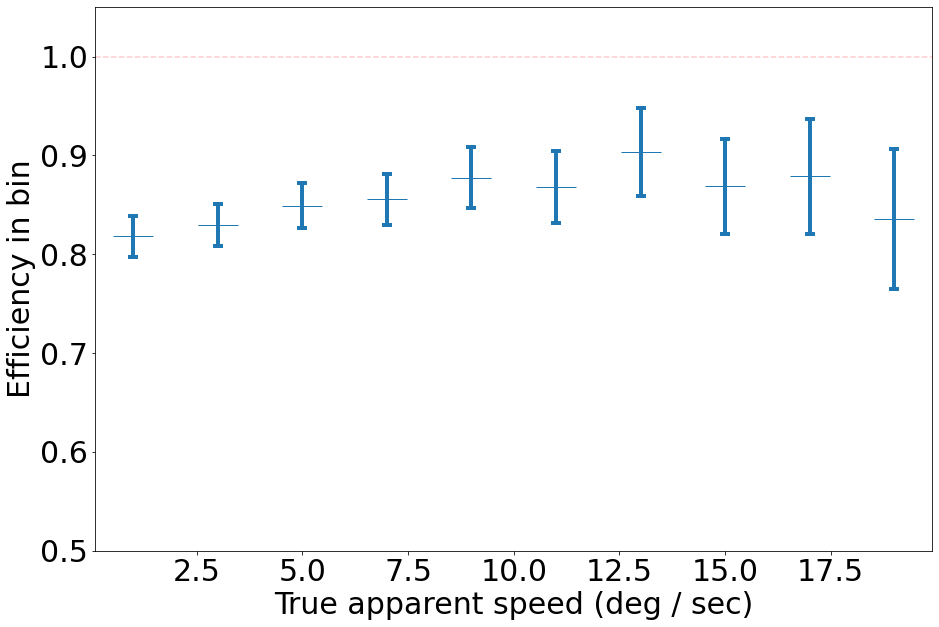}
\end{adjustwidth}
\caption{(\textbf{Left}): A 2D histogram of reconstructed and true apparent speed, computed from the apparent 2D distance between two consecutive bounding box centers, for matched true and reconstructed trajectories. (\textbf{Middle}): Purity as a function of reconstructed apparent speed.  (\textbf{Right}): Efficiency as a function of true apparent speed. Error bars are statistical uncertainties reflecting the bin population~count.\label{fig:synth_apparent_speed}}
\end{figure}

Since we lacked distance estimation during this commissioning period, distance, size, velocity, and acceleration of an object are degenerate quantities. Instead, we focus on the apparent area of the bounding box and the apparent projected 2D speed of its center.
Figure~\ref{fig:synth_apparent_speed} examines the trajectory efficiency and purity as a function of apparent speed, which is uniform along each trajectory in this dataset. We compute the apparent speed from the apparent distance between two consecutive bounding box centers, for matched true and reconstructed trajectories. As seen in the left of Figure~\ref{fig:synth_apparent_speed}, a very low (<1 deg/sec) reconstructed apparent speed seems to characterize pathological trajectories, since they can correspond to a wide range of true apparent speeds. Conversely, a very low true apparent speed can be reconstructed into a wide range of apparent speeds. In the middle and right of Figure~\ref{fig:synth_apparent_speed}, purity and efficiency have no obvious correlation to reconstructed and true apparent speed. The mean efficiency trends down for higher true apparent speed, for which we have fewer true points to detect and reconstruct a trajectory, but smaller uncertainties would be required to confirm this speculation.

\begin{figure}[H]
\begin{adjustwidth}{-\extralength}{0cm}
\centering
\includegraphics[width=5.5cm]{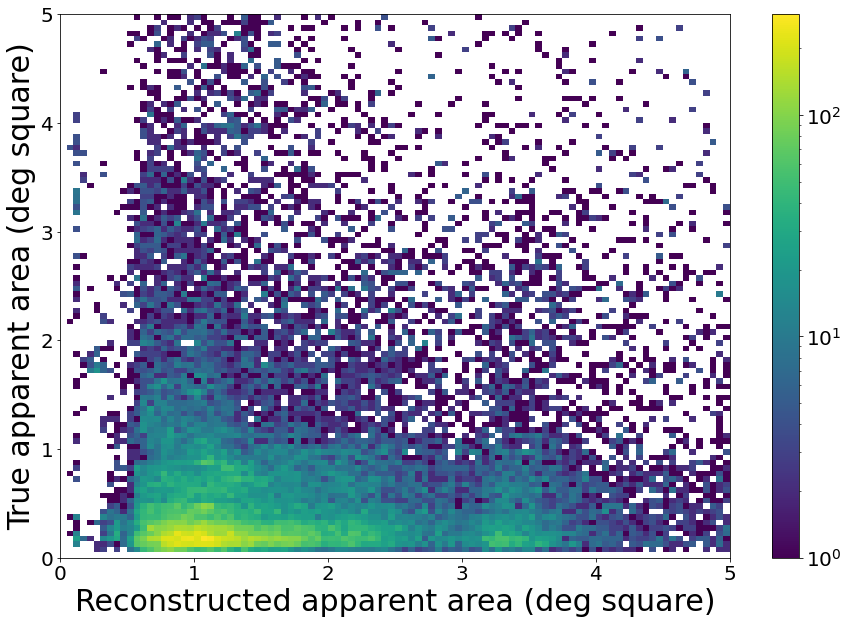}\quad
\includegraphics[width=5.5cm]{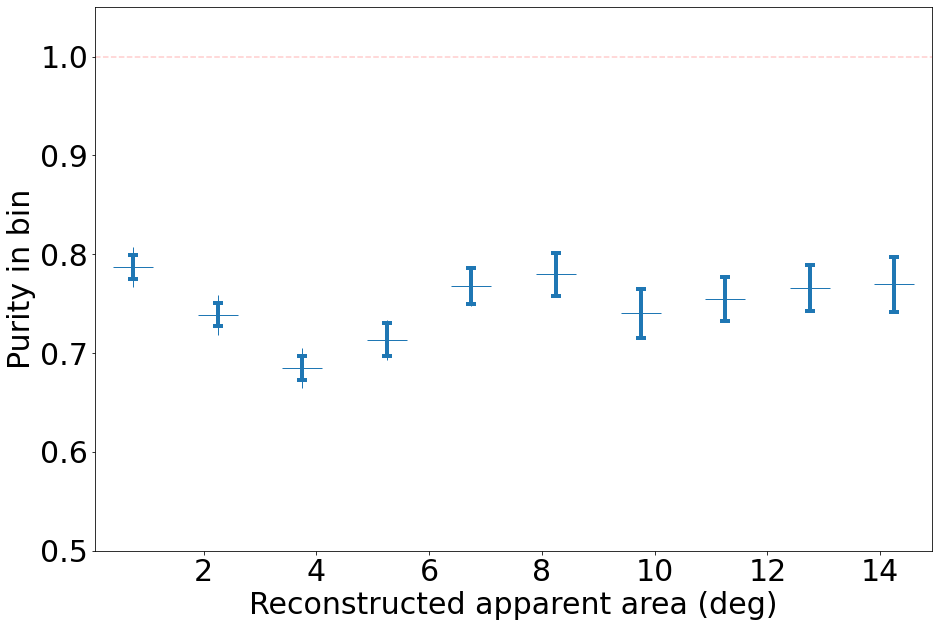}\quad
\includegraphics[width=5.5cm]{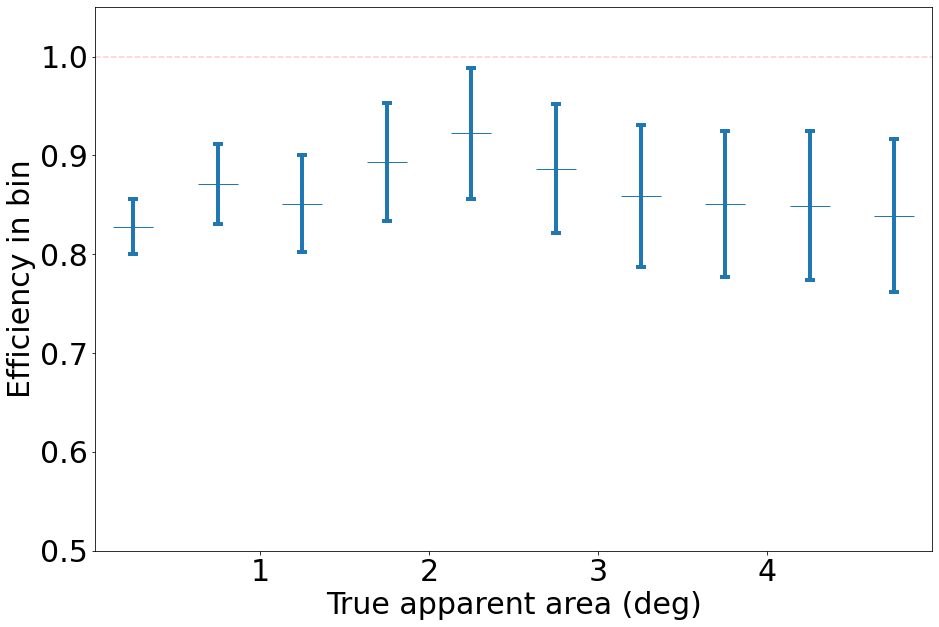}
\end{adjustwidth}
\caption{(\textbf{Left}): A 2D histogram of reconstructed and true apparent area. (\textbf{Middle}): Purity as a function of reconstructed apparent area.  (\textbf{Right}): Efficiency as a function of true apparent area. Error bars are statistical uncertainties reflecting the bin population count.\label{fig:synth_apparent_area}}
\end{figure}

 Using the same metrics, Figure~\ref{fig:synth_apparent_area} examines the trajectory efficiency and purity as a function of apparent bounding box area in angular degrees squared. Purity shows variations beyond the statistical error bars as a function of reconstructed bounding box apparent area, which may suggest that false positives are unevenly distributed in this variable. Efficiency shows a slight improvement when the true apparent area increases. 
 
\begin{figure}[H]
\includegraphics[width=6.5cm]{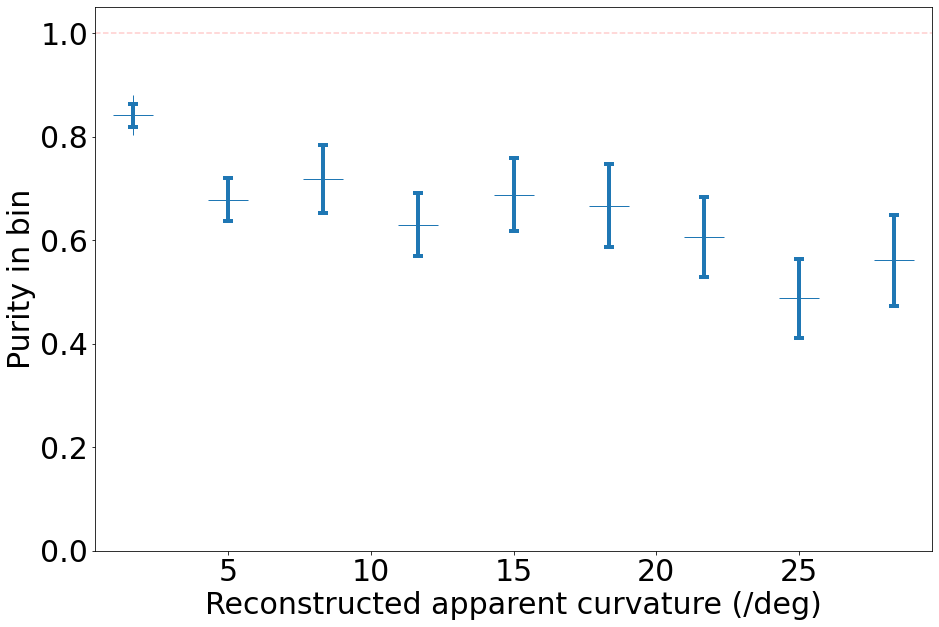}\quad
\includegraphics[width=6.5cm]{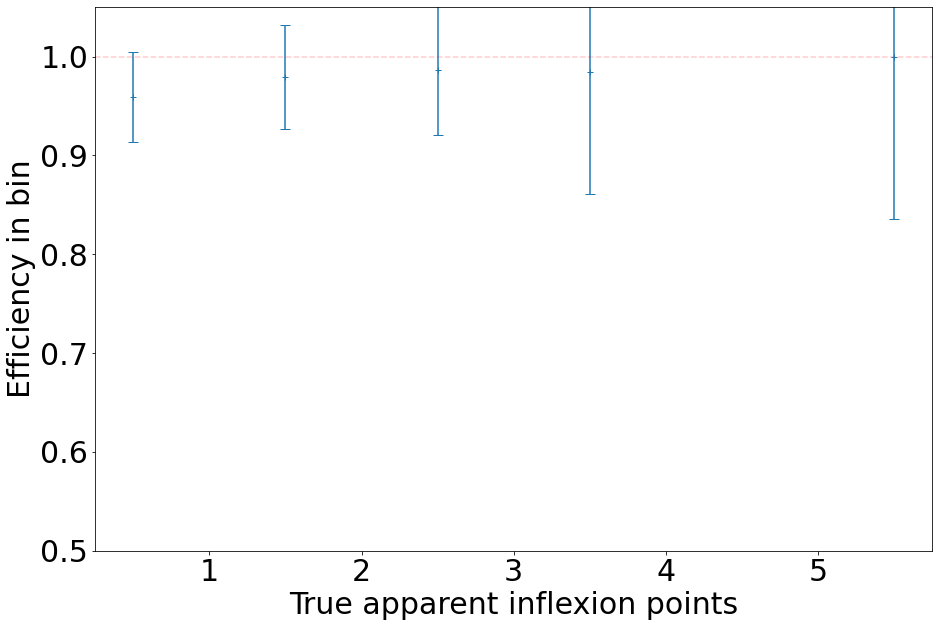}
\caption{(\textbf{Left}): Purity as a function of reconstructed curvature for curved trajectories.  (\textbf{Right}): Efficiency as a function of inflection point counts for piecewise trajectories. Error bars are statistical uncertainties reflecting the bin population count.\label{fig:synth_curv_infl}}
\end{figure}

 Figure~\ref{fig:synth_curv_infl} focuses on the curved and piecewise trajectories. The left-most plot shows that for curved trajectories, a large curvature correlates with lower purity, i.e., there are fewer true positives. 
 For the piecewise trajectories, one might intuitively expect the efficiency to drop when the inflection point count increases, but the right side of Figure~\ref{fig:synth_curv_infl} suggests that there is no significant variation; only the uncertainty of the metric increases. We attribute this to the non-unique trajectory-matching scheme, which allows multiple reconstructed trajectories to be matched to the same true trajectory. In other words, the data presented in this figure do not account for trajectory fragmentation. 

\vspace{-9pt}
\vspace{-9pt}

\begin{figure}[H]
\includegraphics[width=6.5cm]{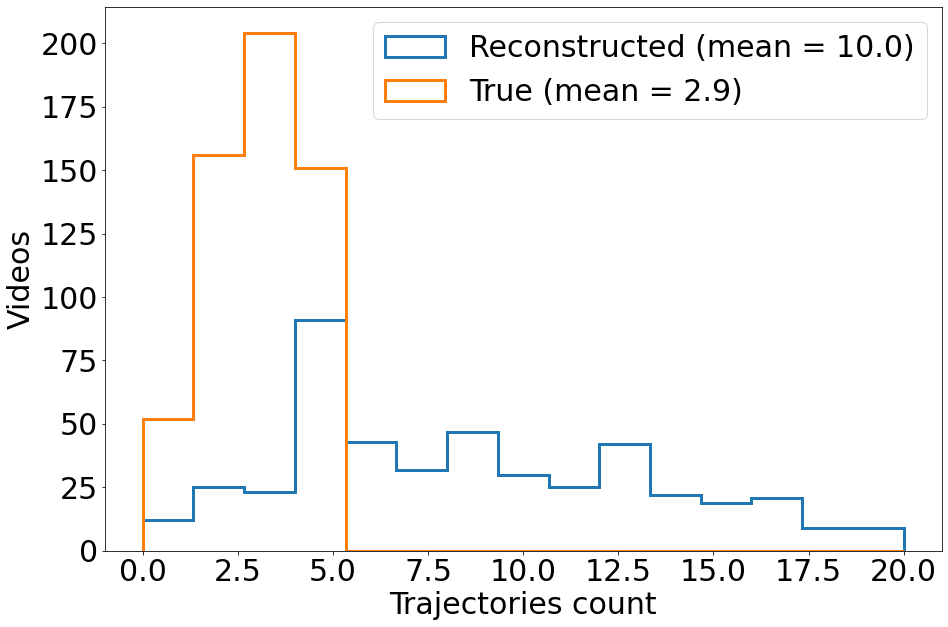}\quad
\includegraphics[width=6.5cm]{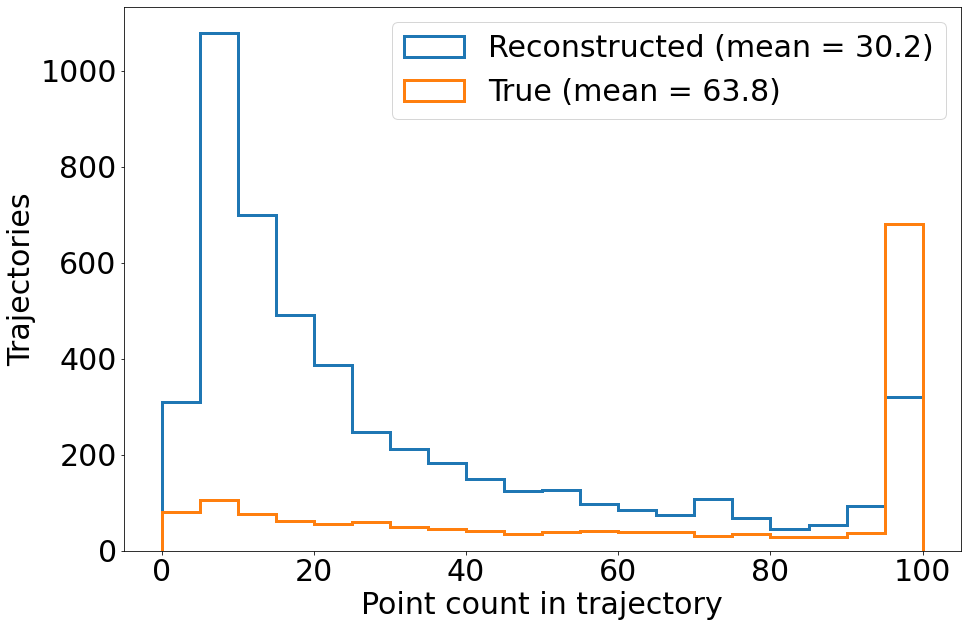}
\caption{(\textbf{Left}): Distribution of individual trajectory counts per video. The reconstructed trajectory count is higher on average due to trajectory fragmentation. (\textbf{Right}): Distribution of trajectory point counts. True trajectories can have at most 100 points due to the dataset generation parameters, and reconstructed trajectories have fewer points on average along the trajectory compared to true trajectories due to missed or dropped detections.\label{fig:synth_traj_fragmentation}}
\end{figure}

Figure~\ref{fig:synth_traj_fragmentation} examines how much fragmentation the reconstructed trajectories display. On the left is shown the number of synthetic videos, with 
the true and reconstructed trajectory count shown on the x-axis. There are approximately three times as many reconstructed trajectories as true ones. This indicates that, on average, true trajectories are broken into about three 
fragments after passing through the pipeline YOLOv5 + SORT. On the right of the same figure, we compare the number of points in true versus reconstructed trajectories. A given true trajectory can have at most 100 points (since the synthetic videos in this dataset have at most 100 frames), which is why the distribution of true trajectory point count peaks at 100. The reconstructed trajectories are split in two distributions: some are successfully complete at 100 points, and some are presumably broken into fragments, each with a peak point count around 10.

In summary, among the true trajectories generated for this synthetic dataset, about 15\% lack either frame-level object detection or trajectory reconstruction and are thus entirely missed. About 13\% of the reconstructed trajectories are spurious and do not correspond to any true trajectory. The reconstruction of apparent speed is severely hampered by the fragmentation of trajectories, which causes a true trajectory to be broken on average into three smaller fragments of trajectory.

\subsection{Unimodal Aerial Census}

\hl{Ref.}~\cite{watters2023scientific} 
described the key steps to search for scientific anomalies in the sky. The first step is a census of all aerial phenomena, whether man-made or natural. Although the Dalek data collected during this commissioning period are not yet science-quality, only involved a single instrument, and did not have access yet to distance estimation for kinematics measurements, we look at the objects reconstructed by our pipeline in order to pave the way for the coming full-scale and multi-modal aerial census. We conduct a deliberately simplistic or \textit{toy} search for outliers, in order to illustrate how we will apply a likelihood statistical analysis to any upcoming full-scale aerial census.

\subsubsection{Toy Outlier Search}
\label{sec:observations}
\paragraph{Dataset: Reconstructed Trajectories}
For this toy outlier search, we look only at the reconstructed trajectories from the five months of commissioning the Dalek. The search is only conducted on the hemispherical seven cameras of the Dalek, excluding recordings from the zenith camera (camera 8).
First, we exclude trajectories with fewer than four recorded data points, which do not provide sufficient information for apparent trajectory analysis. Secondly, we filter out the data points from reconstructed trajectories that intersect with the masked area of each camera, i.e., the area at or below the treeline, eliminating potential excessive detections caused by distant objects or environmental noise such as a growing treeline, leaves, or birds perching in the treeline.
After the masking and initial cleaning, we fit each reconstructed trajectory with cubic splines to reduce noise and improve the accuracy of calculated trajectory parameters. Between January and May 2024, from the seven hemispherical cameras, we collected and reconstructed 502,015 trajectories. 


\setlength{\fboxsep}{0pt}  
\setlength{\fboxrule}{1pt}  

\begin{figure}[H]
\begin{adjustwidth}{-\extralength}{0cm}
\centering 
 \fbox{\includegraphics[width=1\textwidth]{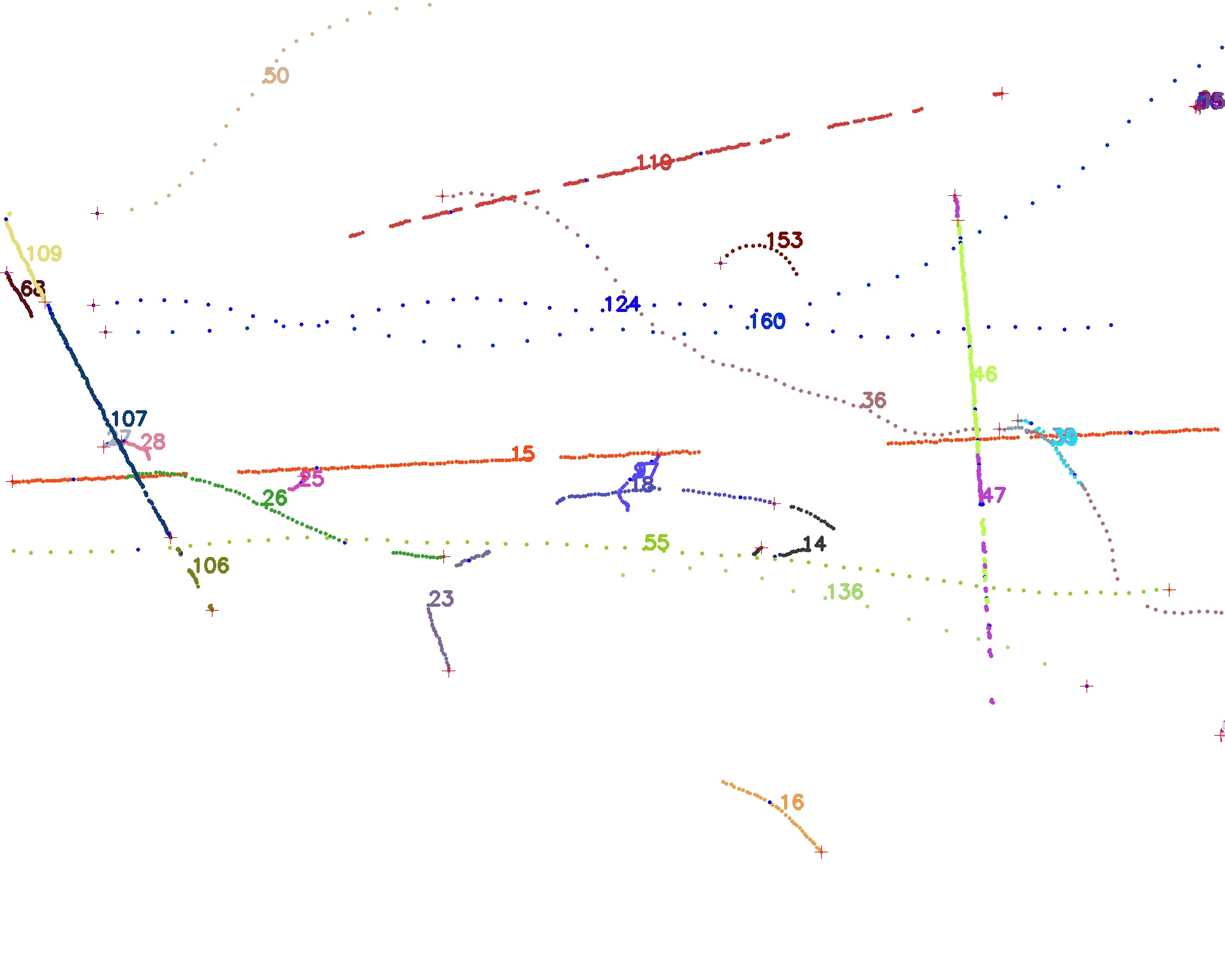}}
\end{adjustwidth}
\caption{\hl{Overlay of multiple} 
 reconstructed trajectories randomly sampled from different videos in the commissioning dataset. Each trajectory is plotted as a set of consecutive 2D points, representing the unique detections underlying the trajectory, and assigned a unique color. The numbers next to each trajectory are unique identifiers assigned by the SORT algorithm.\label{fig:trajectory_comparison}}
\end{figure}

In Figure~\ref{fig:trajectory_comparison}, we show examples of real-world reconstructed trajectory data points. Each color represents a reconstructed object’s trajectory, and the numbers next to each trajectory are unique identifiers generated by the tracking algorithm (SORT). A true trajectory may be broken down into several reconstructed trajectories, each with their own unique identifier and color. Each data point is separated in time by at least one frame interval, i.e., 0.1 s. The distance between data points reflects the apparent reconstructed kinematics of the object and the performance of the frame-by-frame YOLO detection stage. 
The red and orange lines (labeled 118 and 15) in Figure~\ref{fig:trajectory_comparison} belong to an object moving horizontally across the image. The trajectories contain segments with uninterrupted detections, indicated by the high spatial density of data points, as well as large empty gaps; these objects likely display a steady, linear motion disrupted by an occasionally low detection rate. The green and blue trajectories (labeled 36, 124, 160, 50) are curvy and include gradual changes in direction. Looping and curvy trajectories, such as the purple (labeled 18) and the brown (labeled 153) ones, suggest more complex or erratic motions, which could be generated by birds, for example. With the SORT parameters that we tuned in Section~\ref{sec:sort_benchmark}, we are able to reconstruct some trajectories where an occlusion happened, such as those labeled 36 and 46, which intersect each other. Figure~\ref{fig:trajectory_comparison} also includes several long trajectories (labeled 46, 47, 107, and 110) generated by an airplane at different times. Notably, trajectories 46 and 47 were produced by the same object. Due to the loss of detection for a significant number of frames, the tracking ID was switched, causing the trajectory to appear segmented. This illustrates the challenges posed by intermittent detection loss during tracking, resulting in the separation of a continuous path into multiple labeled segments.

    

\begin{figure}[H]
\includegraphics[width=0.8\linewidth]{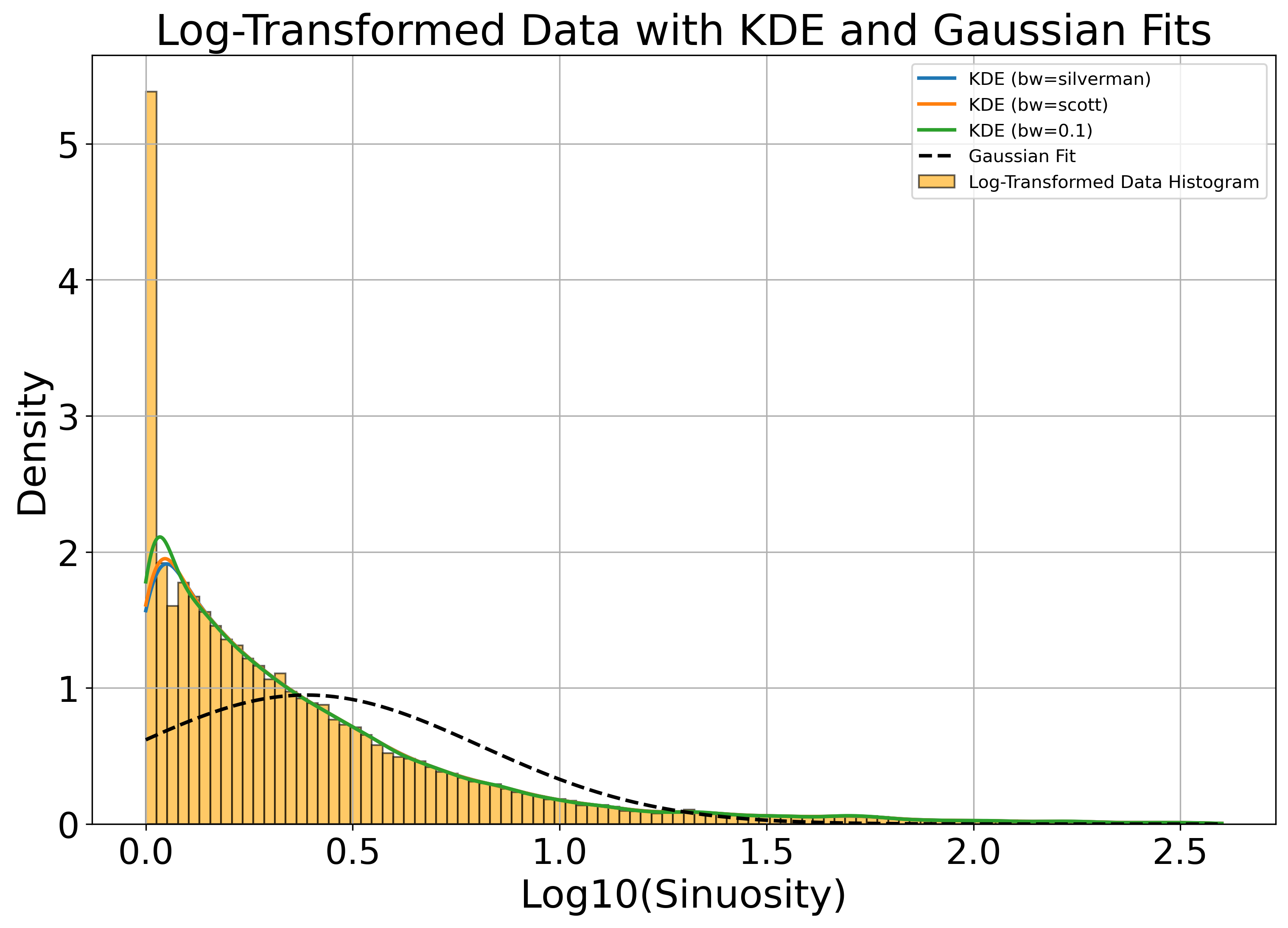}
\caption{Normalized histogram of the log-transformed distribution of sinuosity values computed for all reconstructed trajectories from the five months of commissioning. The lines show the fit with four models for probability density estimation: Kernel Density Estimation (KDE), with different bandwidths in colored solid lines, and a Gaussian fit in a dashed line.\label{fig:histogram}}
\end{figure}

\paragraph{Toy Outlier Criteria Based on Sinuosity}
Our toy analysis will look for outliers in the simple metric called the \textit{sinuosity} of each 2D reconstructed trajectory, which is defined as the total trajectory length divided by the length of the straight line from the trajectory start to end points. Figure~\ref{fig:histogram} shows the distribution of sinuosity values for randomly selected trajectories. We estimate the probability density of sinuosity values in order to identify high-sinuosity outliers, using both non-parametric and parametric approaches. Given the distribution's apparent non-Gaussian probability density, we compare Kernel Density Estimation (KDE) models with varying bandwidths,  with a Gaussian distribution fit as a baseline.
The distribution of sinuosity values is log-transformed to reduce skewness. We show in Figure~\ref{fig:histogram} the fit of the different models to the underlying probability density. Bandwidth selection is critical in KDE as it controls the smoothing level: finer bandwidths capture greater details but risk overfitting, while larger bandwidths provide greater smoothness at the expense of capturing distribution nuances. We test KDE with different bandwidths, including Scott's and Silverman's methods as well as a custom fixed bandwidth of 0.1, to balance the smoothness and fit quality.

In Table~\ref{tab:log_likelihood_comparison} below, we compute the log-likelihood of each model's fit with the real distribution of sinuosity values. Higher (less negative) values indicate a better fit. KDE with a 0.1 bandwidth yields the highest log-likelihood, outperforming both the other KDE with a different bandwidth and the Gaussian fit. The KDE model performs best at capturing the underlying density of this dataset. The Gaussian fit’s low log-likelihood suggests it was unable to accommodate the distribution’s skewness.

\begin{figure}[H]
\centering
\includegraphics[width=6.5cm]{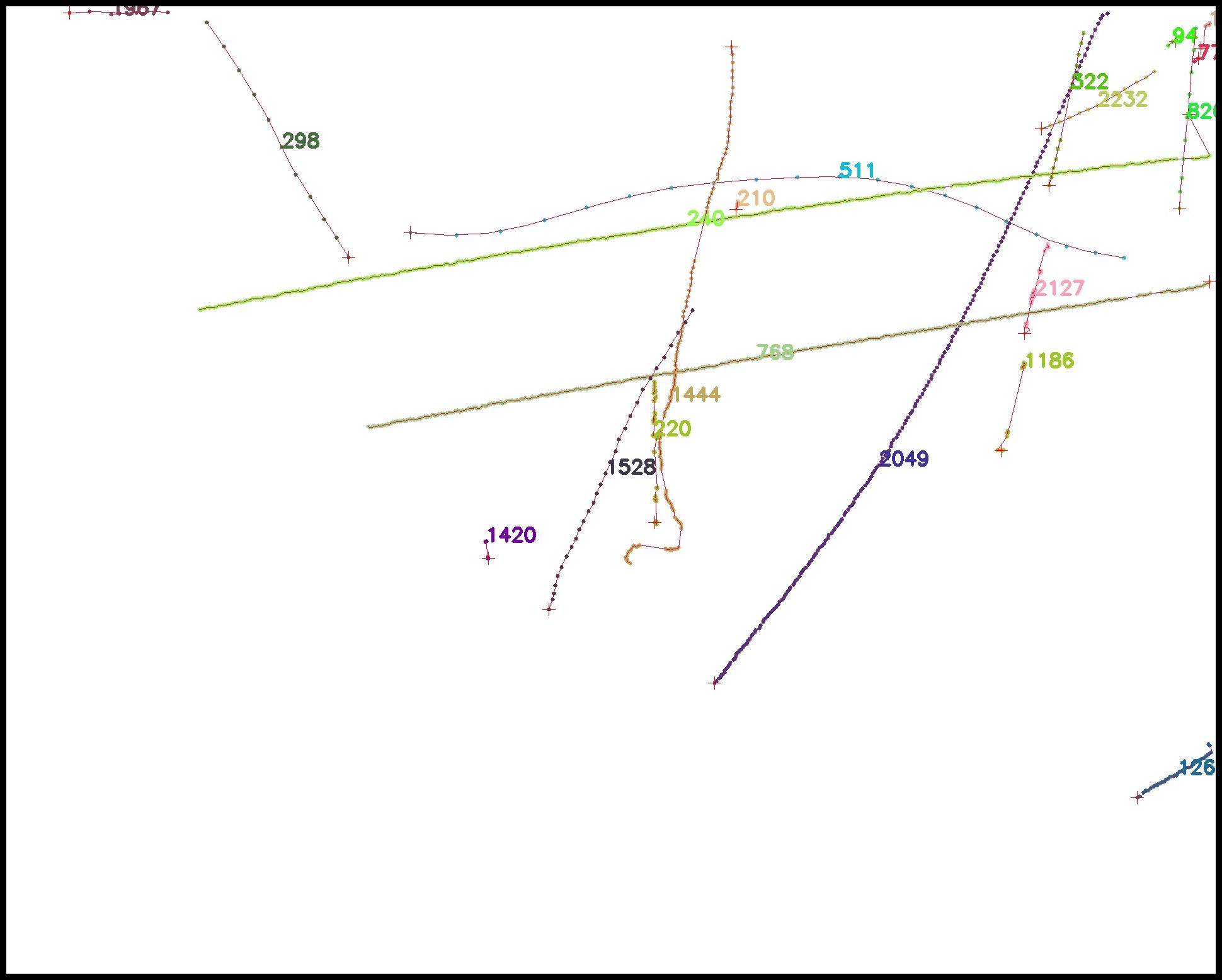}\quad
\includegraphics[width=6.5cm]{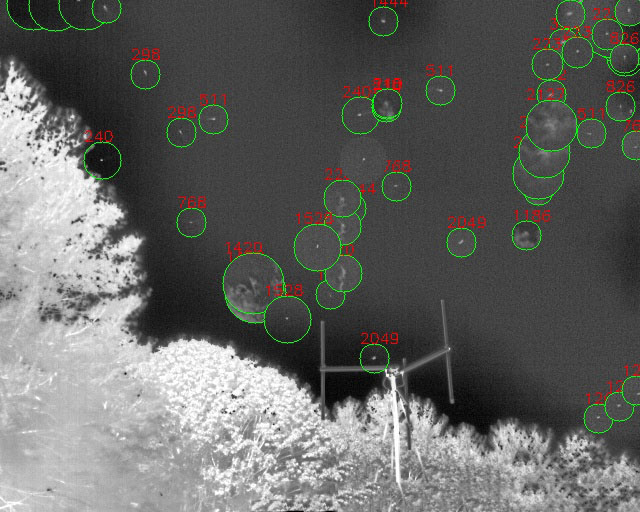 }\quad
\includegraphics[width=6.5cm]{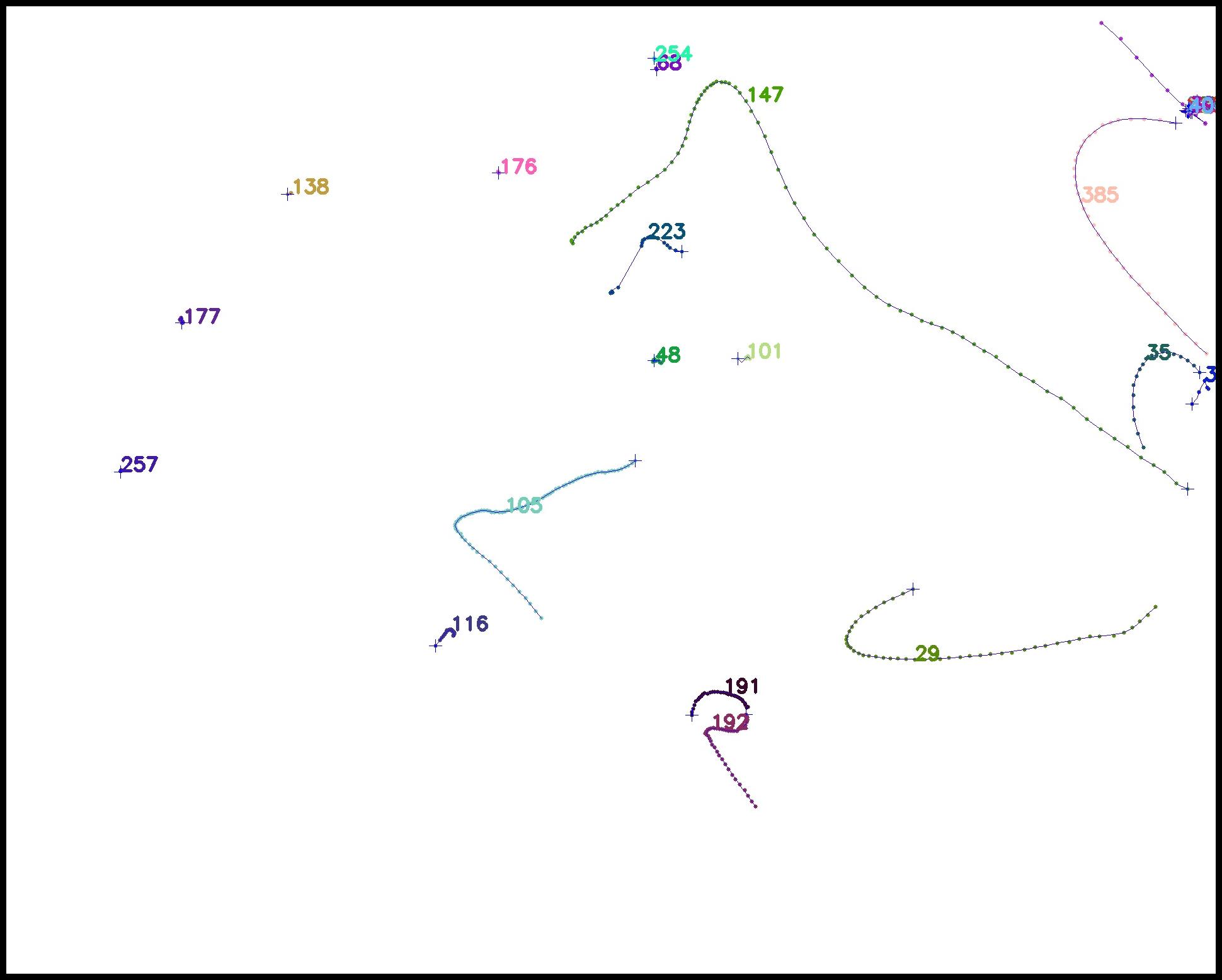}\quad
\includegraphics[width=6.5cm]{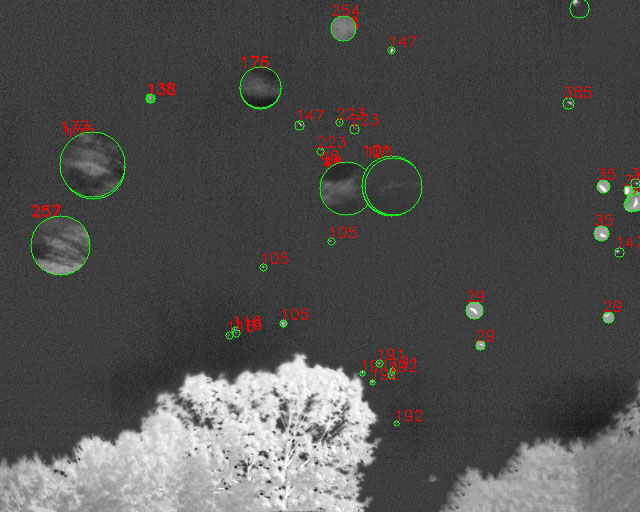}\quad
\includegraphics[width=6.5cm]{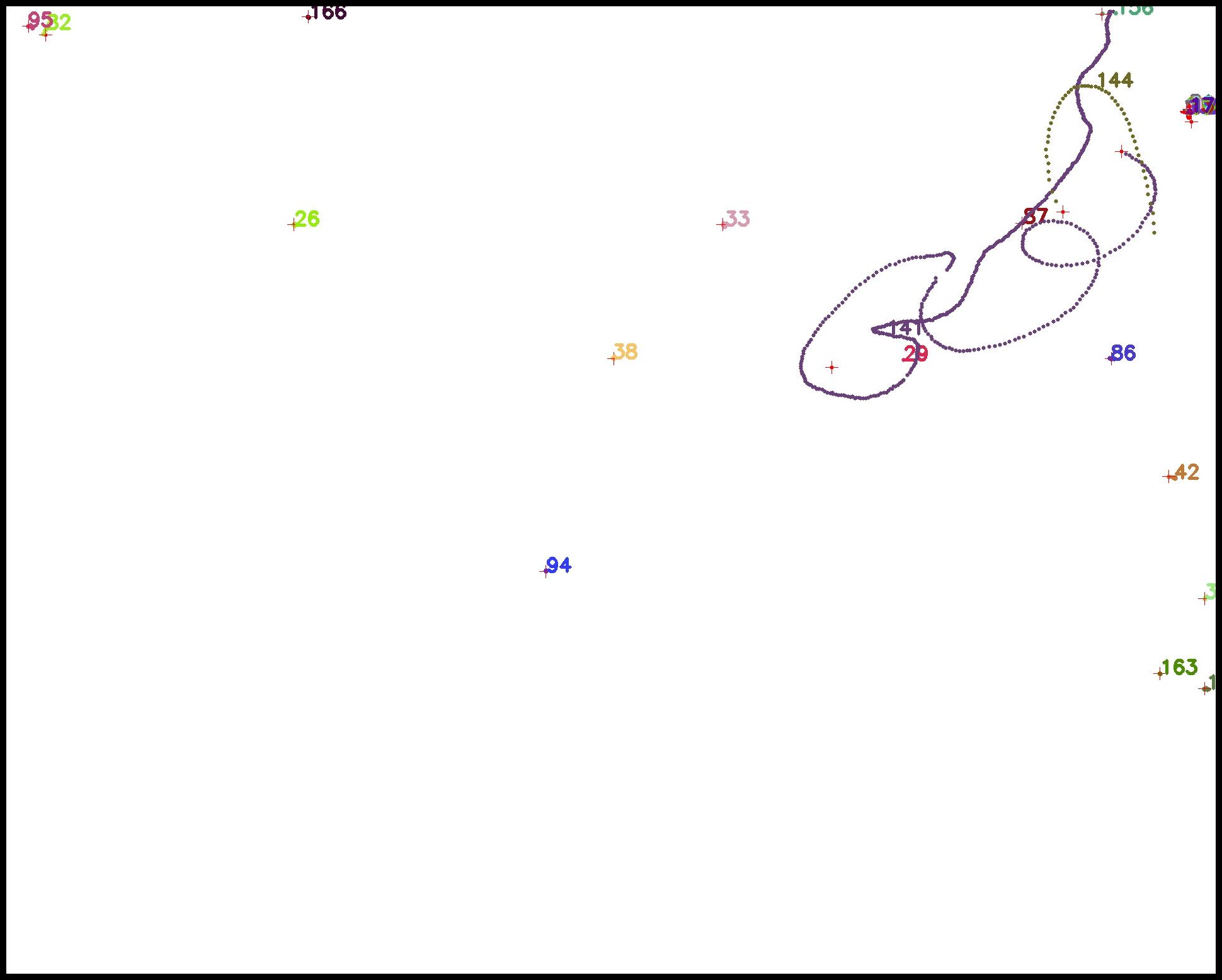}\quad
\includegraphics[width=6.5cm]{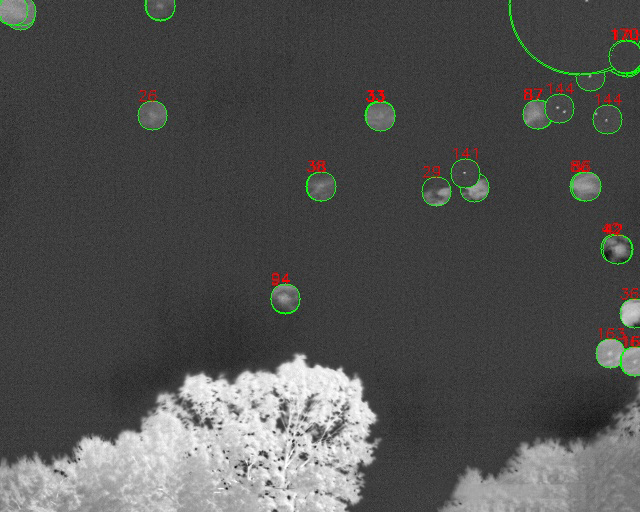}
\caption{\hl{(\textbf{Left}): Examples} of reconstructed trajectory data points. (\textbf{Right}): For each corresponding reconstructed object, we overlay three snapshots of the frame-by-frame object YOLO detections inside green outlines and annotated with red text (unique identifiers). The snapshots are taken and located at the start, middle, and end of each trajectory. The background is a frame of the corresponding camera which does not contain any detected objects, overlaid for reference. (\textbf{Top row}): Objects with low reconstructed trajectory sinuosity, between 1.0 and 1.2. (\textbf{Middle row}): Objects with reconstructed trajectory sinuosity ranging from 1.2 to 3.0. (\textbf{Bottom row}): Objects with reconstructed trajectory sinuosity >3.0.\label{fig:straigth}}
\end{figure}

For this toy outlier search, we arbitrarily decide to use the 75th percentile as a selection threshold, which corresponds to a sinuosity value of 3.0. The sinuosity distribution can be partitioned into three categories based on the observed behavior in these sinuosity ranges. Trajectories with a sinuosity between 1.0 and 1.2 tend to be straight and predominantly associated with planes and birds, with minimal contamination from clouds. Trajectories with a sinuosity between 1.2 and 3.0 are curvilinear. Trajectories exceeding a sinuosity of 3.0 are primarily attributed to clouds, with some contribution from birds and a negligible presence of planes. Clouds do not move very fast and they can appear almost stationary for a period of time: clouds' reconstructed trajectories tend to have a very short Euclidean distance (i.e., length of a straight line from start to end of the trajectory), which drives up the sinuosity value computed for these trajectories. Among the original dataset of 502,015 trajectories, we find that $\sim$16\% of them, or 81,873 trajectories, have sinuosity values greater than 3.0. This threshold makes it feasible for us to manually examine the distribution's upper tail, i.e., reconstruct trajectories with high sinuosity. For the purpose of this manual examination, to sift through the selected high-sinuosity trajectories, we generate two types of visualizations which are daily summaries of reconstructed trajectories. One is similar to Figure~\ref{fig:trajectory_comparison}, explained previously. The other uses a background camera image computed from an average of 100 empty frames. We overlay on this background, for each trajectory, three snapshots inset in a green circle, which show the object's detection bounding box at the start, middle, and end of the trajectory. This visualization reveals regular patterns of trajectories accumulated daily. Figure~\ref{fig:straigth} illustrates the partition of sinuosity values and randomly sampled examples from each of the three ranges of sinuosity. The left and right columns display the two types of daily summaries we just described. The top, middle, and bottom rows of Figure~\ref{fig:straigth} show examples of reconstructed trajectories with sinuosity in the ranges of $[1.0; 1.2]$ (predominantly identified as planes and birds), $[1.2; 3.0]$ (mostly birds and clouds), and $[3.0; \infty )$ (mostly clouds), respectively.

\vspace{-9pt}

\begin{table}[H]
    \caption{Log-likelihood comparison of KDE and Gaussian models for the distribution of trajectory sinuosity values. A larger log-likelihood indicates a better fit. KDE with a fixed bandwidth of 0.1 performs best.}
    \begin{tabularx}{\textwidth}{lC}
        \toprule
        \textbf{Model} & \textbf{Log-Likelihood on Test Data} \\
        \midrule
        KDE (bw = Silverman) & $-$1097.19 \\
        KDE (bw = Scott) & $-$1020.60 \\
        KDE (bw = 0.1) & $-$751.99 \\
        Gaussian & $-$6860.76 \\
        \bottomrule
    \end{tabularx}
    \label{tab:log_likelihood_comparison}
\end{table}







\begin{figure}[H]
    \begin{adjustwidth}{-\extralength}{0cm}
    \centering
    \includegraphics[width=5.5cm]{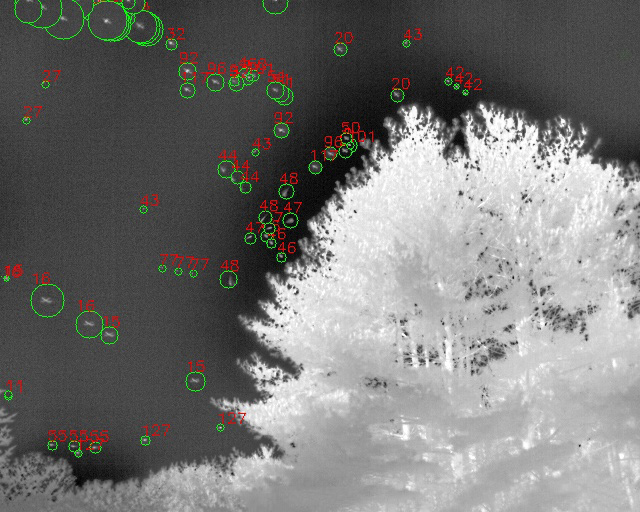}\quad
    \includegraphics[width=5.5cm]{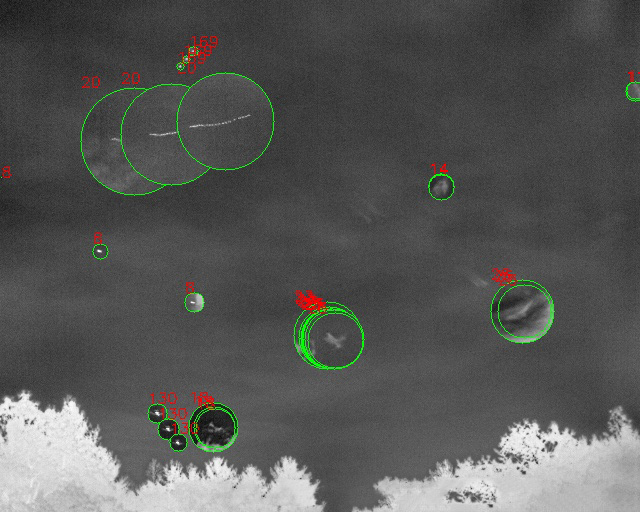}\quad
    \includegraphics[width=5.5cm]{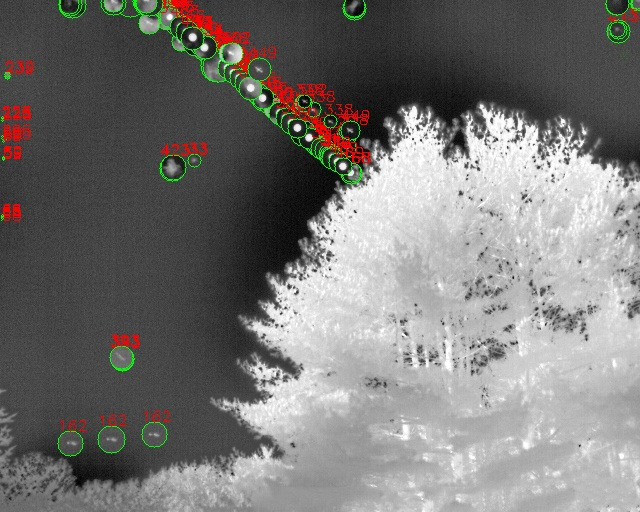}
    \end{adjustwidth}
    \caption{(\textbf{Left}): Airplanes on camera 7. (\textbf{Middle}): Birds and clouds on camera 4. (\textbf{Right}): Helicopters and Moon on camera 7. All are examples of detected objects that can be visually identified by inspecting the detection bounding box and the corresponding video recording. We show here examples of airplane, bird, cloud, helicopter, and Moon detections. For each reconstructed trajectory, we overlay three snapshots of the frame-by-frame object YOLO detections inside green outlines and annotated with red text (unique identifiers). The snapshots are taken and located at the start, middle, and end of each trajectory. The background is a frame of the corresponding camera, which does not contain any detected objects.}  
    \label{fig:trajectory_summary}  
\end{figure}

\paragraph{Manual Examination of High-Sinuosity Trajectories}
The approach described above helps to identify objects following unusually curved paths or exhibiting erratic movement patterns, distinct from the smoother motion of most tracked objects. It is followed by a manual review of the individual IR images making up the trajectories that passed the high-sinuosity criteria. The majority of trajectories are associated with mundane objects. 
Figure~\ref{fig:trajectory_summary} shows examples of reconstructed trajectories where the object can be identified by looking at the detection bounding boxes and the original recordings. Some can be identified as airplanes flying across the sky (e.g., trajectories 88, 21, 70, 19, and 57 in the left of Figure~\ref{fig:trajectory_summary}), others are identified as flocks of birds flying together in formation (e.g., trajectory 20 in the middle of Figure~\ref{fig:trajectory_summary}, or trajectory 107 in the left of Figure~\ref{fig:trajectory_summary}), rotorcraft (e.g., trajectory 162 in the right of Figure~\ref{fig:trajectory_summary}), as well as clouds (14, 25 in the middle of Figure~\ref{fig:trajectory_summary}). 

\begin{figure}[H]
\includegraphics[width=6.5cm]{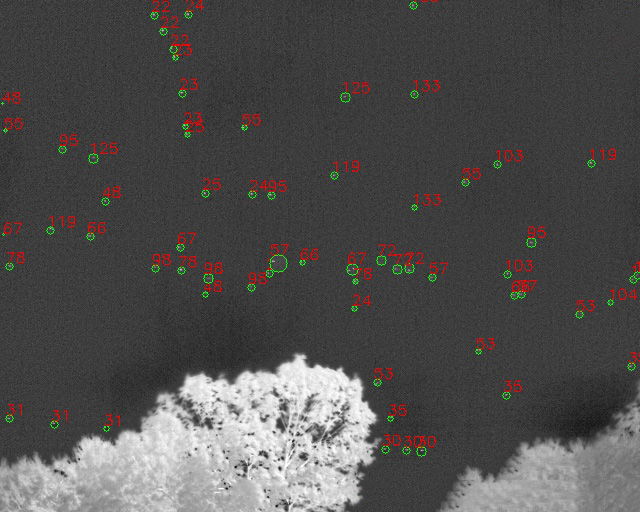}\quad
\includegraphics[width=6.5cm]{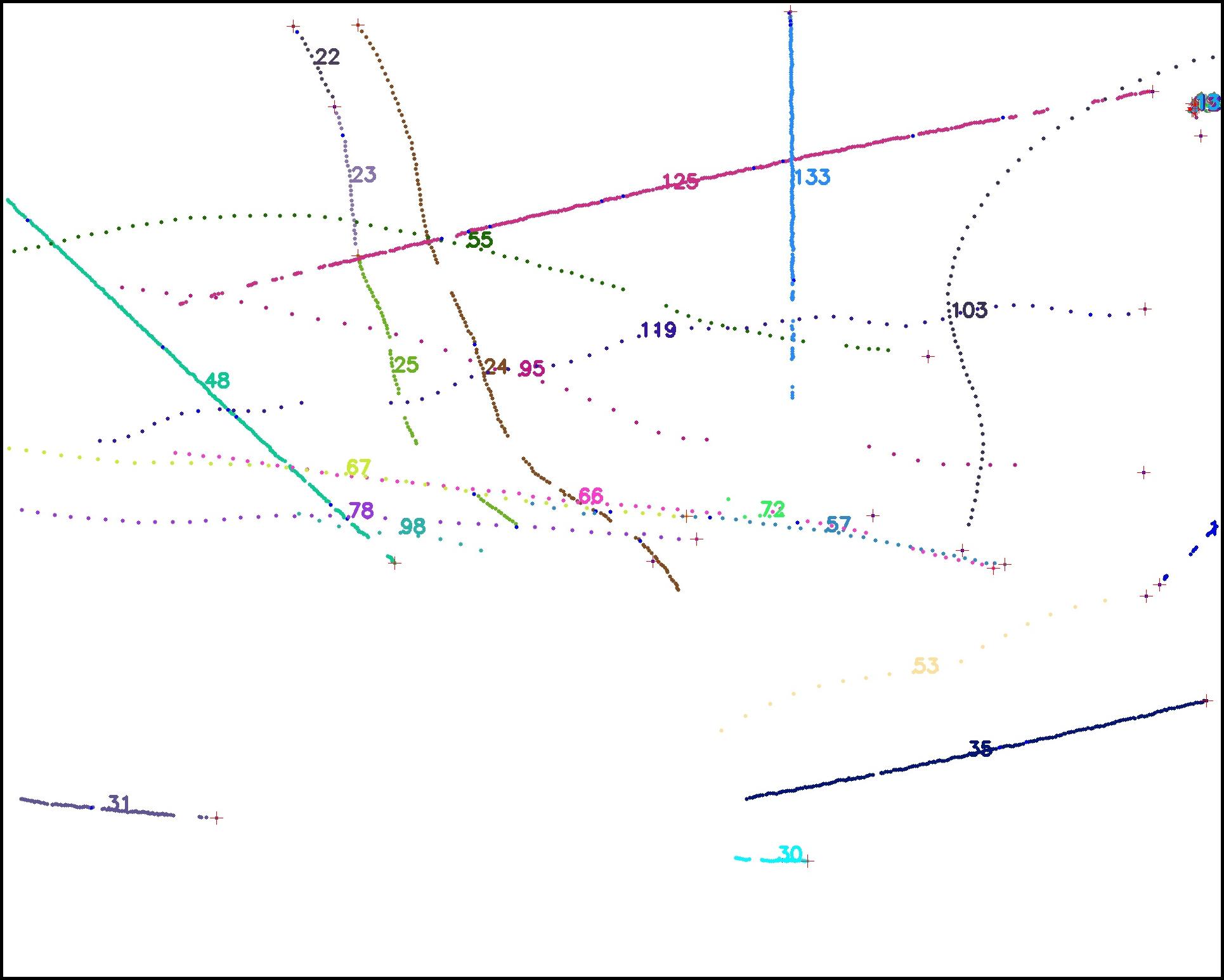}
\caption{Examples of trajectories created by objects with small apparent size. Left: For each corresponding reconstructed trajectory, we overlay three snapshots of the frame-by-frame object YOLO detections inside green outlines and annotated with red text (unique identifiers). The snapshots are taken at different points along the trajectories. Right: Corresponding examples of reconstructed trajectory data points. \label{fig:small_objects}}
\end{figure}

While some objects can be visually identified, especially when they are large in apparent size, many objects remain visually indistinct due to their smaller size. This is particularly evident from the bounding box sizes shown in Figure~\ref{fig:detection_time} (middle), which demonstrate that most detected objects are small. As a result, the majority of small objects remain unidentified through visual inspection alone. Despite having the reconstructed trajectory, the small object sizes make it less feasible to discern specific characteristics or object types, as illustrated in Figure~\ref{fig:small_objects}. Thus, having additional modalities and range estimation techniques such as optical triangulation~\cite{szenher2023hardware} or passive radar~\cite{randall2023skywatch} is needed to identify objects in many cases. Fortunately, for our toy outlier search, few small objects pass the high-sinuosity criteria. During our manual examination we were able to associate mundane objects with the majority of trajectories with high sinuosity. The visual inspection of the recordings helped to group trajectories with high sinuosity into several categories: single birds (presumably hawks), flocks of birds, planes, leaves, and clouds. The Supplementary Materials shows typical reconstructed trajectories for all of these high-sinuosity categories.

After manually scanning all trajectories flagged as outliers by the high-sinuosity criteria, 144 trajectories remained ambiguous even after a manual check. 

\begin{figure}[H]
\includegraphics[width=0.8\linewidth]{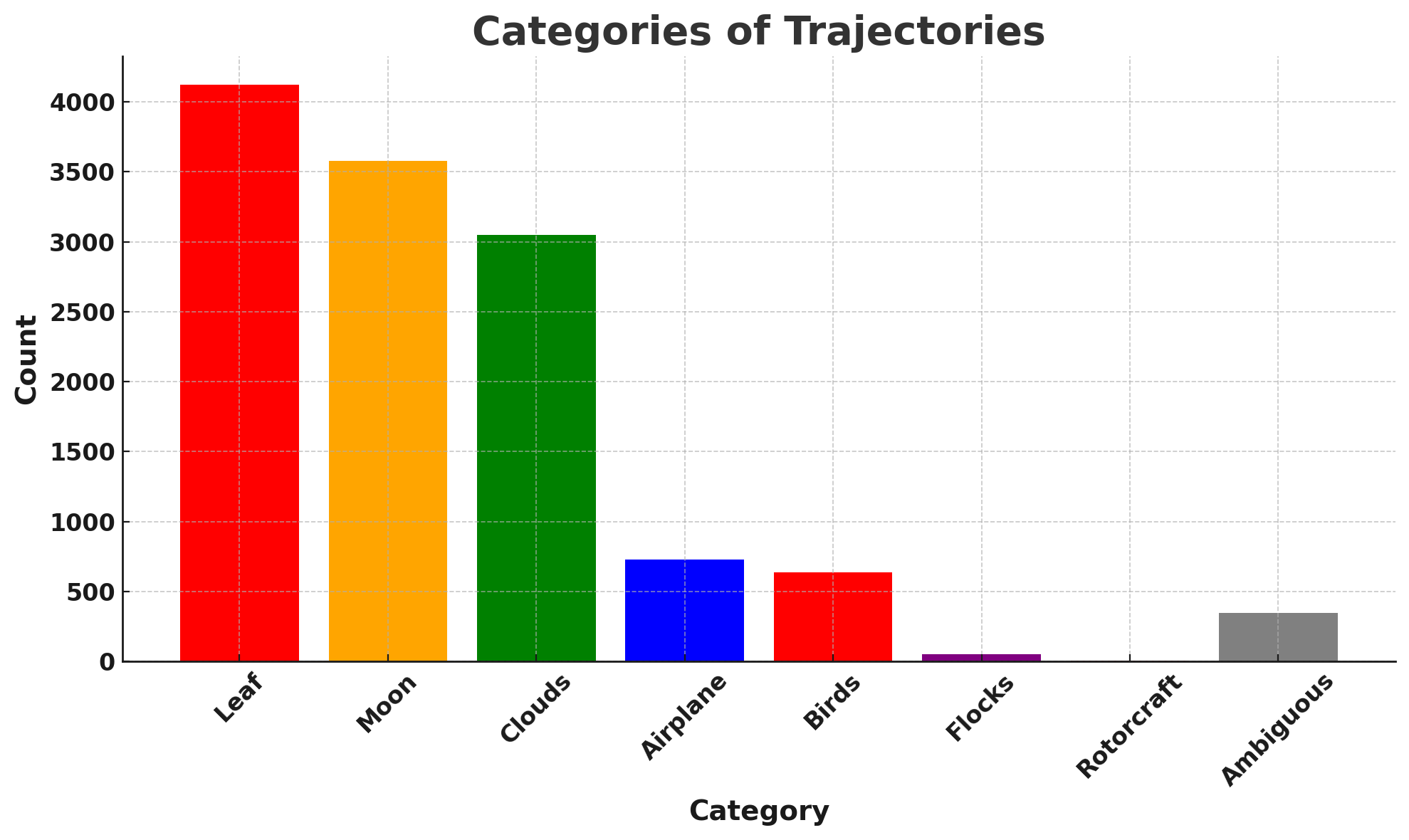}
\caption{Manual classification of reconstructed trajectories from Dalek recordings, a small subset of the full commissioning dataset
.\label{fig:orientation}}
\end{figure}

\begin{figure}[H]
\includegraphics[width=0.8\linewidth]{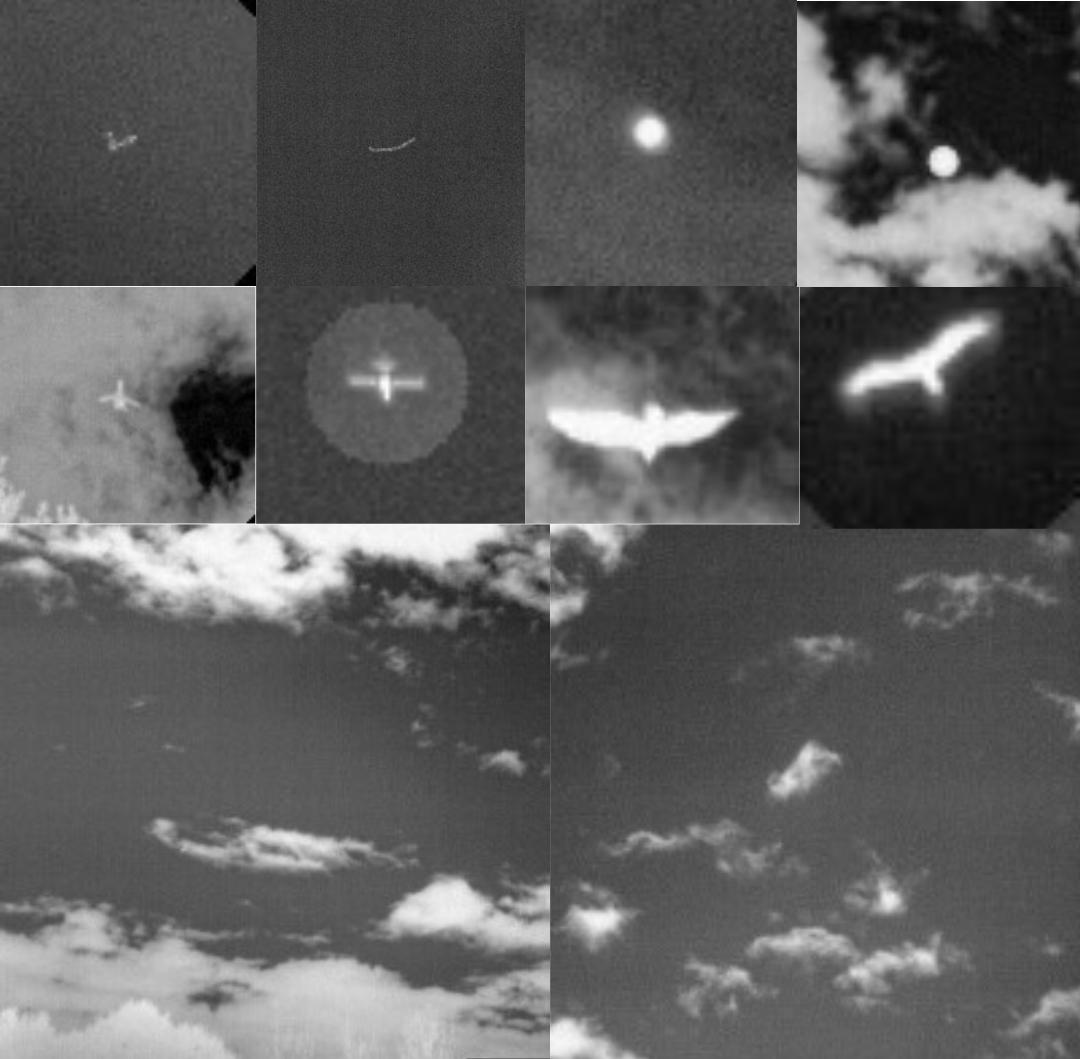}
\caption{After manual classification of reconstructed trajectories, we sample typical objects (in pairs) from each category. These images are crops of the objects for illustration purposes. First row:  flocks of birds and the Moon; second row: planes and single birds; third row: clouds.\label{fig:typical_bjects}}
\end{figure}

Finally, to establish the list of potential object categories present in the Dalek dataset, we manually labeled 12,000 trajectories and identified eight distinct categories tallied in Figure~\ref{fig:orientation}: leaf, Moon, clouds, airplane, birds, flocks, rotorcraft, and ambiguous. Among these, the category of ``ambiguous'' is defined as objects that were flying at a distance, making visual classification challenging. Despite the difficulty in categorizing these ambiguous objects, they did not exhibit any abnormal behavior. Figure~\ref{fig:typical_bjects} also shows zoomed in infrared images of representative objects of each category observed in this labeled dataset of trajectories.

\vspace{-9pt}

\setlength{\fboxsep}{0pt}  
\setlength{\fboxrule}{1pt}  

\vspace{-9pt}

\vspace{-9pt}

\vspace{-9pt}

\subsubsection{Likelihood Analysis}
\label{sec:likelihood}

Our goal in this section is to demonstrate the use of a likelihood-based test to characterize the statistical significance of the results of a quantitative outlier search. We use as our test case the toy outlier search conducted in the previous section. This is a toy case because it is unimodal (one sensor type: IR cameras), degenerate (apparent 2D trajectory), and using only a single characteristic of the apparent trajectory (``sinuosity'').  In the future, we will have data from multiple modalities, use range estimation to remove the spatial degeneracy (e.g., with triangulation), and introduce multiple characteristics from each modality (e.g., speed and size) using data from a full multi-modal, multi-spectral observatory. We follow the seminal paper~\cite{cowan2011asymptotic} on likelihood-based tests of new physics.

In our case, we define the null hypothesis, $H_0$, as describing a known dataset with no outliers. We test $H_0$ against the alternative hypothesis, $H_1$, that our dataset has both this known background and outliers. In the toy outlier analysis described in the previous section, our observed dataset consists of measurements of apparent sinuosity of each reconstructed trajectory. 

The $p$-value quantifies the statistical significance of the observations assuming the null hypothesis $H_0$, i.e., the probability of observations of equal or worse compatibility with $H_0$. A small $p$-value indicates that it would be unlikely for the null hypothesis alone to have generated the observed data.


Significance is defined in physics by a Z-score: $Z = \psi^{-1}(1-p)$, where $\psi^{-1}$ is the quantile of the standard Gaussian distribution of the values of the characteristic of interest, $\phi$; in other words, it is the inverse of its cumulative distribution. 




The probability of observing a specific dataset consistent with a particular hypothesis is called the \textit{likelihood}. In our toy case, $\phi$, the physical parameter of interest for us, is a trajectory's 
apparent sinuosity.
Note that when we go beyond our simple counting experiment here, we will make a histogram $\vec{n}$ of detected and reconstructed object characteristics, such as acceleration or trajectory inflexion points.

If we assume the signal $s$ and background $b$ have properties that are known perfectly, we have an ideal experiment and the likelihood is trivial:
\[
\mathcal{L}(\vec{n}, \phi) = \prod_{\text{bins}} \text{Poisson}(n_i | \phi s_i + b_i)
\]

In a realistic measurement however, we have to account for non-negligible statistical and systematic uncertainties. 
Sources of systematic uncertainties in our measurements can include, for example, calibrations such as the conversion from 16-bit to temperature units scale; efficiencies in identification and reconstruction; and resolution, for example, of the reconstructed acceleration. Let us consider for the sake of this simple example only one systematic uncertainty: assume we know the uncertainty of our efficiency for reconstructing trajectories of objects, and assume that this uncertainty has a Gaussian distribution. We represent it with a parameter $\alpha$ of scale such that the values $\pm 1$ correspond to this uncertainty. Then, we have to modify the likelihood to model this uncertainty:
\[
\mathcal{L}(\vec{n} | \phi, \alpha) = \prod_{\text{bins}} \text{Poisson}(n_i | \phi s_i + b_i (\alpha)) \cdot \text{Gauss} (0 | \alpha, 1)
\]
where $b_i (\alpha)$ is a response function which tells us how much the background changes as a result of $\alpha$, and $\text{Gauss} (0 | \alpha, 1)$ represents a Gaussian of mean 0 and variance 1 evaluated in $\alpha$. In principle, we can build an approximation of $b_i (\alpha)$ using the synthetic dataset, for example, running our selection with $\pm 1\sigma$ variations of $\alpha$. From now on we refer to this full likelihood as a \textit{profile likelihood}.

\[
\mathcal{L}(\vec{n} | \phi, \alpha) = \prod_{i=1}^N \frac{(\phi s_i + b_i (\alpha))^{n_i} e^{-(\phi s_i + b_i (\alpha))}}{n_i!} \cdot \text{Gauss} (0 | \alpha, 1)
\]

Having defined our profile likelihood based on our measurement and its uncertainties, we can move on to a statistical test to compare the two hypotheses $\mathcal{H}_0$ and $\mathcal{H}_1$. A classical frequentist approach is to use the likelihood-ratio test. For $\mathcal{H}_0$, the profile likelihood-ratio is defined as
\[
\lambda(\phi) = \frac{\mathcal{L}(\vec{n} | \phi, \hat{\hat{\alpha}})}{\mathcal{L}(\vec{n} | \hat{\phi}, \hat{\alpha})} 
\]
$\lambda$ has values in $[0, 1]$. Being close to 1 reflects a good agreement between the measurement and the given value of $\phi$. The numerator is a conditional maximal-likelihood estimator of $\alpha$: $\hat{\hat{\alpha}}$ is the value of $\alpha$ for which $\mathcal{L}$ is maximal given a specific value of $\phi$. $\hat{\phi}$ and $\hat{\alpha}$ are maximum-likelihood estimators.
In practice, the test statistics used for setting an upper limit on $\phi$ are 
\[
q_{\phi} = \begin{cases}
-2 \ln \lambda(\phi) & \hat{\phi} \leq \phi \\
0 & \hat{\phi} > \phi
\end{cases}
\]

The higher $q_{\phi}$ is, the less agreement there is between the measurement and the given value of $\phi$. 
In order to find the sensitivity of our measurement for an exclusion limit, we have to look at the median significance with which we could reject $\phi > 0$, assuming a dataset generated under the null hypothesis of background only.
The approximated median exclusion significance, or Asimov discovery significance, is, assuming the Wald approximation~\cite{cowan2011asymptotic},
\[
\text{med}[Z_{\phi} | 0] = \sqrt{q_{\phi, A}}
\]
where $q_{\phi, A} = -2\ln \lambda_A(\phi)$ is the test statistic for the Asimov dataset, defined as the dataset where the bin counts are equal to their expectation values. If the corresponding p-value is below a threshold $\beta$, we say that the value of $\phi$ is excluded at a confidence level of $1-\beta$. In practice, the upper limit is found numerically as the value for which $p = \beta$.

We apply this method to the toy search for outliers described in the previous section, focusing on a simple counting experiment instead of a histogram experiment. We start by defining our background and signal counts, including uncertainties. Out of $N^{obs} = $502,015 reconstructed trajectories, $N_{flag} = $81,873 were flagged with a sinuosity greater than 3.0, and after manual inspection, $N_s^{obs} = $144 trajectories remained ambiguous, although with plausible mundane explanations. Before we compute a profile likelihood, we have to estimate the uncertainties on these counts.

On the manually labeled real-world dataset containing a variety of objects, we found a recall for the object detection stage of $\sim$63\%. 
An object detection recall of 63\% means that we should assign an uncertainty of at least $\sim$37\% to the reconstructed sinuosity: we are potentially missing one out of three points in any given trajectory. The worst impact on the Euclidean distance from trajectory start to end happens when we are missing points at the start or at the end of a straight trajectory, i.e., a fractional uncertainty of 37\%. For the trajectory curvilinear length estimation, the worst impact happens for high-sinuosity trajectories; for example, for a sawtooth-shaped trajectory, the total length of the trajectory may be off by as much as 50\%. Sinuosity is the ratio of the two, to which we assign a fractional uncertainty of $62\%$ after propagating the two fractional uncertainties. In the toy outlier analysis, instead of selecting trajectories with sinuosity greater than 3 we should be using a threshold of 1.14 if we account for this uncertainty. 
Thus, the number of flagged trajectories is assigned an uncertainty of $\sim$400\%. The manual inspection identifies a fraction $\alpha = \frac{N_s^{obs}}{N_{flag}}$ as ambiguous.
Since manual processes are not flawless, we assign  $\alpha$ an uncertainty of $10\%$. When propagating errors we use $\Delta N_s^{obs} = N_{flag} \Delta \alpha + \alpha \Delta N_{flag}$. We find that the signal count is $N_s^{obs} = 144 \pm 8763$.

\begin{figure}[H]
\includegraphics[width=9cm]{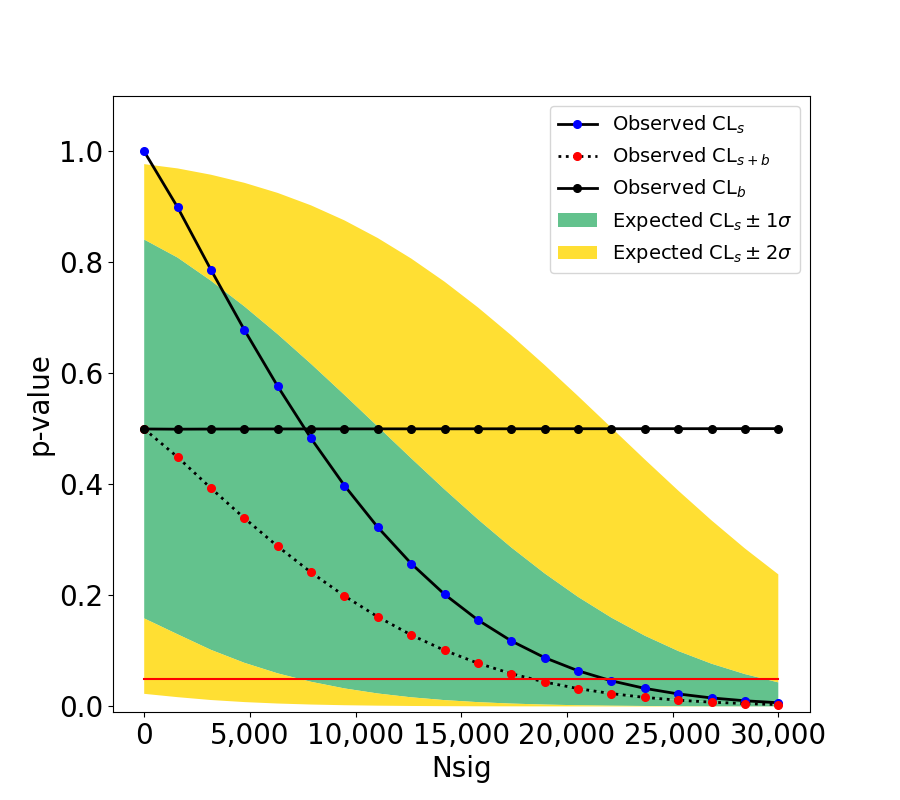}
\caption{\hl{Graphical} 
 representation of the observed and expected upper limit calculation. This shows the upper limit on the ambiguous outlier count at a 0.05 significance level (shown as a red line) for a counting experiment where $N_{b}= 627,339 \pm 324,668$ and $N_s = 180 \pm 10,999$, assuming Gaussian uncertainties. The observed upper limit at a 95\% confidence level is 18,271.\label{fig:brazil}}
\end{figure}

On a synthetic dataset without any clouds, rain, or high relative humidity, the identification recall metric for SORT was estimated at 87\% in Section~\ref{sec:sort_benchmark}. To be conservative, we set the overall efficiency of the pipeline (the fraction of true objects within our acceptance that are reconstructed) at $\epsilon = 80\pm 20$\%. After propagating the uncertainty from $\epsilon$ we unfold the number of background trajectories to $N_{b} = \frac{N^{obs} - N_s^{obs}}{\epsilon} = 627,339 \pm 324,668$ and the number of outlier trajectories to $N_s = \frac{N_s^{obs}}{\epsilon} = 180 \pm 10,999$. This is our estimate of the true background and the count of outlier objects (as defined by our toy outlier criteria) in the parameter space of apparent 2D trajectory sinuosity, within the Dalek's acceptance, if the detection pipeline was perfect.
Finally, we use the Python package \texttt{hepstats}~\cite{hepstats} to implement a likelihood inference using the Asimov dataset. We assume that the uncertainties on $N_s$ and $N_b$ have a Gaussian shape. We find an observed upper limit of 18,271 at a 95\% confidence level, for the number of apparent 2D trajectories with outlier sinuosity from objects having passed through the Dalek's detection volume and time window during the commissioning period, as shown in Figure~\ref{fig:brazil}.

\section{Discussion}


The $\sim$500,000 reconstructed aerial trajectories that we report from this commissioning period of five months give us insight into the volume of object parameter space that the Dalek is sensitive to. From the commissioning studies, we expect that the Dalek acceptance includes objects of apparent size larger than three~pixels, which corresponds to a standard airliner of wingspan 50~m at a distance of 10~km or a bird of wingspan 1.5~m at a distance of 300~m. Kinematics-wise, objects with a speed larger than 200 m/s at a distance of 1~km or a speed larger than 2 km/s at a distance of 10~km are outside the Dalek's acceptance based on both the frame rate and our analysis requirement of at least four points in a trajectory. The reconstructed aerial trajectories are analyzed with a toy outlier selection based on high sinuosity. About 16\% of trajectories are flagged as outliers and manually examined. A total of 144 ambiguous objects remain, which are likely mundane objects but cannot be elucidated without distance estimation and multi-modality at this stage of development. We demonstrate the application of a likelihood-based statistical test to evaluate the significance of this toy outlier analysis. Our observed count of ambiguous outliers combined with systematic uncertainties yields an upper limit of 18,271 outliers (toy outliers flagged as unusual) having passed through the Dalek's detection volume and time window during the commissioning period at a 95\% confidence level. This upper limit corresponds to $\sim$4\% of the objects we reconstructed and $\sim$22\% of the objects flagged as outliers based on their apparent trajectory sinuosity. For context, classified studies performed by governmental agencies such as the All-domain Anomaly Resolution Office (AARO), which are often able to leverage distance estimations from radar data and data from multiple modalities, reported that $\sim$3\% of the cases brought to their attention remained ambiguous. We expect that our upper limit will decrease in the future after we improve our detection pipelines and include multiple instruments and range estimation in the survey. This paper also paves the way for conducting a similar commissioning process for each of our observatory's instruments before conducting an in-depth, multi-modal aerial survey. 
This likelihood-based analysis is generalizable to quantifying uncertainties in classifying outliers in any metric obtained from the data collected; for example, for the infrared camera array, classifying outliers in apparent velocity, object size, or frequency of flashing lights. It can also be applied to data collected and metrics derived from other sensor modalities, such as color in visible light cameras or sound pressure levels in audio data. 
The observatory that this infrared camera array belongs to will conduct a long-term multi-modal aerial census of at least one year to capture seasonal variations. After that, we will conduct outlier searches in the dataset collected by the observatory in search of measurable anomalous phenomena. Any outlier search will be accompanied by an estimate of its statistical significance, including an evaluation of systematic uncertainties, and we plan to use the likelihood-based method that is applied in this paper.

\section{Conclusions}

This paper details the design and commissioning of an all-sky infrared camera array that is one component of a sky monitoring, multi-modal, ground observatory that we are developing to conduct a long-term aerial census in search of measurable anomalous aerial phenomena. We demonstrate novel uses of ADS-B-equipped aircraft to extrinsically calibrate each infrared camera and to quantify our reconstruction pipeline performance for aircraft. We evaluate our current reconstruction pipeline for detection and tracking of aerial objects, finding an overall reconstruction rate of 13.4\% for aircraft (of 27,467 aircraft in range, 3,678 
were detected) which constitutes a first baseline, to be improved on in the future. On synthetic datasets, we find that the YOLOv5 detection stage has a precision of 74\% and a recall of 63\%, while the SORT tracking stage has an identification precision of 95\% and an identification recall of 87\%. On a real-world dataset derived from ADS-B-equipped aircraft recordings, we find an acceptance rate (fraction of airplanes passing in the effective field of view of at least one camera that are recorded) of 41\% for ADS-B-equipped aircraft, and a mean frame-by-frame aircraft detection efficiency (fraction of recorded airplanes in individual frames which are successfully detected) of 36\%.  We see that (a) acceptance increases with field of view and camera uptime and (b) efficiency decreases with increased precipitation, reduced atmospheric visibility, increased relative humidity, and is reduced near the treeline and by dust and raindrops on the lens. The latter effects are not uniform over the cameras nor the images due to reduced detections from boundary effects of the detection algorithm and interference from treeline growth. On synthetic videos, we find that with the current detection and tracking pipeline (a) about 15\% of the true trajectories lack either frame-level object detection or trajectory reconstruction and are thus entirely missed, while (b) about 13\% of the reconstructed trajectories are spurious and do not correspond to any true trajectory. Additionally, the reconstruction of apparent speed is severely hampered by the fragmentation of trajectories, which causes a true trajectory to be broken on average into three smaller fragments of trajectory. Finally, we conduct a unimodal aerial census over the five months of commissioning data. We reconstructed approximately $\sim$500,000 trajectories of various aerial objects from this period. We demonstrated a method for the robust quantification of uncertainties in our search for aerial anomalies. The generalizable likelihood-based method can be used to measure the significance of an outlier search using any given data metric. In our toy example, we used the sinuosity of apparent 2D reconstructed object trajectories from the IR images. The commissioning studies outlined in this paper are a blueprint for the other instruments of our observatory, as well as for any other scientific aerial observatories. Once all of our instruments are commissioned, we will conduct a multi-modal long-term aerial survey and the significance  of any future outlier search during this aerial survey will be assessed using a generalized version of the likelihood-based method demonstrated in this paper.

\vspace{6pt} 
\supplementary{\hl{The following supporting} 
 information can be downloaded at:  \linksupplementary{s1}, Figure S1: Left: Camera background image, overlaid with snapshots from the objects' detections along their trajectories for two specific trajectories with IDs 39 and 306, identified as hawks after manual examination. Right: All reconstructed trajectories in a given day, including trajectories 39 and 306, summarized by assigning a unique color and identifier to each point of these trajectories; Figure S2: Left: Camera background image, overlaid with snapshots from the objects' detections along their trajectories for two specific trajectories with IDs 14 and 28, identified as hawks after manual examination. Right: All reconstructed trajectories in a given day, including trajectories 14 and 28, summarized by assigning a unique color and identifier to each point of these trajectories; Figure S3: Zoomed view of the individual object detections overlaid along different reconstructed trajectories. The red number is the unique identifier assigned to each trajectory by the tracking algorithm (SORT). All of these examples were identified as flocks of birds after manual examination; Figure S4: Zoomed view of the individual object detections overlaid along different reconstructed trajectories. The red number is the unique identifier assigned to each trajectory by the tracking algorithm (SORT). All of these examples were identified as flocks of birds after manual examination; Figure S5: Left: Camera background image, overlaid with snapshots from the objects' detections along their trajectories, for three specific trajectories with IDs 18, 37, and 115, which were all identified after manual examination as leaves. Right: All reconstructed trajectories in a given day, including trajectories 18, 37, and 115, summarized by assigning a unique color and identifier to each point of these trajectories.}

\authorcontributions{Conceptualization, A.D., L.D., S.L., and W.A.W.; methodology, E.K. (Eric Keto) and M.S.; software, E.M., R.C., A.F., M.S., and W.A.W.; formal analysis, L.D., R.C., A.F., and A.B.; resources, A.D., F.S., and A.W.; data curation, L.J.; writing---original draft preparation, L.D., M.S., R.C., A.F., L.J., S.L., and F.S.; writing---review and editing, M.P., M.S., S.L., and W.A.W.; supervision, A.L., W.A.W., and S.L.; project administration, E.K. (Ezra Kelderman) and A.L.; funding acquisition, A.L. All authors have read and agreed to the published version of the manuscript.}

\funding{This research was funded by private donations to the Galileo Project.}

\dataavailability{The raw data supporting the conclusions of this article will be made available by the authors on request.} 

\acknowledgments{The authors thank Cameron Pratt for his help with review and editing.}

\conflictsofinterest{\hl{Author F.S. was employed by the company Atlas Lens Co. The remaining authors declare that the research was conducted in the absence of any commercial or financial relationships that could be construed as a potential conflict of interest. The funders had no role in the design of the study; in the collection, analyses, or interpretation of data; in the writing of the manuscript; or in the decision to publish the results.}} 

 \appendixtitles{no} 
\appendixstart
\appendix
 \section[\appendixname~\thesection]{}
\vspace{-9pt}

\begin{table}[H]
\caption{Hours of the day when the sunshades are mechanically closed in summer (yellow---May, June, July), winter (blue---November, December, January), and equinox (green---all other months) modes, in UTC-4 
timezone. In orange, the schedule shows the hours when we are \textit{not} recording, in local timezone, which is UTC-4 
in the summer and UTC-5 
in the winter due to daylight saving. In future, we will align both schedules to be identical, but during this commissioning period of five months each camera had a different amount of ``lost'' recording time, i.e., recordings of blank frames because of closed sunshades. \label{fig:dalek_schedule}}

\renewcommand{\arraystretch}{1.2}
\renewcommand{\aboverulesep}{.1pt}
\renewcommand{\belowrulesep}{.1pt}
\begin{adjustwidth}{-\extralength}{0cm}
\begin{tabularx}{\fulllength}{lCCCCCCCCCCCCCCCCCCCCCCCC}
    \toprule
    \textbf{Hours} & \textbf{0} & \textbf{1} & \textbf{2} & \textbf{3} & \textbf{4} & \textbf{5} & \textbf{6} & \textbf{7} & \textbf{8} & \textbf{9} & \textbf{10} & \textbf{11} & \textbf{12} & \textbf{13} & \textbf{14} & \textbf{15} & \textbf{16} & \textbf{17} & \textbf{18} & \textbf{19} & \textbf{20} & \textbf{21} & \textbf{22} & \textbf{23} \\
    \midrule
    Camera 1 & \multicolumn{24}{c}{}\\
    \textit{Summer} & & & & & & & & & & & & & & \multicolumn{8}{c}{\cellcolor{yellow!25}} & & & \\ 
    \textit{Winter} & \multicolumn{24}{c}{} \\ 
    \textit{Equinox} & \multicolumn{24}{c}{} \\ 
    \textit{Recording} & & & & & & & & & & & & & & \multicolumn{8}{c}{\cellcolor{orange!25}} & & & \\
    \midrule    
     Camera 2 & \multicolumn{24}{c}{} \\ 
    \midrule
    Camera 3 & \multicolumn{24}{c}{}\\
    \textit{Summer} & & & & & & \multicolumn{5}{c}{\cellcolor{yellow!25}} & & & & & & & & & & & & & & \\
    \textit{Winter} & & & & & & & & \multicolumn{3}{c}{\cellcolor{blue!25}} & & & & & & & & & & & & & & \\
    \textit{Equinox} & & & & & & & & \multicolumn{3}{c}{\cellcolor{green!25}} & & & & & & & & & & & & & & \\
    \textit{Recording} & & & & & & \multicolumn{5}{c}{\cellcolor{orange!25}} & & & & & & & & & & & & & & \\
     \midrule
     Camera 4 & \multicolumn{24}{c}{}\\
    \textit{Summer} & & & & & & & \multicolumn{7}{c}{\cellcolor{yellow!25}} & & & & & & & & & & & \\
    \textit{Winter} & & & & & & & & \multicolumn{5}{c}{\cellcolor{blue!25}} & & & & & & & & & & & & \\
    \textit{Equinox} & & & & & & & & \multicolumn{5}{c}{\cellcolor{green!25}} & & & & & & & & & & & & \\
    \textit{Recording} & & & & & & & \multicolumn{7}{c}{\cellcolor{orange!25}} & & & & & & & & & & & \\
    \midrule
    Camera 5 & \multicolumn{24}{c}{}\\
    \textit{Summer} & & & & & & & & \multicolumn{7}{c}{\cellcolor{yellow!25}} & & & & & & & & & & \\
    \textit{Winter} & & & & & & & & \multicolumn{8}{c}{\cellcolor{blue!25}} & & & & & & & & & \\
    \textit{Equinox} & & & & & & & & \multicolumn{9}{c}{\cellcolor{green!25}} & & & & & & & & \\
    \textit{Recording} & & & & & & & & \multicolumn{7}{c}{\cellcolor{orange!25}} & & & & & & & & & & \\
    \midrule
    Camera 6 & \multicolumn{24}{c}{}\\
    \textit{Summer} & & & & & & & & & & & \multicolumn{10}{c}{\cellcolor{yellow!25}}  & & & & \\   
    \textit{Winter} & & & & & & & & & & & \multicolumn{6}{c}{\cellcolor{blue!25}}  & & & & & & & & \\
    \textit{Equinox} & & & & & & & & & & \multicolumn{8}{c}{\cellcolor{green!25}}  & & & & & & & \\
    \textit{Recording} & & & & & & & & & & & \multicolumn{10}{c}{\cellcolor{orange!25}}  & & & & \\
    \midrule
    Camera 7 & \multicolumn{24}{c}{}\\
    \textit{Summer} & & & & & & & & & & & & \multicolumn{10}{c}{\cellcolor{yellow!25}}  & & & \\ 
    \textit{Winter} & & & & & & & & & & & & & \multicolumn{4}{c}{\cellcolor{blue!25}}  & & & & & & & & \\
    \textit{Equinox} & & & & & & & & & & & & \multicolumn{6}{c}{\cellcolor{green!25}}  & & & & & & & \\
    \textit{Recording} & & & & & & & & & & & & \multicolumn{10}{c}{\cellcolor{orange!25}}  & & & \\
    \midrule
    Camera 8 & \multicolumn{24}{c}{}\\
    \textit{Summer} & & & & & & & & & \multicolumn{12}{c}{\cellcolor{yellow!25}}  & & & & \\ 
    \textit{Winter} & \multicolumn{24}{c}{} \\
    \textit{Equinox} & & & & & & & & & & & \multicolumn{6}{c}{\cellcolor{green!25}}  & & & & & & & &\\ 
    \textit{Recording} & & & & & & & & & & & \multicolumn{10}{c}{\cellcolor{orange!25}}  & & & & \\
    \noalign{\hrule height 1.0pt}
\end{tabularx}
\end{adjustwidth}

\end{table} 



\begin{adjustwidth}{-\extralength}{0cm}

\reftitle{References}





\PublishersNote{}
\end{adjustwidth}
\end{document}